
\documentstyle{amsppt}
\TagsOnRight
\catcode`\@=11
\def\logo@{}
\catcode`\@=13
\parindent=8 mm
\magnification 1200
\hsize = 6.25 true in
\vsize = 8.7 true in
\hoffset = .2 true in
\parskip=\medskipamount
\topmatter
\title
   Darboux Coordinates and Liouville-Arnold Integration
   in Loop Algebras
\endtitle
\author
  M.R. Adams${}^1$, J. Harnad${}^2$ and J. Hurtubise${}^3$
\endauthor
 \rightheadtext{Darboux Coordinates}
 \leftheadtext{Adams, Harnad and Hurtubise}
\endtopmatter
\footnote""{\noindent${}^\dag$Research
  supported in part by the Natural Sciences and Engineering Research
  Council of Canada and by U.S. Army Grant DAA L03-87-K-0110}
\footnote""{\noindent${}^1$Department of Mathematics, University of Georgia,
  Athens, Georgia}
\footnote""{\noindent${}^2$Department of Mathematics and
  Statistics, Concordia University, Montr\'eal, Qu\'e. and \newline
  Centre de recherches math\'ematiques,
  Universit\'e de Montr\'eal, C.P. 6128-A,
  Montr\'eal, Qu\'e. H3C 3J7}
\footnote""{\noindent${}^3$Department
  of Mathematics, McGill University, Montr\'eal, Qu\'e. and \newline
  Centre de recherches math\'ematiques,
  Universit\'e de Montr\'eal, C.P. 6128-A,
  Montr\'eal, Qu\'e. H3C 3J7}\def\rom{\roman}

\def \smaller {\eightpoint}
\def \wt {\widetilde}
\def \wh {\widehat}
\def \mt {\mapsto}
\def \ra {\rightarrow}

\def \lra {\longrightarrow}
\def \lmt {\longmapsto}
\def \a {\alpha}
\def \d {\delta}
\def \e {\epsilon}
\def \k {\kappa}

\def \l {\lambda}

\def \m  {\mu}

\def \s {\sigma}

\def \t {\tau}
\def \o {\omega}

\def \z {\zeta}

\def \ss {\subset}

\def \Lgp {\widetilde{\frak g}^+}

\def \AB {\bold{A}}

\def \BB {\Cal B}
\def \CC {\Cal C}
\def \DD {\Cal D}
\def \FF {\Cal F}
\def \II {\Cal I}
\def \JJ {\Cal J}
\def \KK {\Cal K}
\def \LL {\Cal L}
\def \MM {\Cal M}
\def \NN {\Cal N}
\def \LL {\Cal L}
\def \OO {\Cal O}
\def \PP {\Cal P}
\def \QQ  {\Cal Q}
\def \SS {\Cal S}
\def \TT {\Cal T}
\def \di {\partial}

\centerline{\bf Abstract}
\bigskip
\baselineskip=10pt
\centerline{
\vbox{
\hsize= 5.5 truein
{\smaller Darboux coordinates are constructed on rational coadjoint orbits of
the positive frequency part $\wt{\frak{g}}^+$ of loop algebras. These are
given by the values of the spectral parameters at the divisors corresponding to
eigenvector line bundles over the associated spectral curves, defined within a
given matrix representation. A Liouville generating function is obtained in
completely separated form and shown, through the Liouville-Arnold integration
method, to lead to the Abel map linearization of all Hamiltonian flows induced
by the spectral invariants. The results are formulated in terms of
sheaves to allow for singularities due to a degenerate spectrum. Serre duality
is used to define a natural symplectic structure  on the space of line bundles
of suitable degree over a permissible class of spectral curves, and this is
shown
to be equivalent to the Kostant-Kirillov symplectic structure on  rational
coadjoint orbits, reduced by the group of constant loops. A similar
construction involving a framing at infinity is given for the nonreduced
orbits.
The general construction is given for $\frak{g}=\frak{gl}(r)$ or
$\frak{sl}(r)$,
with reductions to orbits of subalgebras determined as invariant fixed point
sets under involutive automorphisms. As illustrative examples, the case
$\frak{g=sl}(2)$, together with its real forms, is shown to reproduce the
classical integration methods for finite dimensional systems defined on
quadrics, with the Liouville generating function expressed in hyperellipsoidal
coordinates, as well as the quasi-periodic solutions of the cubically nonlinear
Schr\"odinger equation. For $\frak{g=sl}(3)$, the  method is applied to the
computation of quasi-periodic solutions of the two component coupled nonlinear
Schr\"odinger equation. This case requires a further symplectic constraining
procedure in order to deal with singularities in the spectral data at
$\infty$.}}}
 \bigskip
\bigskip
\vfill

\newpage
\document
\baselineskip=14pt

\subheading{Introduction}
\medskip

   In a series of recent papers {\bf[AHP, AHH1, AHH2]}  isospectral
Hamiltonian flows on rational coadjoint orbits of loop algebras arising from
the Adler, Kostant, Symes (AKS) theorem were studied. A
systematic way of representing both finite dimensional
integrable Hamiltonian systems and quasi-periodic solutions of
integrable PDE's within this framework was developed using moment
map embeddings. A representation of such flows in terms of  rational matricial
functions of the loop parameter determines an invariant spectral curve and
associated linear flows of eigenvector line bundles {\bf [vMM, AvM]}. The
Dubrovin-Krichever-Novikov technique {\bf [K, KN, Du]}
may be used to  solve the equations of motion in terms of theta functions
{\bf [RS, AHH1]}, but the Hamiltonian content of this approach is not evident.

The link between the algebro-geometric method of integration and the
Hamiltonian  point of view  was made within the context of differential
algebras by Gel'fand and Dickey {\bf[GD, D]}, the linearization being based
upon the Liouville quadrature method. Within the loop algebra setting, however,
the symplectic content of the algebro-geometric integration method  has until
now only been developed in specific examples, without a general theory
encompassing  all generic cases.  Certain aspects, of course, are easy to see;
the constants of motion, for example, are essentially the coefficients of the
defining equations of the spectral curves, and their level sets are affine
subvarieties of the associated Jacobians. The algebro-geometric approach thus
gives a foliation of the phase space by tori, which may be seen in the
symplectic framework as Lagrangian. However the Hamiltonian content of the
actual integration procedure requires further clarification within the general
framework of loop algebras. (For related developments in the case of rapidly
decreasing boundary conditions, see  {\bf [FT, BS1, BS2]}, where the link
between the scattering transform and Darboux coordinates on Poisson Lie groups
is developed.)

   In this paper we use the Lie algebraic and algebro-geometric structures
associated with isospectral flows of the  AKS type on the dual
$\wt{\frak{g}}^{+*}$ of the positive frequency half of a loop algebra
$\wt{\frak{g}}$ to introduce a set of Darboux coordinates on
rational coadjoint orbits: the ``spectral divisor coordinates'' associated with
line bundles over spectral curves embedded in a specific ruled surface. These
essentially give a  complete separation of variables in  the Hamilton-Jacobi
problem for all AKS flows on such orbits - or, more precisely, a Liouville
generating function for the canonical transformation to linearizing coordinates
in completely separated form. The transformation is expressed in terms of
abelian integrals and hence this procedure derives the Abel map linearization
within the Hamiltonian setting provided by loop algebras, with the
Liouville-Arnold
torus identified with the Jacobi variety of the underlying spectral curve.
The general approach is developed only for the loop
algebras $\wt{\frak{gl}}(r)^{+}, \ \wt{\frak{sl}}(r)^{+}$, but the integrated
AKS flows for more general loop algebras may be obtained by restricting to
invariant symplectic submanifolds obtained as fixed point sets under involutive
automorphisms.

   Section 1 gives the  classical Hamiltonian  approach to the
introduction of spectral divisor coordinates on rational coadjoint
orbits and the construction of the completely separated Liouville generating
function (eq.~(1.76)). Certain partial reductions of the orbits that  arise in
most applications  are also dealt with. This section is essentially classical
in spirit, and may in principle  be understood  without going  beyond the tools
of nineteenth century mathematics (Riemann surfaces, abelian
integrals, Lie Poisson structures and Hamilton-Jacobi theory). The only
additional notion needed as a unifying factor is that of finite dimensional
Poisson subspaces of the dual of a loop algebra, consisting of elements
that are rational in the loop parameter. Such spaces can, however, be given
a natural Poisson structure without any further familiarity with loop algebras.

   More specifically, the spectral curve $\Cal S$  of $\Cal N(\l)$, an
$r \times r$ matrix valued  rational function  of the complexified loop
parameter $\l$, is determined by the characteristic equation
$$
 \text{det}(\Cal N(\l) - \l \z I) =0  ,   \tag{0.1}
$$
and so sits naturally in ${\Bbb C}^2$ or, after compactification, in an ambient
ruled surface $\Cal T$. The line bundle corresponding to $\Cal N(\l)$ is
associated to
a divisor on the curve $\Cal S$, given by the finite zeroes of a component of
the eigenvector, and hence determines a set of functions given by
the corresponding coordinate pairs $(\l_\mu,\z_\mu)_{\mu=1, \dots g}$
($g=$ genus of $\Cal S$) on the ambient surface.  (The bundle in question is
actually of degree $g+r-1$, but the associated divisor can be normalized with
$r-1$ points chosen over $\l=\infty$.) The key point in understanding the link
between the symplectic and algebraic geometry is that the set of functions
$(\l_\mu,\z_\mu)_{\mu = 1, \dots g}$ turn out in general to form a Darboux
coordinate
system on a reduced version of the coadjoint orbits  (Theorem 1.4),
and can very simply be augmented using invariants related to the spectrum
over $\l=\infty$ to form a Darboux system on  the full orbits (Theorem 1.5).
{}From this fact, the canonical linearization of flows induced by the spectral
invariants leads, through the Liouville-Arnold method (eqs.~(1.82a,b)),
 to the Abel map (Theorem 1.6) and hence, to
$\theta$-function formulae for the integrated flow (Corollary 1.7).

    The development  of Section 1 is essentially
self-contained and sufficiently explicit  to lead directly to
the examples of Section 3, but a deeper understanding of the underlying
approach requires the  constructions of Section 2. These are aimed at an
intrinsic explanation of the underlying symplectic structure from an
algebro-geometric viewpoint and involve  some  more modern machinery,
such as  Serre duality. It also is convenient to formulate the results in
terms of sheaves, in order to have a  setting in
which smooth and singular curves may be treated on an equal footing. Such
a generalization is required to allow for the types of spectra
occurring in the rational matrix valued  functions that arise in some of  the
more
interesting applications, such as the coupled nonlinear Schr\"odinger (CNLS)
equation.

	The naturally defined class of spectral divisor coordinates, leads
to the following intrinsic characterization, identifying
the (Lie algebraic) Kostant-Kirillov symplectic form on rational coadjoint
orbits in  terms of purely algebro-geometric data. On the reduced orbits, the
infinitesimal variations of the spectral curves correspond to sections of the
normal bundle of $\Cal S \ss \Cal T$, constrained to vanish at the poles of
$\Cal N(\l)$. The space of such sections may be identified with
$V:= H^0(\Cal S, K_{\Cal S})$; i.e. to sections of the canonical bundle.
On the other hand, variations of the line bundles, representing the tangent
space to the isospectral foliation, are given by the cohomology
group $W := H^1(\Cal S, \Cal O)$, and Serre duality tells us that
$W=V^{*}$.  The tangent space to the reduced orbit is thus identified
with $V \oplus V^*$,  which leads to a natural symplectic form
corresponding to this decomposition. The remarkable fact is that this
coincides with the Kostant-Kirillov form (Theorem 2.7) on the orbits reduced
under the action of the subgroup of constant loops. A similar
construction, involving supplementary data over $\l = \infty$, holds on the
nonreduced orbits as well (Theorem 2.8). The AKS theorem, together with this
identification between Lie algebraic and algebro-geometric symplectic forms,
seems to be at the root of the ubiquitous presence of algebro-geometric
constructions in the theory of integrable systems.

     Although the main contents of Sections 1 and 2 concern the loop algebras
$\wt{\frak{gl}}(r)^+$ or $\wt{\frak{sl}}(r)^+$, reductions
to other algebras, obtained as fixed point sets under involutive automorphisms,
are also placed in the symplectic framework (Theorems 2.9-2.11). This allows
a determination of the flows induced by spectral invariants through
restriction to the corresponding invariant symplectic submanifolds, but does
not produce an intrinsic formulation in terms of separating Darboux coordinates
on the coadjoint orbits of these subalgebras. The analogous coordinates
 have yet to be derived in the general case, though one obtains, by restriction
from the case  $\wt{\frak{gl}}(r)^+$, extensions of the results of Section 2
applicable to these subalgebras.

In Section 3,  these results are illustrated in  a number of
examples. As a first application, Darboux coordinates are computed on generic
coadjoint orbits of the Lie algebras $\frak{sl}(2,\Bbb C)$ and
 $\frak{sl}(3,\Bbb C)$, viewed as rational coadjoint orbits of the
corresponding loop algebras having only one simple pole. Next, the
$\wt{\frak{sl}}(2)$ case with $n$ simple poles is shown to reproduce the
standard linearization results for well-known classical examples of
finite dimensional systems (cf\. {\bf [M]}).
For this case, the ``spectral divisor coordinates'' are essentially just the
hyperellipsoidal coordinates, the reductions corresponding to fixing spectral
data at $\infty$ lead to constraints defined by quadrics, and the spectral
curve
 is always hyperelliptic.  This case also includes the ``finite gap''
quasi-periodic solutions of familiar sytems of PDE's such as the cubically
nonlinear Schr\"odinger (NLS) equation. The higher rank  ``spectral divisor
 coordinates'' are thus really generalizations of hyperellipsoidal coordinates.

Finally, as an illustration of the $\wt{\frak{sl}}(3)$ case, involving trigonal
curves, the finite gap solutions of the coupled $2$-component nonlinear
Schr\"odinger (CNLS) equation are also obtained by the Liouville-Arnold
integration technique. In this case the  particular structure of the
spectrum at infinity leads to further singularities in the curve,
and hence a decrease of the arithmetic genus  relative to the
generic case, and an incomplete set of Darboux coordinates on the
coadjoint orbit. This example is used to indicate how such problems may be
dealt
 with by restricting to an invariant symplectic submanifold on which the
spectral curves share the same generic type of singularities. The linearization
on the constrained manifold then proceeds in the same way as in the
unconstrained case.
\bigskip
\noindent {\it  Background and Acknowledgements:} For a more complete account
of
the moment map construction leading to isospectral flows on rational coadjoint
orbits and the algebro-geometric method of integration, the reader
should consult {\bf [AHP, AHH1, AHH2]}. The present work is the first complete
account of the spectral Darboux coordinate construction and the
Liouville--Arnold integration method on loop algebras, but earlier summaries
and announcements of the main results communicated at various conferences and
workshops may be found in  {\bf [H, AHH3, AHH4, AHH5]}. The authors are pleased
to acknowledge helpful discussions  with  L. Dickey, B. Dubrovin, H. Flaschka,
P. van Moerbeke, E. Previato and A. Reyman relating to this material.

\bigskip
\bigskip
\vfill
\newpage
\subheading{1. Darboux Coordinates and Linearization of Flow}
\medskip \noindent
{\it 1a.\quad Rational Orbits and Spectral Curves \hfill}

  The Hamiltonian systems to be considered here involve isospectral flows of
matrices determined by equations of Lax type:
$$
\frac{d\NN(\l)}{dt} =[\BB(\l), \NN(\l)], \tag{1.1}
$$
where $\NN(\l),\ \BB(\l)$ are $r \times r$ matrices depending on a complex
parameter $\l$. The matrix $\NN(\l)$ is taken to be of the form
$$
\NN(\l) = \l Y + \l \sum_{i=1}^n \frac{N_i}{\l -\a_i},  \tag{1.2}
$$
where $Y \in \frak{gl}(r)$, and $\{\a_i \in \Bbb C\}_{i=1, \dots,n}$ are
constants.
Thus, we are considering rational $\NN(\l)$ with fixed, simple poles at the
finite points $\{\a_i\}$ and possibly at $\infty$. Rational
matrices with higher order poles may be dealt with similarly, but will not be
considered here for the sake of notational simplicity.

The particular form (1.2) arises naturally as the translate by $\l Y$
of the image
$$
\NN_0(\l) = \l \sum_{i=1}^{n}\frac{N_i}{\l -\a_i}   \tag{1.3}
$$
of a moment map from a symplectic vector space parametrizing rank--$r$
perturbations of a fixed $N \times N$ matrix with eigenvalues
$\{\a_i\}_{i=1, \dots n}$ into the dual $(\Lgp)^*$ of a loop
algebra, represented by $r\times r$ matrix functions of the complexified
loop parameter $\l$, holomorphic in a suitable domain [{\bf AHP, AHH2}].
This serves to embed a large class of integrable systems as Lax pair flows
in $(\Lgp)^*$. The image space for such maps is a Poisson subspace of
$(\Lgp)^*$, with respect to the Lie Poisson structure, the symplectic leaves
(coadjoint orbits) consisting of rational functions of $\l$. Since a specific
$r \times r$ matrix representation is involved, we view $\frak{g}$ as a
subalgebra of $\frak{gl}(r,\Bbb{C})$ or $\frak{sl}(r,\Bbb{C})$, obtained
generally by reductions under involutive automorphisms (cf.~ Section 2c).

          The loop algebra elements $X\in \wt{\frak{gl}}(r)$ are viewed as
smooth maps $X:S^1 \mt \frak{ gl}(r)$ from a fixed circle $S^1$ in the
complex $\l$ - plane, containing the points $\{\a_i\}$ in its interior, and the
subalgebra $\wt{\frak{gl}}(r)^+$ consists  of those $X(\l)$ that extend as
holomorphic functions to the interior of $S^1$.  The loop group
$\wt{Gl}(r)$  similarly consists of smooth maps $g:S^1 \mt Gl(r)$, while the
subgroup $\wt{Gl}(r)^+$ consists again of those $g(\l)$ that extend
holomorphically inside $S^1$.  The subspace
$\wt{\frak{gl}}(r)_-\ss\wt{\frak{gl}}(r)$ of loops extending holomorphically
outside $S^1$ to $\infty$ is identified with a dense subspace of the dual space
$\wt{\frak{gl}}(r)^{+*}$ through the  dual pairing:
$$
 \align <\mu, X> &:=
\frac{1}{2\pi i} \oint_{S^1}\text{tr}\left(\mu(\l)X(\l)\right)\frac{d\l}{\l} ,
\tag{1.4}
\\ \mu \in \wt{\frak{gl}}(r)_- &\ , \ X\in \wt{\frak{gl}}(r)^+.
\endalign
$$
The matrix $\BB(\l)$ has the form:
$$
\BB(\l) = (d\Phi(\NN(\l)))_+, \tag{1.5}
$$
where $\Phi \in \II(\wt{\frak{gl}}(r)^*)$ is an element of the ring of
$Ad^*$-invariant polynomials on $\wt{\frak{gl}}(r)^*$ and the subscript $+$
 means projection to the subspace $\wt{\frak{gl}}(r)^+$. In general, no
notational distinction will be made between $\wt{\frak{gl}}(r)^{+*}$ and
$\wt{\frak{gl}}(r)_-$. The coadjoint action of $\wt{Gl}(r)^+$ on rational
elements $\NN_0$ of the form (1.3) is given by:
$$
\align
g : \wt{\frak{gl}}(r)_- &\lra \wt{\frak{gl}}(r)_- \\
g : \l\sum_{i=1}^{n}\frac{N_i}{\l -\a_i} & \lmt \l\sum_{i=1}^{n}
\frac{g(\a_i)N_ig(\a_i)^{-1}}{\l -\a_i} .   \tag{1.6}
\endalign
$$

Equation (1.1) is Hamilton's equation on the coadjoint orbit
${\QQ}_{\NN_0} \ss \wt{\frak{gl}}(r)^{+*} $, with respect to the orbital
(Kostant-Kirillov) symplectic form $\o_{\text{orb}}$, corresponding to the
Hamiltonian:
$$
 \phi(\mu) = \Phi(\mu + \l Y).    \tag{1.7}
$$
The Poisson commutative ring of such functions on $\QQ_{\NN_0}$ will be denoted
$\FF_Y$. According to the ``shifted'' version {\bf [FRS]} of the
Adler-Kostant-Symes theorem [{\bf A, Ko, S}], such systems generate
commuting Lax pair flows. Moreover, they may be shown to be completely
integrable on ``generic'' coadjoint orbits [{\bf RS, AHP, AHH1}] in
 $\wt{\frak{gl}}(r)^{+*}$. On such orbits, the (AKS)  ring of commuting
invariants
is generated by the coefficients of the characteristic polynomial of $\NN
(\l)$.

In analyzing the spectrum, it is convenient to deal with matricial polynomials
in $\l$, so
we define $$
\align
\hat{\LL} &:= \frac{a(\l)}{\l}\NN (\l) \\
{}&=Y a(\l) + L_0\l^{n-1} + \dots + L_{n-1},  \tag{1.8}
\endalign
$$
where
$$
a(\l) := \prod_{i=1}^{n}(\l-\a_i). \tag{1.9}
$$
The matrix
$$
L_0 = \lim_{\l \to \infty} \NN_0 (\l)  =\sum_{i=1}^{n}N_i  \tag{1.10}
$$
may be viewed as a moment map generating the conjugation action of $Gl(r)$ on
$\wt{\frak{gl}}(r)^{+*}$:
$$
\align
Gl(r) \times \wt{\frak{gl}}(r)^{+*} &\lra \wt{\frak{gl}}(r)^{+*}\\
(g, \ X(\l)) &\lmt gX(\l) g^{-1} .   \tag{1.11}
\endalign
$$
The matrix $\hat{\LL}$ satisfies the same Lax equation (1.1) as $\NN (\l)$,
and the coefficients of its characteristic polynomial:
$$
\PP (\l, z) := \text{det} (\hat{\LL} (\l) - z I) \tag{1.12}
$$
generate the same ring of invariants as that of $\NN (\l)$.

\noindent {\it Remark:}\ It is also possible
to view
$$
\l^{-n+1}[\hat{\LL} - Y a(\l)] :=\LL(\l)  \tag{1.13}
$$
directly as an element of an  orbit in $\wt{\frak{gl}}(r)^{+*}$ (polynomial
in $\l^{-1}$). Since the ring of invariants is the same, the results are
equivalent, with a suitable redefinition of the Hamiltonians and
parametrization of the spectral curve (cf.~ [{\bf AHP}]). We retain our present
conventions, with $\NN_0(\l)$ viewed as the point in $\wt{\frak{gl}}(r)^{+*}$
undergoing Hamiltonian flow, since these are adapted to examining the
particular spectral constraints occurring at the finite values
$\{\l = \a_i\}$ that appear in specific examples (cf.~Section 3).

The spectral curve $\SS_0 \ss \Bbb{C}^2$ defined by the characteristic equation
$$
\PP (\l, z) = 0 \tag{1.14}
$$
is invariant under the AKS Hamiltonian flows. Let $m$ be the degree of
$\hat{\LL}(\l)$, ($m=n$ if $Y \neq 0$  or $m=n-1$ if $Y=0$) and  let $\{k_i\}$
denote
the ranks of the matrices $\{N_i\}_{i=1, \dots n}$ in (1.2) (coadjoint
invariants, and
hence invariants of any Hamiltonian flow in $\QQ_{\NN_0})$.
\proclaim {Lemma 1.1} The spectral polynomial $\PP (\l,z)$ has the form:
$$
\PP (\l,z) =  (-z)^r + z^{r-1}\PP_1(\l) + \sum_{j=2}^rA_j(\l)\PP_j(\l)z^{r-j},
\tag{1.15}
 $$
where
$$
A_j(\l) :=\prod_{i=1}^n(\l-\a_i)^{\text{max}(0, j-k_i)}  \tag{1.16}
$$
and
$$
\text{deg}\ \PP_j(\l) = \sum_{i=1}^n\text{min}(j,k_i)\  - j(n-m)
=: \k_j  \tag{1.17}
$$
\endproclaim
\noindent {\it Remark:} \ This means that $\PP(\l,z) $, and all its partial
derivatives in $\l$ or $z$ up to order $r-k_i-1$ vanish at $(\a_i,0)$.
\demo{Proof} This
follows immediately by expanding  $\text{det}(\hat{\LL}(\l) -zI)$ and using the
fact
that  $\hat{\LL}(\a_i) = N_i\prod_{j=1, j\neq i}^n (\a_i - \a_j)$ has rank
$k_i$.
\hfill$\square$
\enddemo
 The structure of $\PP (\l, z)$ implies that on $\SS_0$, $z \sim O(\l^m)$ as
$\l \ra
\infty$. This suggests assigning $z$ a homogeneity degree $m$,  thereby
giving $\PP (\l, z)$  an overall degree $rm$. We may then compactify $\SS_0$,
regarding it as the affine part of an $r-$sheeted branched cover of $\bold
P^1$, by embedding it in the total space $\TT$ of $\OO (m)$, the $m$th
power of the hyperplane section bundle over $\bold P^1$, whose sections are
homogeneous functions of degree $m$ (cf.~ [{\bf AHH1}] and Sec.~ 2).
The pair $(\l, z)$ is viewed as the base and fibre coordinates over the affine
neighborhood $U_0 := \pi^{-1}( \bold P^1 - \{\infty\})$. Over $U_1 :=
\pi^{-1}(\bold P^1 - \{0\})$, we have coordinates  $(\tilde{\l}, \tilde{z})$
related to $(\l,z)$ on $U_0 \cap U_1$ by:  $$
 \tilde{\l} = \frac{1}{\l}\ , \quad
\tilde{z} = \frac{z}{\l^m} . \tag{1.18}
$$
Re-expressing (1.14) as a polynomial equation in $(\tilde{\l}, \tilde{z})$
extends $\SS_0$ to $U_1$, thereby defining its   compactification
$\SS \ss \TT$.

Let us assume that $\SS$ has no multiple components. Let $(\l_0,z_0)$
belong to $\SS$, and suppose that the multiplicity of the eigenvalue
$z_0$ of $\hat{\LL}(\l_0)$  is $k>1$. It follows from the constructions of
[{\bf AHH1}] (cf.~ also Section 2) that there is a partial  desingularisation
$\wt {\SS}$ of $\SS$ such that, generically, the number of points  (with
multiplicity) in $\wt{ \SS}$ over $(\l_0,z_0)$  equals the number of Jordan
blocks of $\hat{\LL}(\l_0)$ with eigenvalue $z_0$, and $\wt{ \SS}$ is smooth
over $(\l_0,z_0)$. If, for
example, $\hat{\LL}(\l_0)$ has only one Jordan block of size $k$ with
eigenvalue $z_0$, then $\wt{ \SS} = \SS$ and, generically, $\SS$ has a smooth
$k$-fold branch point  over $\bold P^1$. In the opposite extreme, if
$\hat{\LL}(\l_0)$ has $k$ independent eigenvectors with eigenvalue $z_0$
then generically there are  $k$ points  (with multiplicity) over $(\l_0,z_0)$
  and  $\SS$ has a k-fold node.

These remarks are of particular importance when $\l_0= \a_i$, since the Jordan
form
 of $\hat{ \LL}(\a_i)$ is an invariant of the coadjoint orbit. Thus, if
$\hat{\LL}(\a_i)$
is diagonalisable with multiple eigenvalues, the generic spectral curve for the
orbit will
be singular.

\noindent {\it Genericity Conditions}

\nopagebreak
 In what follows, we  only consider the singularities that follow from the
specific structure (1.2), (1.8) assumed for  $\hat{\LL}(\l_0)$. We shall make
the simplifying assumption that the $N_i$ (and hence $\hat{\LL}(\a_i))$) are
diagonalizable, with the only multiple eigenvalue being $z=0$, with
multiplicity
$r-k_i$. This property is, of course, ``generic'' for orbits with
$\text{rank}(N_i)=k_i$, but is only assumed in order to simplify the
exposition.
 If other Jordan forms are allowed for the $N_i$'s, the only effect is to
change the specific form (1.22) for the spectral polynomial $\PP(\l,z)$, (1.27)
for the genus formula determining the dimension of $\QQ_{\NN_0}$ and the
explicit expressions (1.83), (1.85) for the abelian differentials. All these
can easily be modified to hold for other cases. The main results, contained in
Theorems 1.3-1.6, Corollary 1.7 and the subsequent sections, remain valid
{\it mutatis mutandis}.

We also assume that one of the following two conditions hold:

 \noindent
{\it Case (i):} $Y=0$ and $L_0$ has a simple spectrum ($m=n-1$).

 \noindent
{\it Case (ii):} $Y\neq 0$ and has a simple spectrum ($m=n$).

\noindent
 Again, these conditions are generic and invariant on coadjoint orbits, but in
section 3 it will be indicated how they may be relaxed.

 Finally, we make a further spectral genericity assumption regarding the
singularities of the curve $\SS$; namely, that the only singularities occur at
the points $(\a_i,0)$, where there is an $r-k_i$-fold node with $r-k_i$
distinct branches intersecting transversally. This amounts to requiring that
the eigenspaces of $\hat{\LL}(\l)$ all be $1$-dimensional except at $\l=\a_i$,
 where, by the structure of $\NN_0(\l)$, the eigenvalue $z=0$ has an eigenspace
of dimension $r-k_i$. The desingularization $\wt{\SS}$ is then smooth and is
isomorphic to $\SS$ away from $(\a_i,0)$. This condition  is generic in the
space of $\NN_0$'s of the form  (1.3) and, if satisfied at any point of
${\QQ}_{\NN_0}$, it is also valid in a neighborhood of the isospectral
manifold through that point. (In particular, it is invariant under the AKS
flows.)

The coefficients of the polynomials $\PP_j(\l)$ generate the AKS ring on each
coadjoint orbit $\QQ_{\NN_0} \ss \wt{\frak{gl}}(r)^{+*}$ and should be viewed
as functions on the Poisson submanifold consisting of rational elements of the
form (1.3) (with rank$(N_i) = k_i$). Note that
$$
\PP_1(\l) = \text{tr} \hat{\LL}(\l), \tag{1.19}
$$
and hence its coefficients are Casimir invariants (i.e. constants on all
coadjoint orbits). The nonzero eigenvalues
$\{z_{i\k}\}_{\ i= 1, \dots n, \k=1,\dots k_i}$ over the points
$\{\l = \a_i\}_{i= 1, \dots n}$ are also Casimir invariants, since they are
determined as the nonzero roots of the characteristic equation:
$$
\text{det}[N_i\prod_{j=1, j\neq i}^n (\a_i - \a_j) \  -z I] = 0,   \tag{1.20}
$$
which is invariant under the coadjoint action (1.6). The $N :=\sum_{i=1}^n
k_i$
trivial invariants $\{z_{i\k}\}$ determine, in particular,  the coefficients of
$\PP_1(\l)$, since $$
\sum_{\k=1}^{k_i}z_{i\k} = \text{tr}\ \hat{\LL}(\a_i) = \PP_1(\a_i), \quad
i=1, \dots n .   \tag{1.21}
$$
(For case (i), this is sufficient to determine the degree $n-1$ polynomial
$\PP_1(\l)$; for case (ii), the degree $n$ coefficient is just tr $Y$.)
This may all be summarized by noting that the spectral curves $\wt {\SS}$ on
the orbit $\QQ_{\NN_0}$ are constrained to pass through the $N+n$ points
$\{(\a_i, z_{i\k}), (\a_i, 0)\}$, with $r-k_i$ branches intersecting at the
singular points $\{(\a_i, 0)\}$, the values $\{z_{i\k}\}$ being fixed.
 It should also be noted that for $Y \neq 0$ the leading (deg $\k_i$) terms
in the polynomials $\PP_j(\l)$ are constants, determined entirely by the
symmetric invariants of $Y$. For $Y=0$, the leading terms are not constants,
but they are determined as symmetric invariants of $L_0$, and hence are
constant on its level sets.

  A way to express $\PP(\l,z)$ in terms of independent, non-Casimir
invariants is to choose a reference point $\NN_R \in \QQ_{\NN_0}$ on the orbit
and parametrize the difference between $\PP(\l,z)$ and its value $\PP_R(\l,z)$
at $\NN_R$.
\proclaim {Proposition 1.2} In a neighborhood of the point $\NN_R \in
\QQ_{\NN_0}$, the characteristic polynomial has the form:
$$
\PP(\l,z) \equiv \PP_{R}(\l,z) + a(\l)\sum_{j=2}^{r}a_j(\l)p_j(\l)z^{r-j}
\tag{1.22}
$$
where
$$
\align
a_j(\l) & = \prod_{i =1}^{n}(\l-\a_i)^{\text{max}(0, j-k_i-1)},  \tag{1.23}
\\
p_j(\l) & =: \sum_{a=0}^{\d_j}P_{ja} \l^a  \tag{1.24}
\endalign
$$
and $\{p_j(\l)\}_{j=1,\dots r}$ are polynomials of degree:
$$
\align
\d_j &\equiv \text{deg}\ p_j(\l) = \cases d_j-j  \quad &\text{if} \quad Y=0 \\
 d_j  \quad & \text{if} \quad Y \neq 0
\endcases \tag{1.25a} \\
d_j& \equiv \sum_{i=1}^{n}\text{min}(j-1, k_i ).   \tag{1.25b}
\endalign
$$

For $Y=0$, the leading coefficients $P_{j\d_j}$ are constant translates of the
elementary symmetric invariants of $L_0$, while for $Y\neq 0$, the leading
coefficients $P_{j\d_j}$ are all constants; namely, the elementary symmetric
invariants of $Y$ (translated by the corresponding leading terms in
$\PP_{R}(\l,z)$). The number of spectral parameters $\{P_{ja}\}$,
($a=0,\dots \d_j + n-m - 1,\  j=2, \dots r$) defining the polynomials $p_j(\l)$
on generic orbits is thus: $$ \align
d &\equiv \sum_{j=2}^{r} \left (d_j - (n-m)(j - 1)\right) \\
&= \wt{g} +r - 1,  \tag{1.26}
\endalign
$$
where
$$
\wt{g} = \frac{1}{2}(r-1)(mr-2)- \frac{1}{2}\sum_{i=1}^{n}(r-k_i)(r-k_i -1).
\tag{1.27}
$$
In a neighborhood of any generic point on $\QQ_{\NN_0}$, these spectral
invariants are all independent.
\endproclaim
\demo {Proof} The structure of $\PP(\l)$ follows Lemma 1.1, plus the fact
that $\PP(\l,z) - \PP_R(\l,z)$ vanishes at each $\l=\a_i$, while $z$ vanishes
at least linearly in $\l -\a_i$ along each branch through $(\a_i,0)$.
{}From formula (1.6) and the above genericity conditions regarding the
residues $N_i$, the dimension of the coadjoint orbit $\QQ_{\NN_0}$ is
easily computed to be $2d$. From the proof of complete integrability of the AKS
flows on such orbits
given in {\bf [AHH1]}, it follows that the isospectral foliation is Lagrangian,
 and hence the $d$ spectral parameters $\{P_{ja}\}$ are independent. The
expression of $P_{j\d_j}$ in terms of the elementary symmetric invariants of
 $L_0$ or $Y$ follows directly from the fact that the leading term in
$\hat{\LL}(\l)$ in eq.~ (1.12) is either $L_0\l^{n-1}$ or $Y\l^n$.
\hfill$\square$
 \enddemo
It follows from the adjunction formula applied to the curve $\wt{ \SS }$
obtained by blowing up $\TT$ once at each point $(\a_i,0)$  (cf.~[{\bf AHH1},
 {\bf GH}]) that $\wt g$  in eq.~(1.27) is also equal to the (arithmetic)
genus of $\wt{ \SS }$.  If we reduce such an orbit under the $Gl(r,\Bbb C)$
action (1.9) for case (i), or the action of the stabilizer
$G_Y \ss Gl(r, \Bbb C)$ of $Y$ for case (ii), the dimension of the reduced
space is precisely $2\wt g$, and the projected spectral invariants again
define completely integrable Hamiltonian systems [{\bf AHH1}].
These facts suggest exploiting the orbital symplectic structure further
so as to explicitly integrate the isospectral flows via Hamiltonian methods.
This
will be the content of the following subsections.
\medskip \noindent
{\it 1b. Divisor Coordinates on Reduced Orbits  \hfill}

  Define
$$
\KK(\l,z) := \hat{\LL}(\l)- z I, \tag{1.28}
$$
and let $\wt{\KK}(\l,z)$ denote its classical adjoint (matrix of cofactors).
Let $V_0 \in \Bbb C^r$ be an eigenvector of $L_0$ in case (i), or of $Y$ in
case
(ii). From the results of [{\bf AHH1}], it follows that the set of polynomial
equations:
$$
\wt{\KK}(\l,z)V_0 = 0  \tag{1.29}
$$
have, away from $(\a_i,0)$, precisely $\wt{g}$ generically distinct finite
solutions $\{(\l_{\mu}, z_{\mu})\}_{\mu = 1, \dots \tilde{g}}$ that may be
viewed as functions on the coadjoint orbit $\QQ_{\NN_0}$. (Changing to the
coordinates $(\tilde{\l},\tilde{z})$, there are also $r-1$ further solutions
with  $\tilde{\l} = 0$, i.e., $\l=\infty$. If $V_0$ is not chosen as an
eigenvector of $L_0$ or $Y$, the remaining $r-1$ solutions will generically
also be at finite values of $(\l,z)$.)

  The significance of these functions in terms of the algebraic geometry of the
 spectral curves $\wt{\SS}$ may be summarized as follows (cf.~[{\bf AHH1}] and
Section 2a below for the detailed construction). To each matricial polynomial
$\hat{\LL}(\l)$ is associated a degree $\wt{g} + r - 1$ line bundle
$\wt E \ra \wt{\SS}$ over the partly desingularized spectral curve $\wt{\SS}$.
Away from the degenerate eigenvalues this coincides with the dual of the bundle
 of eigenvectors of $\hat{\LL}^{T}(\l)$ over $\SS$. At a smooth point $(\l,z)$
of $\SS$, the fibre of $\wt E$ is the cokernel of the map $\KK(\l,z)$:
 $$
0 \lra \Bbb C^r @>{\KK (\l,z)}>> \Bbb C^r \lra \wt E \lra 0 .    \tag{1.30}
 $$
More generally,  this exact sequence defines the direct image of
$\wt E$ over $\wt{\SS}$ (cf.~ Sec.~ 2a).  Vectors $V_0$ in $\Bbb C^r$ then give
 sections of $\wt E$  by projection. These sections vanish precisely at the
points where $V_0$ is in the image of $\KK (\l,z)$. Since
$\KK (\l,z)\wt {\KK} (\l,z) = \PP(\l,z) I$, this is  equivalent to (1.29), at
least over the open set of points in $\SS$  corresponding to
nondegenerate eigenvalues,  for  which the corank of ${\KK (\l,z)}$ is one.
{}From [{\bf AHH1}], the degree of  $\wt E$ is $\wt g + r-1$,  so sections of
$\wt E$ have $\wt g+r-1$ zeroes. The choice of $V_0$ as an eigenvector of
the leading term in $\hat{\Cal L}(\l)$ implies that $r-1$ of
these are over $\l=\infty$, and the coordinates of the remaining $\wt g$
points are the finite solutions $\{ (\l_\mu,z_\mu)\}_{\mu=1,..,\wt g}$.

   In evaluating Poisson brackets, it is preferable to introduce another
normalization, corresponding to the eigenvalues of $\NN(\l)\over{\l}$ rather
than $\hat{\LL}(\l)$, by defining:
$$
\z := \frac{z}{a(\l)}  \tag{1.31}
$$
and
$$
\MM(\l, \z) := \frac{\NN(\l)}{\l} - \z I,  \tag{1.32}
$$
with classical adjoint $\wt{\MM}(\l, \z)$. Then
$$
\wt{\KK}(\l,z) = [a(\l)]^{r-1} \wt{\MM}(\l, \z)  \tag{1.33}
$$
and eq.~ (1.29) is equivalent to:
$$
\wt{\MM}(\l,\z)V_0 = 0.  \tag{1.34}
$$
The $\wt{g}$ solutions $\{(\l_\mu, z_\mu)\}_{\mu = 1, \dots \tilde{g}}$ are
thus related to the solutions $\{(\l_{\mu}, \z_\mu)\}_{\mu = 1, \dots
\tilde{g}}$ of (1.34) by:
$$
\z_{\mu} = \frac{z_\mu} {a(\l_\mu)}.   \tag{1.35}
$$

   Viewing $\{(\l_\mu, \z_\mu)\}_{\mu = 1, \dots \tilde{g}}$ as functions on
$\QQ_{\NN_0}$, we may evaluate their Poisson brackets with respect to the
orbital (Kostant-Kirillov) symplectic structure $\o_{orb}$.
\proclaim {Theorem 1.3} The Poisson brackets of the functions
$(\l_\mu, \z_\mu)_{\mu = 1, \dots \tilde{g}}$ are:
$$
\{\l_\mu, \l_\nu\}= 0,\quad \{\z_\mu, \z_\nu\}= 0,\quad \{\l_\mu, \z_\nu\}=
\d_{\mu\nu}. \tag{1.36}
$$
\endproclaim
\demo {Proof}
   Choose a basis in which the leading term in $\hat{\LL}(\l)$ (i.e. $L_0$ for
case (i) and $Y$ for case (ii)) is diagonal, and let  $V_0 = (1,0 \dots 0)^T$.
 Let $\widetilde{\MM}_{ij}(\l,\z)$ denote the $ij^{th}$ component of
$\widetilde{\MM}(\l,\z)$. The points $(\l_{\nu},\z_{\nu})$ are then determined
by the conditions
$$
\wt{\MM}_{k1}(\l_{\nu},\z_{\nu}) =0  \tag{1.37}
$$
 for all $k$.  Generically, these
points are cut out by only two of these equations, say
$$
\wt{\MM}_{11} =\wt{\MM}_{21} = 0  .  \tag{1.37a}
$$
  That is, generically the matrix  $$
F_{\nu} \
:=
\pmatrix{\partial \wt{M}_{11}\over \partial
\lambda}&{\partial \wt{M}_{11}\over \partial
\z}\cr
{\partial \wt{M}_{21}\over \partial \lambda}& {\partial
\wt{M}_{21}\over \partial
\z}
\endpmatrix
\ (\lambda_{\nu}, \z_{\nu})  \tag{1.38}
$$
is invertible.  By implicit differentiation, the Poisson brackets of the
 functions $(\lambda_{\nu}, \z_{\nu})$ are
then:
$$
\multline
\pmatrix
\{\lambda_{\nu},\lambda_{\mu}\} &
\{\lambda_{\nu},\z_{\mu}\}
\cr\{\z_{\nu},\lambda_{\mu}\} & \{ \z_{\nu},\z_{\mu}\}
\endpmatrix
= \\
(F_{\nu})^{-1}
\pmatrix
\{\wt{\MM}_{11}(\l_{\nu},\z_{\nu}),\
\wt{\MM}_{11}(\l_{\mu},\z_{\mu})\} &
\{\wt{\MM}_{11}(\l_{\nu},\z_{\nu}),\
\wt{\MM}_{21}(\l_{\mu},\z_{\mu})\}\cr
\{ \wt{\MM}_{21}(\l_{\nu},\z_{\nu}),\
\wt{\MM}_{11}(\l_{\mu},\z_{\mu})\} &
\{\wt{\MM}_{21}(\l_{\nu},\z_{\nu}),\
\wt{\MM}_{21}(\l_{\mu},\z_{\mu})\}\cr
\endpmatrix
 \ \ (F_{\mu})^{T-1} .\endmultline \tag{1.39}
$$

   To determine the brackets in the matrix on the right hand side of equation
(1.39) we first recall that if  $f$ and $g$ are functions on the orbit
$\QQ_{\NN_0}$, their Poisson bracket at a point $\mu \in {\QQ}_{\NN_0} \ss
\widetilde{\frak{gl}}(r)_-$ is given by $$
\{F,G\} = <\mu,[\frac{\delta f}{\delta \mu},\frac{\delta g}{\delta
\mu}]>  \   ,\tag{1.40}
$$
where $\frac{\delta f}{\delta \mu}$ is the differential of $f$  at $\mu$,
considered as an element of $\widetilde{\frak{gl}}(r)^+$, and the pairing
$<\ ,\ >$
is defined by eq.~ (1.4).  The $ij^{th}$ coefficient of $\MM(\l,\z)$ evaluated
 at the point $(\l_0,\z_0)$, viewed as a function of
 $\mu \in \widetilde{\frak{gl}}(r)_-$,  may be written
$$
\MM_{ij}(\l_0,\z_0) = -<\mu,\frac{e_{ji}}{\l - \l_0}>  - \ \z_0 I  ,
  \tag{1.41}
$$
where $e_{ji}$
is the matrix with a 1 in the $ji$th place and zeroes elsewhere.  It follows
that
$$
\frac{\delta \MM_{ij}(\l_0,\z_0)}{\delta \mu} = -\frac{e_{ji}}{\l - \l_0}
\tag{1.42}
$$
and hence, dropping the $0$ subscripts,
$$
\align
&\{\MM_{ij}(\l,\z),\MM_{kl}(\sigma,\eta)\}\\
 & = \frac{1}{\l -\sigma}\bigl[\bigl(\MM_{il}(\l,\z) -
\MM_{il}(\sigma,\eta)\bigr)\delta_{jk} - \bigl(\MM_{kj}(\l,\z) -
\MM_{kj}(\sigma,\eta)\bigr)\delta_{il}\bigr]  . \tag{1.43}
\endalign
$$
    Since $\widetilde{\MM}(\l,\z)$ is the classical adjoint of $\MM (\l,\z)$ we
have
$$
\widetilde{\MM}(\l,\z)\MM(\l,\z) = \det (\MM(\l,\z))I.  \tag{1.44}
$$
Differentiating with respect to a parameter $t$ yields
$$
\frac{d\widetilde{\MM}(\l,\z)}{dt} = \frac{\widetilde{\MM}(\l,\z)
\text{tr}\bigl((\frac{d}{dt}\widetilde{\MM}(\l,\z))
\widetilde{\MM}(\l,\z)\bigr) -
\widetilde{\MM}(\l,\z)(\frac{d}{dt}\MM(\l,\z))
\widetilde{\MM}(\l.\z)}{\det(\MM(\l,\z))}  \tag{1.45}
$$
away from points $(\l,\z) $ where $\det(\MM(\l,\z)) =0$; i.e., points on the
spectral curve.
     Thus, away from the spectral curve,
$$
\frac{\partial\widetilde{\MM}_{ij}(\l,\z)}{\partial \MM_{pq}(\l,\z)} =
\frac{\widetilde{\MM}_{qp}(\l,\z) \widetilde{\MM}_{ij}(\l,\z) -
\widetilde{\MM}_{ip}(\l,\z) \widetilde{\MM}_{qj}(\l,\z)}{\det{ \MM(\l,\z)}}.
\tag{1.46}
$$
  The derivation property of the bracket
$$
\{\widetilde{\MM}_{ij}(\l,\z),\widetilde{\MM}_{kl}(\s,\eta)\} =
\sum_{pqrs}\frac{\partial \widetilde{\MM}_{ij}(\l,\z)}{\partial
\MM_{pq}(\l,\z)}\frac{\partial \widetilde{\MM}_{kl}(\s,\eta)}{\partial
\MM_{rs}(\s,\eta)}\{\MM_{pq}(\l,\z),\MM_{rs}(\s,\eta)\} \tag{1.47}
$$
then gives
$$
\align
 \{\wt{\MM}_{i1}(\l,\z),\wt{\MM}_{k1}(\s,\eta)\} =
(\frac{1}{\l-\s})\biggl[&\frac{1}{\det{\MM(\s,\eta)}}
\bigl[ (\wt{\MM}(\s,\eta)\wt{\MM}(\l,\z))_{k1}
\wt{\MM}_{i1}(\s,\eta)
\\
& -(\wt{\MM}(\s,\eta)\wt{\MM}(\l,\z))_{i1}
\wt{\MM}_{k1}(\s,\eta)\bigr]
 \\
&+ \frac{1}{\det{\MM(\l,\z)}} \bigl[ (\wt{\MM}(\l,\z)\wt{\MM}(\s,\eta))_{i1}
\wt{\MM}_{k1}(\l,\z)
 \\
& -(\wt{\MM}(\l,\z)\wt{\MM}(\s,\eta))_{k1}
\wt{\MM}_{i1}(\l,\z)\bigr]\biggr]  .  \tag{1.48}
\endalign
$$
      By eq.~ (1.37),  $\wt{\MM}_{k1}(\l_{\nu},\z_{\nu})$ vanishes for all
$k,\nu$.
Taking the limits $(\l,\z) \ra  (\l_{\mu},\z_{\mu})$,  $(\s,\eta) \ra
(\l_\nu,\z_\nu)$ along any path {\it transversal} to the curve $\SS$,   the
right hand side
of equation (1.48) has limit zero for $\nu\neq\mu$ (the simple zero in
det$\MM$ is cancelled by a double zero in the numerator), implying
$$
\align
\{\wt{\MM}_{11}(\l_{\nu},\z_{\nu}),\
\wt{\MM}_{11}(\l_{\mu},\z_{\mu})\} &=
\{\wt{\MM}_{11}(\l_{\nu},\z_{\nu}),\
\wt{\MM}_{21}(\l_{\mu},\z_{\mu})\}  \\
=\{\wt{\MM}_{21}(\l_{\nu},\z_{\nu}),\
\wt{\MM}_{21}(\l_{\mu},\z_{\mu})\} &=0  \tag{1.49}
\endalign
$$
when $\mu \ne \nu$.  Hence $\{\l_{\nu},\l_{\mu}\},\
\{\z_{\nu},\z_{\mu}\}$ and $\{\l_{\nu},\z_{\mu}\}$  all vanish when $\nu \ne
\mu$.

    To compute the bracket for $\nu = \mu$ we first note that the brackets on
the diagonal of the matrix on the right hand side of equation (1.39) are zero
in this case.  Thus, to show that $\{\l_{\nu},\z_{\nu}\} = 1$ it suffices to
show that
$$
\{\wt{\MM}_{11}(\l_{\nu},\z_{\nu}),\wt{\MM}_{21}(\l_{\nu},\z_{\nu})\} =
\det(F_{\nu}).   \tag{1.50}
$$
To compute the left hand side of (1.50) we first take the limit $(\l,\z) \ra
(\s,\eta)$ in (1.43) using  the derivation property (1.47) of the bracket to
show
$$
\{\wt{\MM}_{11}(\l,\z),\wt{\MM}_{21}(\l,\z)\} =
\sum_{prs}(\frac{\partial \wt{\MM}_{11}}{\partial \MM_{pr}} \frac{\partial
\wt{\MM}_{21}}{\partial \MM_{rs}} - \frac{\partial \wt{\MM}_{11}}{\partial
\MM_{rs}}\frac{\partial \wt{\MM}_{21}}{\partial
\MM_{pr}})\frac{d\MM_{ps}}{d\l}.
\tag{1.51}
$$
On the other hand
$$
\det \pmatrix{\partial \wt{M}_{11}\over \partial
\lambda}&{\partial \wt{M}_{11}\over \partial
\z}\cr
{\partial \wt{M}_{21}\over \partial \lambda}& {\partial
\wt{M}_{21}\over \partial
\z}
\endpmatrix
(\l,\z) = \sum_{prq}(\frac{\partial \wt{\MM}_{11}}{\partial \MM_{pr}}
\frac{\partial \wt{\MM}_{21}}{\partial \MM_{qq}} - \frac{\partial
\wt{\MM}_{11}}{\partial \MM_{qq}}\frac{\partial \wt{\MM}_{21}}{\partial
\MM_{pr}})\frac{d\MM_{pr}}{d\l}.
 \tag{1.52}
$$
   Equation (1.50) now follows by substituting eq.~ (1.46) into eqs.~
(1.51) and (1.52) and using the fact that $\wt{\MM}(\l_{\nu},\z_{\nu})$ has
rank $1$.
\hfill$\square$
\enddemo

  The implication of Theorem 1.3 is that the functions
$\{(\l_{\mu}, \z_\mu)\}_{\mu = 1, \dots\tilde{g}}$ {\it nearly} provide a
Darboux coordinate system on the coadjoint orbit $\QQ_{\NN_0}$. However,
the dimensions are not quite right. For case (i), we have
$$
\text{dim}\ \QQ_{\NN_0} = 2\wt{g} + (r+2)(r-1)  \tag{1.53a}
$$
for generic orbits, while for case (ii)
$$
\text{dim}\ \QQ_{\NN_0} = 2(\wt{g} + r -1).  \tag{1.53b}
$$
(Note that in these formulae, it is the value of $\wt g$ that is different,
 according to eq.~(1.27), {\it not} the dimension of $\QQ_{\NN_0}$ which, of
course,
 is the same.)

  On the other hand, for case (i), the Marsden-Weinstein reduced coadjoint
orbit $\QQ_{\text{red}}$, obtained by fixing the value of the $Gl(r)$ moment
map $L_0$ and quotienting by its stabilizer $G_{L_0} \ss Gl(r)$, is  of
dimension $2\wt{g}$. Similarly, for case (ii)  we may reduce by the
stabilizer $G_Y \ss Gl(r)$ of $Y$, since  the shifted AKS Hamiltonians of the
form (1.7) are invariant under this subgroup and the restriction of $L_0$ to
the corresponding subalgebra $\frak{g}_Y$ is conserved under the flows.
The reduced orbit under this action, also denoted $\QQ_{\text{red}}$, is
again of dimension $2\wt{g}$. (Note again that the value of $\wt{g}$ for the
latter case is,
by eq.~(1.27), $\frac{1}{2}r(r-1)$ greater than for the former.) Thus, if the
coordinates
$(\l_{\mu}, \z_\mu)$ could be shown to be projectable to the reduced spaces,
and if the
reduced Poisson brackets remain the same as in eq.~ (1.36), we would have
Darboux coordinates on $\QQ_{\text{red}}$.

   For case (ii) this may be seen immediately. Since $V_0$ was assumed to be an
eigenvector of $Y$, with no degeneracy allowed, the defining equation
(1.34) is invariant under the stabilizer $G_Y \ss Gl(r)$ (an $r-1$ dimensional
abelian group under our hypotheses). Thus
$(\l_{\mu}, \z_\mu)_{\mu = 1, \dots\tilde{g}}$ are all invariant under the
Hamiltonian $G_Y$- action, and the Poisson brackets of their projection to
$\QQ_{\text{red}}$ are the same as on  $\QQ_{\NN_0}$.

   For case (i), we cannot quite apply Hamiltonian symmetry reduction under
$Gl(r)$, since the functions $(\l_{\mu}, \z_\mu)$ are only invariant under the
stabilizer
subgroup $G_{L_0}$. However, we may still compute the Poisson brackets on the
reduced space by the procedure used for constrained Hamiltonian systems.
Let us first choose the reduction condition given by the level set:
$$
L_0 = \text{diag}\{l_i\},  \tag{1.54}
$$
where the eigenvalues $\{l_i\}$ are, by our genericity assumption,
distinct. The diagonal terms in eq.~ (1.54) are the first class constraints,
which  generate the Hamiltonian $G_{L_0}$ - action, and  the terms with $i >1$
may be chosen
as the independent generators. Applying the standard procedure of modifying the
Hamiltonian by adding a linear combination of the remaining, second class
constraints, we see that the following modified functions generate flows that
are tangential to the constrained submanifold:
$$
\align \hat{\l}_{\mu} &= \l_{\mu} - \sum_{i,j=1, i\neq j}^{r} \frac{\{\l_\mu,
(L_0)_{ij}\}}{l_i - l_j}(L_0)_{ji},  \tag{1.55a} \\
 \hat{\z}_{\mu} &= \z_{\mu} - \sum_{i,j=1,
i\neq j}^{r} \frac{\{\z_\mu, (L_0)_{ij}\}}{l_i - l_j}(L_0)_{ji}.   \tag{1.55b}
\endalign
$$
Evaluating their Poisson brackets, we find, again:
$$
\{\hat{\l}_\mu, \hat{\l}_\nu\}= 0,\quad \{\hat{\z}_\mu, \hat{\z}_\nu\}= 0,
\quad\{\hat{\l}_\mu, \hat{\z}_\nu\}= \d_{\mu \nu} \   \tag{1.56}
$$
since, by implicit differentiation of the defining equations (1.37),  the
second
factor in (1.55a,b)  involves terms of the form  $\{\wt{\MM}_{k1}(\l,\z),
(L_0)_{ij}\}$ which, applying the  chain rule and  eq.~(1.46), vanish  unless
$i=1$. The cross terms in the Poisson brackets (1.56) therefore all contain
terms proportional to  $\{\wt{\MM}_{k1}(\l,\z), (L_0)_{i1}\}$, \ $i\neq 1$,
which vanish at $(\l ,\z) =(\l_{\mu},\z_{\mu})$. Since the functions
$(\hat{\l}_\mu, \hat{\z}_\mu)$ coincide with $(\l_\mu, \z_\mu)$ on
the constrained manifold and generate tangential flow, it follows that the
projections of $(\l_\mu, \z_\mu)$ to $\QQ_{\text{red}}$ (the quotient of
the constrained manifold by $G_{L_0}$) satisfy the same Poisson bracket
relations as (1.56). Finally, for other values of $L_0$ than (1.54), we just
repeat the same argument with respect to a diagonalizing basis of
eigenvectors.

  Combining these results we obtain, for both cases (i) and (ii):
\proclaim {Theorem 1.4}  The projections of
$(\l_\mu, \z_\mu)_{\mu = 1, \dots\tilde{g}}$ to the reduced orbit
$\QQ_{\text{red}}$, in both case (i) ($Y = 0$) and case (ii) ($Y \neq 0$, with
distinct eigenvalues), are Darboux coordinates; that is, the reduced symplectic
form is: $$
\o_{\text{red}} = \sum_{\mu =1}^{\tilde{g}} d\l_{\mu}\wedge d\z_{\mu}.
\tag{1.57}
$$
\endproclaim

\noindent {\it Remark${}^{1}$:} \ The proof of Theorem 1.3 did not depend on
the fact that there are $\wt{g}$ finite points in the spectral divisor. If
the vector $V_0$ is not chosen as an eigenvector of $Y$, the number of such
finite points, and corresponding coordinate pairs $(\l_\mu,\z_\mu)$, may be
between $\wt g$ and $\wt g +r-1$. The number of points over $\l=\infty$ equals
the number of eigenvalues $\tilde z$ of the asymptotic form of
$\hat{\LL}(\l)$ (i.e., $Y$ for case (ii) and $L_0$ for case (i)), for which
$V_0$ is in the image of $Y- \tilde z I$ for case (ii)
(resp. $L_0-\tilde z I$ for case (i)). This is
zero for generically chosen (non-diagonal) $Y$ (or $L_0$) or, equivalently, if
 $Y$ is taken as a diagonal matrix, and $V_0$ chosen as a vector with no
vanishing components (e.g.~ $V_0=(1, 1, \dots 1)^T$). In this case, the number
of
finite spectral divisor coordinate pairs $(\l_\mu,\z_\mu)$ will actually be
$\wt g +r -1$, sufficient to provide a Darboux coordinate system for the full
orbit in case (ii) and an $r(r-1)$ co-dimensional symplectic submanifold
in case (i) (cf.~ Section 1c).
\footnote""{\noindent${}^1$
Thanks are due to B.~Dubrovin for raising the point discussed in this remark.}
 However, for the examples involving integrable systems that will be of
interest to us (cf.~ Section 3), it is not this type of spectral Darboux system
that is needed for directly determining solutions, but those derived in the
following subsection. The problem lies with the invertibility of the Abel map
(cf.~ Section 1d), which requires a degree $\wt g$ divisor. The remaining $r-1$
points of the spectral divisor are related to the singular differentials having
pole singularities over $\l = \infty$.

   There remains then the question of the nonreduced orbits $\QQ_{\NN_0}$. Can
the functions $(\l_\mu, \z_\mu)_{\mu = 1, \dots\tilde{g}}$ somehow be
completed to provide a Darboux coordinate system on $\QQ_{\NN_0}$? The
answer is: yes, for case (ii), and partially for case (i). The construction is
given in the following subsection.

\medskip \noindent
{\it 1c. Darboux Coordinates on Unreduced Orbits \hfill}
\nopagebreak

In case (i) we shall obtain Darboux coordinates, not  on the complete coadjoint
orbit
$ \QQ_{\NN_0}$, but on a constrained submanifold $\QQ_{\NN_0}^0\ss \QQ_{\NN_0}$
consisting of elements for which the off-diagonal elements of $L_0$ vanish:
$$
(L_0)_{ij} =0 \quad \text{if} \quad i \neq j .  \tag{1.58}
$$
By our earlier genericity assumptions, the diagonal elements $(L_0)_{ii}$
are hence distinct, and  it is easily verified that
$\QQ_{\NN_0}^0 \ss \QQ_{\NN_0}$ is a symplectic  submanifold of dimension
$$
\text{dim}\ \QQ_{\NN_0}^0 = 2(\wt{g} + r - 1).  \tag{1.59}
$$
(Note that $m=n-1$ in the genus formula (1.27) and we are dealing with case
(1.53a), {\it not} (1.53b).) For case (ii), we choose a basis in which
 $Y$ is  diagonal:
$$
Y = \text{diag}\{Y_i\}  .  \tag{1.60}
$$

 Thus in both cases, the leading term of  $\hat{\LL}(\l)$ is diagonal.  As in
the proof of Theorem 1.3, we  also choose the eigenvector  $V_0$ in (1.29) to
be $V_0 = (1,0,0,...,0 )^T$. In both cases, let
$$
P_i := (L_0)_{ii}, \quad i= 1 , \dots  r.  \tag{1.61}
$$
These generate the action of the group $D$ of diagonal matrices, which equals
$G_{L_0}$ and $G_{Y}$, respectively, for cases (i) and (ii). The generator
$P_1$ is not independent of the others, since the sum:
$$
\sum_{i=1}^r P_i = \text{tr}L_0   \tag{1.62}
$$
is a Casimir. These generators Poisson commute amongst themselves and also
with the $D$--invariant functions $(\l_\mu, \z_\mu)_{\mu = 1,\dots\tilde{g}}$,
since
equation (1.37), which determines them, is $D$--invariant.  In case (i), let
$$
q_i := \ln (L_1)_{i1} +
 \frac{1}{2}\sum_{j\neq i,\ j > 1}^r \ln (P_i-P_j) ,
\tag{1.63}
$$
while for case (ii), let
$$
q_i := \ln (L_0)_{i1}  .
\tag{1.64}
$$
With these definitions, we have:

\proclaim {Theorem 1.5}
The coordinate functions
$(\l_{\mu},\ \z_{\mu},\ q_i,\  P_i)_{\mu = 1, \dots \tilde{g}; i=2, \dots r}$
form a Darboux system on $\QQ_{\NN_0}^0$ in case (i), and $\QQ_{\NN_0}$,
in case (ii); that is, the only nonvanishing Poisson brackets between them
are given by:  $$
\{\l_\mu, \z_\nu\} = \d_{\mu \nu}, \quad \{q_i,P_j\} = \d_{ij}  .  \tag{1.65}
$$
Equivalently,
$$
\o_{orb} =\sum_{\mu =1}^{\tilde{g}}d\l_{\mu}\wedge d\z_{\mu} +
\sum_{i=2}^{r}dq_i \wedge dP_i  ,  \tag{1.66}
$$
where the equality refers to the full orbit $\QQ_{\NN_0}$ in case (ii), and
the restriction of $\o_{orb}$ to $\QQ_{\NN_0}^0$ in case (i).
 \endproclaim

 \demo{Proof} The proof proceeds in two steps. First, as in Theorem 1.3,
the Poisson brackets are computed on the full coadjoint orbits. In case (i),
we then reduce this to  the constrained submanifold, which  is symplectic. From
the Poisson
brackets (1.43) used in the proof of Theorem 1.3 follows:
 $$
\align
\{(L_0)_{ij},(L_s)_{kl}\} & =  (L_{s})_{kj}\d _{il} -
(L_{s})_{il}\d_{jk},    \tag{1.67a}\\
\{\MM_{ij}(\l,\z),(L_0)_{kl}\} & =
(Y-\MM(\l,\z))_{il}\d_{jk}-(Y-\MM(\l,\z))_{kj}\d_{il}   \tag{1.67b}\\
\{\MM_{ij}(\l,\z),(L_1)_{kl}\} & = (\l-
\sum_{m}\a_m)[(Y-\MM(\l,\z))_{il}\d_{jk}-
(Y-\MM(\l,\z))_{kj}\d_{il} ]
 \\ & \qquad  +[(L_0)_{il}\d_{kj}-(L_0)_{kj}]\d_{il}.    \tag{1.67c}
\endalign
$$
This implies, in addition to the relations (1.36), the brackets:
$$\align
\{\l_\m,q_i\} & = \{\z_\m,q_i\} = 0  \tag{1.68a}\\
\{\l_\m,P_i\} &= \{\z_\m,P_i\} = 0  \tag{1.68b}\\
\{P_i,P_j\} &= 0  \tag{1.68c}\\
\{q_i,P_j\} &= \d_{ij}  \tag{1.68d}\\
\{q_i,q_j\} &=\ \cases (P_i-P_j)^{-1} &\text{for case (i)}\\
                      0 &\text{for case (ii)}  ,\endcases   \tag{1.68e}
\endalign
$$
where (1.68a) holds only on the constrained manifold $\QQ^0_{\NN_0}$ for case
(i).
As in the proof of Theorem 1.3, we must use the fact that $\wt{\MM}_{i1}$ is
zero
at $(\l_\m,\z_\m)$. In case (ii) this completes the proof. For case (i), the
constraints  must be taken into account. As in the proof of case (i) of
Theorem 1.4,  we shift the functions $(\l_{\mu},\z_{\nu},q_i,P_j)$ by terms
proportional to the second class constraints
 $(L_0)_{ij} =0, \ i \neq j$ to get functions
$(\hat{\l}_{\mu},\hat{\z}_{\nu},\hat{q}_i,\hat{P}_j)$ which agree with
$(\l_{\mu},\z_{\nu},q_i,P_j)$ on $\QQ_{\NN_0}^0$ and which generate flows
in $\QQ_{\NN_0}$ that are tangential to $\QQ_{\NN_0}^0$.  Since
$$
\{(L_0)_{ij},(L_0)_{kl}\} = (L_0)_{kj}\d _{il} -
(L_0)_{il}\d_{jk},    \tag{1.69}
$$
it suffices, for a general function $f$ on $\QQ_{\NN_0}$,   to take
$$
\hat{f} = f - \sum_{i,j=1 i\neq j}^{r}\frac{\{f, (L_0)_{ij}\}}{P_i
-P_j}(L_0)_{ji}.
\tag{1.70}
$$
As in the proof of Theorem 1.4, the Poisson brackets (1.36) remain unchanged on
the constrained manifold. From eq.~ (1.69), it follows  (as in Theorem 1.4),
that the $P_j$'s already generate tangential flows and hence the Poisson
brackets (1.68b-d) remain unchanged. Eq.~ (1.68a) also is unchanged since, by
the same arguments as in the proof of Theorem 1.4, the additional cross terms
obtained after constraining are all proportional to  terms of the form
$\{q_i, (L_0)_{j1}\}$, which vanish on the constrained manifold. Using
eq.~ (1.67a), we see that the remaining Poisson bracket (1.68e) gets shifted
to zero.
 \hfill$\square$
\enddemo
\noindent {\it Remarks:} \item{{\it i)}}The submanifold
$\QQ_{\NN_0}^0 \ss \QQ_{\NN_0}$ is, in fact, the relevant phase space for many
interesting examples of integrable systems, such as the finite gap solutions
of the cubically nonlinear Schr\"odinger equation (cf.~ [{\bf AHP}, {\bf P}]
and Sections 3c, 3d).
\item{{\it ii)}} If, in formulae (1.2), (1.3), we choose $n=1$, $\a_1=0$
and $Y\neq 0$, then $\Cal Q_{\Cal N_0}$ is really a coadjoint orbit in
$\frak{gl}(r)^{*}$ or $\frak{sl}(r)^{*}$ and Theorem 1.5, together with the
$Ad^*$ invariants (Casimirs), provides Darboux coordinate systems for these
finite dimensional Lie algebras (cf.~ Sec.~ 3a).

{\medskip}\noindent
{\it  1d. Liouville - Arnold Integration and the Abel Map \hfill}

We now turn to the integration of the Hamiltonian systems (1.1)
generated either by elements of the Poisson commutative ring $\FF_Y$ of
functions of the form (1.7), with $\Phi$ in the ring
$I(\wt{\frak{gl}(r)}^*)$  of $Ad^*$ - invariants on  $\wt{\frak{gl}(r)}^*$, or
its extension
 $\FF_Y({\bold P})$ by the generators $\{P_i\}_{i=2 ,\dots r}$. Thus, our
Hamiltonians are all expressible as functions of the invariants
$\{P_{ia},P_i\}$.
The notational conventions of the preceding sections allow us to treat cases
(i) and (ii) simultaneously, although it should be remembered that the
spectral curves and ring of invariants $\FF_Y({\bold P})$ depend on the
choice of $Y$, and the relevant symplectic manifold is $\QQ^0_{\NN_0}$ for
case (i) and the entire orbit $\QQ_{\NN_0}$ for case (ii). The reduced spaces,
though both denoted $\QQ_{\text{red}}$, are also different, their dimensions
$2\wt{g}$ being given by the genus formula (1.27) with $m=n-1$ for case
(i) and $m=n$ for case (ii). For case (i),  $\QQ_{\text{red}}$ signifies the
generic $Gl(r)$ - reduction of $\QQ_{\NN_0}$ or, equivalently, the reduction
of $\QQ^0_{\NN_0}$ by the abelian $r-1$ - dimensional group action
generated by  $\{P_i\}_{i=2, \dots r}$. For case (ii),  $\QQ_{\text{red}}$ is
the
reduction of the full orbit   $\QQ_{\NN_0}$ by the latter action.

   The $\wt{g} + r - 1$ independent spectral invariants for case (i) may be
chosen to be $(P_{ia_i},P_j)_{i,j = 2, \dots r; a_i=0 \dots \d_i -1}$, since
the
coefficients $P_{j\d_j}$ occurring in Proposition 1.2 may be expressed as
translates of the elementary symmetric invariants of $L_0 =
\text{diag}\{P_i\}$
$$
P_{j\d_j} = (-1)^{r-j}\sum_{1\leq i_1<\dots i_j}P_{i_1}\dots P_{i_j} + m_j ,
\tag{1.71}
$$
where the constants $\{m_j\}$ depend on the reference polynomial
$\PP_R(\l,z)$. For case (ii), the leading coefficients $\{P_{j\d_j}\}$ are
constants (translates of the elementary symmetric invariants of $Y$) and the
next to leading coefficients are translates of linear combinations of the
$P_i$'s:
$$
P_{j, \d_j-1} = (-1)^{r-j}\sum_{i=1}^{r} P_i\sum_{1\leq i_1<\dots <i_{j-1} \neq
i}
Y_{i_1} \dots Y_{i_{j-1}} + n_j ,  \tag{1.72}
$$
where again, the constants $n_j$ depend on $\PP_R(\l,z)$ and the
constants $Y_i$. Thus, the $\wt{g}+r-1$ independent invariants may be
chosen to be $\{P_{ia_i},P_j\}_{i,j=2,\dots r, a_i=1, \dots \d_i-2}$. The
Hamiltonians may be viewed in the two cases as functions of the
independent invariants:
$$
h = h(P_{ia_i}, P_j) \qquad i,j = 2, \dots r, \quad a_i=1, \dots \d_i -1-\e.
\tag{1.73}
$$
with $\e=0$ for case (i) and $\e=1$ for case (ii).
We can use the Darboux coordinates of Theorem 1.5 to express the symplectic
forms
$\o_{\text{orb}}$ or $\o_{\text{orb}}|_{\QQ^0_{\NN_0}}$ as (minus) the exterior
derivative of a
 $1$-form:
$$
\theta := \sum_{\mu =1}^{\tilde{g}} \z_{\mu}d\l_{\mu} +
\sum_{i=2}^rP_idq_i.   \tag{1.74}
$$
Restricting to the invariant Lagrangian manifolds $\LL$ obtained by
fixing the level sets of $\{P_{ia}, P_j\}$, there exists (within a suitable
neighbourhood of such $\LL$'s)  a  Liouville generating function
$S(\l_{\mu}, q_i, P_{ia}, P_i)$ such that:
$$
\theta|_{\LL} = dS.   \tag{1.75}
$$
Integrating from an arbitrary initial point thus gives
$$
S(\l_{\mu}, q_i, P_{ia}, P_i) =
\sum_{\mu=1}^{\tilde{g}}\int_{\l_{\mu}^0}^{\l_{\mu}} \frac{z(\l, P_{ia}, P_j)
}{a(\l)}d\l +
\sum_{i=2}^r q_iP_i   ,   \tag{1.76}
$$
where the $\l$ integrals are evaluated within a chosen polygonization of the
spectral curve $\wt{\SS}$ and the function
$$
z = z(\l, P_{ia}, P_j)    \tag{1.77}
$$
is determined implicitly along $\LL$ by the spectral equation:
$$
\PP(\l, z(\l, P_{ia}, P_j)) = 0.   \tag{1.78}
$$
Applying the standard canonical transformation procedure, the coordinates
$(Q_{ia}, Q_j)$ canonically conjugate to the invariants $(P_{ia}, P_j)$ are
then  $$
\align
Q_{ia} &= \frac{\di S}{\di P_{ia}} = \sum_{\mu=1}^{\tilde{g}}
\int_{\l_{\mu}^0}^{\l_{\mu}}\frac{1}{a(\l)}\frac{\di z}{\di P_{ia}}d\l
 \tag{1.79a} \\
Q_{i} &= \frac{\di S}{\di P_{i}} = \sum_{\mu=1}^{\tilde{g}}
\int_{\l_{\mu}^0}^{\l_{\mu}}\frac{1}{a(\l)}\frac{\di z}{\di P_{i}}d\l  + q_i .
 \tag{1.79b}
\endalign
$$
Evaluating the integrands by implicit differentiation of eq. (1.78) with
respect
to the invariants $\{P_{ia}, P_i\}$, and using eqs (1.71), (1.72), we have
$$
 \align
\frac{\di z}{\di P_{ia}} &= -a(\l)\frac{a_i(\l)z^{r-i}\l^a}{\PP_z(\l,z)} ,
\quad i=2, \dots r, \quad a =1, \dots  \d_i -1 -\e
\tag{1.80a} \\
 \frac{\di z}{\di P_{i}} &=
-a(\l)\sum_{j=2}^{r}\frac{R_{ij} a_j(\l)(-z)^{r-j}\l^{\d_j-\e}}{\PP_z(\l,z)},
\quad i=2, \dots  r
\tag{1.80b}
 \endalign
$$
 where
$$
R_{ij} :=
\cases
& (P_1- P_i)\sum_{2\leq i_1 < i_2
\dots < i_{j-2} \neq i} P_{i_1} \dots P_{i_{j-2}}   \\
& \text{and} \quad \e=0\qquad  \text{for case (i) }\\
& (Y_1- Y_i)\sum_{2\leq i_1 < i_2
\dots < i_{j-2} \neq i} Y_{i_1} \dots Y_{i_{j-2}}  \\
& \text{and} \quad \e=1\qquad  \text{for case (ii) }\\
\endcases
\tag{1.81}
 $$

   The flow is then given in implicit form by the linear equations:
$$
\align
\sum_{\mu=1}^{\tilde{g}}\int_{\l_0}^{\l_{\mu}}\frac{a_i(\l)z^{r-i}\l^a}{\PP_z(\l,z)}
d\l   &=  C_{ia} - \frac{\di h}{\di P_{ia}} t   \tag{1.82a} \\
\sum_{\mu=1}^{\tilde{g}} \int_{\l_0}^{\l_{\mu}}
\sum_{j=2}^{r}\frac{R_{ij}a_j(\l)(-z)^{r-j}\l^{\d_j-\e}}{\PP_z(\l,z)}d\l   &=
q_i  + c_i - \frac{\di h}{\di P_{i}} t.   \tag{1.82b}
\endalign
$$
where $\{C_{ia}, c_i\}_{i=2, \dots, r;\ a =1, \dots , \d_i -1 -\e}$
are integration constants and a fixed base point
$\l_0$ has been used in the integration.

\noindent {\it Remark:} \ On any given level set of the $P_i$'s
the Hamiltonians $h(P_{ia}, P_j)$  project to the reduced space
$\QQ_{\text{red}}$  and eq.~ (1.82a) alone gives the corresponding
linearization of the reduced flow.

We note that the linearizing map defined by eqs.~ (1.82a,b) involves $\wt {g}
+r-1$ abelian integrals on $\wt {\SS}$. If we knew that the $\wt {g}$
differentials
$$
\o_{ia} := \frac{a_i(\l)z^{r-i}\l^a}{\PP_z(\l,z)}d\l      \tag{1.83}
$$
appearing as integrands in (1.82a) were all holomorphic (i.e. abelian
differentials of the first kind) and independent then, up to a normalizing
change of basis,  (1.82a) would just be the statement that the Abel map
$$
\AB: S^{\tilde{g}} \wt {\SS} \lra \JJ(\wt {\SS}) \tag{1.84}
$$
taking the unordered set of $\wt g$ points $\{p_\mu\}\in \wt{\SS}$ with
coordinates $(\l_\mu, z_\mu)$ to its image in the Jacobi variety $\JJ(\wt
{\SS})$ linearizes the flow - the familiar type of result usually obtained
from  algebro-geometric methods of integration [{\bf AvM, KN, Du,
AHH1}]. In fact, this is exactly the case. Moreover the remaining $r-1$
integrands
 $$
\o_{i}
:=\sum_{j=2}^{r}\frac{R_{ij}a_j(\l)(-z)^{r-j}\l^{\d_j-\e}}{\PP_z(\l,z)}d\l
\tag{1.85}
$$
appearing in (1.82b) are abelian differentials of the third kind
with simple poles at the $r$ points $\{\infty_i\}$ over $\l=\infty$ with local
coordinates
 $$
\infty_i \Leftrightarrow
\cases
&(\tilde{\lambda} =0, \tilde{z}=P_i) \qquad \text{for case (i)}\\
&(\tilde{\lambda} =0, \tilde{z}=Y_i) \qquad \text{for case (ii)} .
\endcases
 \tag{1.86}
$$
\proclaim
{Theorem 1.6} The $\wt {g}$  differentials $\{\o_{ia}\}_{i=1, \dots
\tilde{g}}$ in eq. (1.83) form a basis for the space $H^0(\wt {\SS}, K_{\wt
{\SS}}) $ of abelian differentials of the first kind (where $K_{\wt {\SS}}$
denotes the canonical bundle). The linear flow  equation (1.82a) may therefore
be expressed
as:
$$
\bold{A}(\DD) = \bold{B} +\bold{U} t ,  \tag{1.87}
$$
where $ \bold{B}, \bold{U} \in \Bbb{C}^{\wt{g}}$ are obtained by applying
the inverse of the $\wt{g} \times \wt{g}$ normalizing matrix $\bold{M}$,
with elements
 $$
\bold{M}_{\mu, (ia)} := \oint_{a_{\mu}} \o_{ia},  \tag{1.88}
$$
to the vectors $ \bold{C},\bold{H} \in \Bbb{C}^{\wt{g}}$ with components
$C_{ia}$  and $ -\frac{\di h}{\di P_{ia}}$, respectively (the pair $(ia)$
viewed
as a single coordinate label in $ \Bbb{C}^{\wt{g}}$).

The $r-1$  differentials $\{\o_i\}_{i =2, \dots
r}$ in  eq.~ (1.85) are abelian differentials of the third kind with simple
poles at $\infty_i$ and $\infty_1$, and residues $+1$ and $-1$,
respectively.  After a suitable translation  by elements of  $H^0(\wt {\SS},
K_{\wt {\SS}})$ to obtain the standard normalization with respect to a
canonical homology basis $\{a_{\mu}, b_{\mu} \in H_1(\wt
{\SS},\Bbb{Z})\}_{\mu =1, \dots \tilde{g}}$, these provide a basis for the
$r-1$ dimensional space of normalized differentials with simple poles over
$\l =\infty$.
 \endproclaim
\noindent {\it Remark}:\ Combining these results with the remark following
eq\. (1.82a,b), we see that for Hamiltonian flows on the reduced orbit (or
equivalently for
Hamiltonians that are independent of the $P_i$'s), the linearization map only
involves
abelian differentials of the first kind. For flows on the unreduced orbit it is
necessary to
introduce the differentials of the third kind in order to determine the time
dependence of
the additional coordinates $\{q_i\}$ ({\it viz\.} Corollary 1.7).
  \demo {Proof}
Every holomorphic  $1-$form on $\wt{\SS}$ can be obtained by evaluating
the Poincar\'e residue of a meromorphic $2-$form on $\TT$ with pole
divisor at $\wt{\SS}$.  Over the affine coordinate neighborhood $U_0$ such
a residue has the form
$$
\o =\frac{ f(\l,z)d\l}{\PP_z(\l,z)}, \tag{1.89}
$$
where, for holomorphicity at $\l=\infty$, the total weighted
degree of the polynomial $f(\l,z)$ must not exceed $m(r-1)-2$, and for
holomorphicity at the points $\{(\a_i,0)\}$,  the function
$f(\l,z)$
must vanish to sufficiently high order so as  to cancel the zeroes of
$\PP_z(\l,z)$. Since at these points $\PP_z(\l,z)$ vanishes like
$(\l-\a_i)^{r-k_i}$, while $z$ vanishes along each intersecting branch like
$\l-\a_i$, $f(\l,z)$ must be a sum of terms of the form
$$
z^{r-i}\prod_{j =1}^{n}(\l-\a_j)^{\text{max}(0, i-k_j-1)}\l^a \qquad j=2,\dots
r,
$$
 where, in order to have total degree at most $m(r-1)-2$,
$$
0\leq a \leq \d_j-1-\e.   \tag{1.90}
$$
But these are precisely the $1-$forms $\o_{ia}$ of eq\. (1.83),  which
therefore span the entire $\wt{g}-$dimensional space of holomorphic
$1-$forms $H^0(\wt {\SS}, K_{\wt {\SS}})$. Since, by Proposition 1.2, there
are exactly $\wt{g}$ such $\o_i$'s, they are necessarily linearly
independent.

  Turning to the remaining $r-1$ differentials $\{\o_i\}$ of eq.(1.85), these
have the same structure near the points $(\a_i,0)$ as the $\o_{ia}$'s, and
hence are holomorphic there, but since the numerator polynomial is of
degree $m(r-1)-1$, they have simple poles over $\l=\infty$. To obtain the
exact location of these poles and their residues, recall that the
numerator of (1.85) was obtained by evaluating
$$
\frac{1}{a(\l)}\frac{\di z}{\di P_i}
 = -\frac{\frac{\di\PP}{\di P_i}}{a(\l)  \PP_z}.  \tag{1.91}
$$
Near  $\l=\infty$, $\PP(\l,z)$ is of the form:
$$
\PP(\l,z) =
\cases
&\frac{1}{\tilde{\lambda}^{(n-1)r}}
\lbrack\prod_{i=1}^r(P_i -\tilde{z})  + O(\tilde{\lambda})\rbrack
\quad \text{for case (i)}    \\
&\frac{1}{{\tilde{\lambda}}^{nr}}
\lbrack\prod_{i=1}^r(Y_i -\tilde{z})
+\tilde{\lambda} \sum_{i=1}^r(P_i -\sum_{j=1}^n \a_j Y_i)
\prod_{k\neq i}(Y_k-\tilde{z})
 +O(\tilde{\lambda}^2)\rbrack  \\
& \text{for case (ii)},
\endcases
\tag{1.92}
 $$
where the change of coordinates (1.18) has been used. Hence, for case
(i), near $\l =\infty$,
$$
\frac{1}{a(\l)}\frac{\di z}{\di P_i}d\l \sim
\frac{\prod_{k\neq i}(P_k -\tilde{z})
-\prod_{k\neq 1}(P_k-\tilde{z})}{\sum_{l=1}^r \prod_{k\neq l}(P_k -
\tilde{z})} \frac{d\tilde{\l}}{\tilde{\l}} ,   \tag{1.93a}
$$
and this has simple poles at $\infty_i \ (\tilde{\l}=0, \tilde{z}=P_i), \ i>1$
and
$\infty_1\ (\tilde{\l}=0, \tilde{z}=P_1)$ with residues $+1$ and $-1$,
respectively. Similarly, for case (ii),
$$
\frac{1}{a(\l)}\frac{\di z}{\di P_i}d\l \sim
\frac{\prod_{k\neq i}(Y_k -\tilde{z})
-\prod_{k\neq1}(Y_k-\tilde{z})}{\sum_{l=1}^r \prod_{k\neq l}(Y_k -
\tilde{z})} \frac{d\tilde{\l}}{\tilde{\l}} ,  \tag{1.93b}
$$
giving again simple poles at $\infty_i \ (\tilde{\l}=0, \tilde{z}=Y_i), \ i>1$
and
$\infty_1\ (\tilde{\l}=0, \tilde{z}=Y_1)$ with residues $+1$ and $-1$.
\hfill$\square$
 \enddemo
It follows, since (1.82a) is essentially the
Abel map,  that any function on the Lagrangian manifold $\LL$ that is
symmetric in the coordinates $(\l_{\mu})$ may be expressed along the
flow lines in terms of  quotients of theta functions on the curve $\wt {\SS}$.
In particular, for  the coordinates $\{q_i(t)\}$ themselves,   we
have
\proclaim
{Corollary 1.7} For a suitable choice of constants $\{e_i,
f_i\}_{i=2, \dots r}$,  the coordinate functions $ \{q_i(t)\}$ satsfying
eq.(1.82b)  are given by:
$$
q_i(t) = \ln \left[\frac{\theta(\bold{B}  +  t \bold{U} -\AB(\infty_i)-\bold
{K})} {\theta(\bold{B} + t \bold{U} -\AB(\infty_1) - \bold{K})}\right] +
e_i t +f_i,  \tag{1.94}
$$
where  $\bold{K} \in \Bbb{C}^{\wt{g}}$ is the Riemann constant.
   \endproclaim
\demo
{Proof} We use the standard method underlying the reciprocity theorems
relating  different types of abelian differentials (cf\. [{\bf GH}]). Namely,
on the polygonization of $\wt {\SS}$ obtained by cutting along  a canonical
basis
$\{a_\mu, b_\mu\}$ of cycles, we define the meromorphic differential
$$
d\psi(p):= d\left(\ln \theta(\AB(\DD) - \AB(p) -\bold{K})\right) ,
\tag{1.95}
$$
where $\DD$ is the divisor $\sum_{\mu=1}^{\tilde{g}} p_\mu$ formed from
the $\wt{g}$  points $(p_\mu)_{\mu=1, \dots \wt{g}}$ with coordinates
$(\l_{\mu}, z_{\mu})$
and $p$  denotes the point of evaluation on $\wt {\SS}$. Since $d\psi$ has
simple poles with residues $1$ at the $p_\mu$'s,  we may express the
abelian sum appearing in eq. (1.82b) as an integral
$$
\sum_{\mu=1}^{\tilde{g}}\int_{\l_{\mu}^0}^{\l_{\mu}} \o_i
=\oint_{\CC}\left[\int_{\l^0}^{p}\o_i\right ]d\psi   \tag{1.96}
$$
around a contour $\CC$ enclosing only these singularities of the integrand,
and not the ones at $p = \{\infty_i\}$, which are logarithmic branch points.
Integrating by parts and deforming the contour to the boundary $\BB$ of the
polygonization of $\wt {\SS}$ gives
$$
\sum_{\mu=1}^{\tilde{g}}\int_{\l_{\mu}^0}^{\l_{\mu}} \o_i
=\sum_{j=2}^{r}\oint_{\CC_j}\ln \theta(\AB(\DD) - \AB(p) -\bold{K})
\o_i -\oint_{\BB}\ln \theta(\AB(\DD) - \AB(p) -\bold{K}) \o_i.
\tag{1.97} $$
where the $\CC_j$ are small loops enclosing the poles at $\infty_j$ and no
other singularities.  If the differentials $\o_i$ were normalized, the
contributions to the boundary integral from the pairs $\pm a_{\mu}$ in
$\BB$ would just be  constants (the discontinuity given by the theta
multiplier over the $b_{\mu}$ cycle), and the contributions from the
$b_{\mu}$ terms would vanish. However,  our differentials $\{\o_i\}$ differ
from the normalized ones by linear combinations of the holomorphic
differentials $\{\o_{ia}\}$ in eqs\. (1.83). It follows from eq\. (1.87) that
these differences contribute linear terms in $t$  with constant coefficients.
The remaining terms in eq\. (1.97) may be evaluated  by taking residues at
$\{\infty_i\}_{i=1, \dots r}$, using the results of Theorem 1.6  and
eq\. (1.87)  to yield the logarithmic theta function term in eq\. (1.94).
The constants and  linear terms in (1.94) are then obtained by summing those
from the normalizing shift with those already present in eqs\. (1.82b).
\enddemo

 \noindent {\it Remark:}\ The LHS of (1.82a,b) may be interpreted as an
extended
Abel map from $S^{\tilde{g}} \wt{\SS}$ to a generalized Jacobi variety
$\JJ(\hat {\SS})$ associated to the singularized curve $\hat {\SS}$ obtained
by  identifying the points $\{\infty_i\}$, where  $\JJ(\hat {\SS})$ is a $(\Bbb
C^{*})^{r-1}$ extension of   $\JJ(\wt {\SS})$. The extended theta function for
$\hat {\SS}$ is obtained by multiplying the ordinary theta function for
the nonsingular curve by exponential factors in the extended
directions [{\bf C}].  This may be viewed as the source of  the additional
linear terms in eq\. (1.94).
\bigskip
\bigskip
\subheading{2. The Algebraic Geometry of the Symplectic
Form}
\medskip \noindent
{\it 2a. \ The Geometric Structure of Coadjoint Orbits \hfill}

In this section, we  give a geometric
description of the coadjoint orbits, based on the results
of {\bf [AHH1]} (cf. also {\bf [B]}).  This will be done both in the reduced
case,
treated above in 1b, and the unreduced case, treated in
1c.  There are two parametrizations, corresponding to the choices $Y=0$
(case (i)) and $Y=$ diag $(Y_i)$, $Y_i$ distinct (case
(ii)).  In the reduced case, we consider the reduced
orbits ${\QQ}_{\text{\rm red}}$ of Theorem 1.4, while  in the
unreduced case, we  consider, for case (i), the
``restricted'' orbit ${\QQ}_{\NN_0}^0$ defined by eq.~(1.58), and for case
(ii), the full orbit ${\QQ}_{\NN_0}$. Note
that with these choices the highest order term of the
matricial polynomial $\hat{\LL}(\lambda)$ is diagonal in both cases.

To each element of the unreduced orbit  ${\QQ}_{\NN_0}$, we can associate a
certain set of geometrical data.  First, there is
the spectral curve $\SS \subset {\Cal T}$, defined in
(1.14), which has the following properties
(see {\bf [AHH1]} for a more detailed discussion):
\proclaim{Lemma 2.1}
Let $\{z_{ia}\}_{i=1, \dots n, a=1, \dots r-k_i}$ be the non zero spectrum
of the residue matrices $N_i\prod_{j=1, j\neq i}^n (\a_i - \a_j)$ at $\l=\a_i$.
(The $\{N_i\}$ are assumed to satisfy the genericity conditions
given in Section 1a.)
\item{(1)}  $\SS$ passes through the points $(\alpha_i,
z_{ia})$, is compact and lies in the linear system $\mid{\Cal O} (r m)\mid$.
\item{(2)} Generically $\SS$ has an $(r-k_i)$-fold ordinary singular
point at $(\alpha_i, 0)$;  desingularising $\SS$ at
these $\{(\alpha_i, 0)\}$ yields a smooth curve $\rho:\tilde{\SS}.
\rightarrow \SS$.
\item{(3)} In case (ii), $\SS$ passes through
$(\tilde{\lambda}, \tilde{z}) = (0, Y_i)$ (see (1.60 )).
In case (i),  $\SS$ generically has $r$ distinct points over
$\tilde{\lambda}=0$.
\endproclaim
A second element has already been alluded to.  Let $\pi:
{\Cal T} \rightarrow {\bold P}_1$ be  the natural
projection, where ${\Cal T}$ is, as above, the total space of ${\Cal O}(m)$.
We can lift the bundles ${\Cal O} (j)$ on ${\bold P}_1$ to ${\Cal T}$, and so
to
$\SS$ and $\tilde{\SS}$.  Let all these bundles be denoted   ${\Cal O}
(j)$ and, if $V$ is a sheaf,  let $V\otimes {\Cal O}(j)$
be denoted $V(j)$. In a natural trivialisation over $U_0 \subset {\Cal T}$,
${\Cal O} (m)$ has a basis of sections $\{z,1,
\lambda,\cdots \lambda^m\}$.  Given
$\hat{\LL}(\lambda)$,  define a sheaf $E$, supported over $\SS$,
by the following exact sequence over ${\Cal T}$ (cf. (1.30))
$$
0\rightarrow {\Cal O} (-m)^{\oplus r} \ \
{\buildrel\KK (z,\lambda)\over \longrightarrow}\ \  {\Cal
O}^{\oplus r}\ \ {\buildrel \mu\over
\rightarrow}\  E\rightarrow 0   ,  \tag{2.1}
$$
where $\KK(z,\lambda) = ( \hat{\LL}(\lambda)-z{\bold 1} )$ (i.e. $\mu$ is
projection to the cokernel of  $(\hat{\LL}(\lambda)-z{\bold 1})$).
  If $\SS$ is reduced one has
\proclaim { Lemma 2.2{\bf\  [AHH1]}}
(1)\  $E$ is a torsion free sheaf over $\SS$ and is
generically the direct image of a line bundle $\tilde{E}$ over
$\tilde{\SS}$, with
$$
deg (\tilde{E})= \tilde{g} + r -1 \tag{2.2a}
$$
(2)\ One has:
$$
H^0 \bigl(\SS, E \otimes {\Cal O} (-1)\bigr)=0  . \tag{2.2b}
$$
In consequence, if $F_{\lambda}$ is a fibre of $\pi: {\Cal
T} \rightarrow {\bold P}_1$, the restriction map
$$
H^0 (\SS,E)\rightarrow H^0 (\SS \cap
F_{\lambda}, E)   \tag{2.2c}
$$
is an isomorphism, and so $H^0 (\SS,E)$ is $r$-dimensional.
\endproclaim

\noindent{\it Remark:}\ For the spectral curve $\Cal S$ of {\it any} element
of ${\QQ}_{\NN_0}$,  we can construct the partly desingularised curve
$\wt{\SS}$ mapping to $\SS$  and the line bundle $\wt {E}$ over $\wt{\SS}$
 whose direct image is $E$ as follows. There is an exact sequence of bundles
over $\Bbb P_{r-1}$:
$$
 0\rightarrow T^*\Bbb P_{r-1}(1) \ \
{\buildrel\rho\over \longrightarrow}\ \  {\Cal O}^{\oplus r}\ \
{\buildrel\phi\over \longrightarrow}\ \  {\Cal O}(1)\rightarrow 0,   \tag{2.3}
$$
where $\phi$ is just the evaluation map of sections.
We consider the map of sheaves over $\TT \times \Bbb P_{r-1}$:
$$
T^*\Bbb P_{r-1}(1) \oplus {\Cal O} (-m)^{\oplus r} \ \
{\buildrel\rho\oplus\KK\over \longrightarrow}\ \  {\Cal
O}^{\oplus r}. \tag{2.4}
$$
The cokernel of this map is supported over the union of a curve $\wt{\SS}$
mapping
to $\SS$ and
some projective spaces over the singular points of $\SS$ where the corank of
$\KK (z,\lambda)$ is greater than one. We take $\wt{E}$ to be the restriction
of this cokernel to $\wt{\SS}$.

A third datum  that can be associated to an element of ${\QQ_{\NN_0}}$
is a trivialisation of $E$ over $\lambda = \infty$.
Let $\{e_i\}$ denote the standard basis of
$H^0 (\Cal T, {\Cal O}^{\oplus r})$. If $\infty_j$ is the
point of the curve over $\lambda=\infty$ corresponding
to the $j$-th eigenvector, then from (2.1), since the leading
order term of $\hat{\LL}(\lambda)$ is diagonal, $\mu (e_i)$
is nonzero  over $\infty_j$ only when $i=j$.  Then $\mu (e_i)(\infty_i)$
 defines a trivialisation $\t$ of the fibre of
$E$ at $\infty_i$, i.e. over $\lambda =\infty$.

We can reobtain $\hat{\LL} (\lambda)$ from the triple
$(\SS,E,\t)$ as follows.  The trivialisation $\t$, along with
condition (2) of  Lemma (2.2)  allows us to fix a basis $\{f_i\}$ of
$H^0 (\SS,E)$ by the condition:
$$
f_i (\infty_j) = \delta_{ij}    . \tag{2.5}
$$
Then  $\hat{\LL}(\lambda)$ is defined  as the endomorphism of
$H^0 (\SS,E)$ (expressed in the basis $\{f_i\}$) defined for each $\l$  by the
diagram:
$$\matrix H^0 (\SS,E)\ \  &{\hbox to 1.5cm{\rightarrowfill}}
& H^0 (\SS\cap F_{\lambda}, E)\\ \\
\Bigg\downarrow\ \hat{\LL}(\lambda)&& \Bigg\downarrow\  \times z\\ \\
H^0 (\SS,E)\ \ &  {\hbox to
1.5cm{\rightarrowfill}}&H^0 (\SS\cap F_{\lambda}, E)
\endmatrix \tag{2.6}$$
\medskip
\noindent where $\times z$ denotes multiplication by the fibre
coordinate $z$. This is equivalent to building a resolution:
$$
0\rightarrow \left[ {\Cal O}(-m)^{\oplus r}\simeq
H^0(\SS,E) \otimes {\Cal O} (-m)\right] \ \ {\buildrel
K(z,\lambda)\over \longrightarrow}\ \ \left[{\Cal O}^{\oplus r}
\simeq H^0(\SS,E) \otimes {\Cal O}\right] \rightarrow E\rightarrow 0 .
  \tag{2.7}
$$
Let $\Cal U$ be the variety of equivalence
classes of triples $(\SS,E,\t)$ such that
\item{(1)} $\SS$ satisfies the conditions (1)-(3) of Lemma
2.1,  is generic in the sense of conditions (2), (3) and is such that the
 curve $\tilde{\SS}$ obtained by desingularising at the $(\a_i, 0)$ is smooth.
\item{(2)} $E$ satisfies conditions (1) and (2) of Lemma 2.2, and is generic in
the sense of condition (1).
\item{(3)} $\t$ is a trivialisation of $E$ over the $r$
points of $\SS$ at infinity.

\bigskip
\proclaim{ Theorem 2.3{\bf \ [AHH1]}}
 There is a biholomorphic
equivalence of a non empty Zariski open subset of $\QQ_{\NN_0}$ in case (i)
and of  $\QQ^0_{\NN_0}$ in case (ii) with $\Cal U$.
\endproclaim
The geometric picture then consists of a
$2(\tilde{g} + r-1)$ dimensional space ${\Cal U}$ of triples
$(\SS,E,\t)$ that projects to a $(\tilde{g}+r-1)$ dimensional
space ${\Cal W}$ of curves $\SS$:
$$
\sigma:{\Cal U} \rightarrow {\Cal W}   .  \tag{2.8}
$$
 From the discussion of the previous section, it follows that  the fibres of
 this map form a Lagrangian foliation of ${\Cal U}$, since the
coefficients of the spectral curve provide a $(\tilde{g}+r-1)$- parameter
family of commuting Hamiltonians defining this foliation.  The infinitesimal
aspects of this picture are described by the following theorem.
\bigskip
\proclaim{ Theorem  2.4}
(1)\  The tangent space at $\SS$ to the space ${\Cal W}$ of
curves is $H^0(\tilde{\SS}, K_{\tilde{\SS}} (1))$, where
$K_{\tilde{\SS}}$ is the canonical bundle of $\tilde{\SS}$.
\newline
(2)\  Fixing $\SS$, the tangent space to the fibre
$\sigma^{-1} (\SS)$ is $H^1 (\tilde{\SS}, {\Cal O}
(-1))$.
\endproclaim
\demo{ Proof} (1) First order variations
of a curve $\tilde{\SS}$ immersed in ${\Cal T}$ are in one to
one correspondence with sections of the normal bundle
$N$ of $\tilde{\SS}$ in ${\Cal T}$.  These variations are
not, however, entirely free:  they are constrained to
vanish at $\{\lambda =\alpha_i\}_{i=1, \dots n}$, and in case (ii) at
$\lambda=\infty$.  In other words, they must vanish at
the zeroes of a section of ${\Cal O}(m+1)$, and so the
permissible variations of $\tilde{\SS}$ correspond to $H^0
(\tilde{\SS}, N\otimes {\Cal O} (-m-1)$). However, $\Cal{ O}(-m-2)$
 is isomorphic to the canonical bundle $K_{\Cal T}$ of the surface
$\Cal T$.
Using the notation of eq.~(1.18), this can be seen from the fact that
$$
d\tilde{z} \wedge d\tilde{\l}=-\l^{-m-2}dz \wedge d\l  . \tag{2.9}
$$
 From the adjunction formula, we have $K_{\tilde{\SS}} = N \otimes K_{\Cal T}$,
and hence the tangent space to $\Cal W$ at $\Cal S$ is
$T_{\SS} (\Cal W) = H^0 (\tilde{\SS}, K_{\tilde{\SS}} (1))$.

\noindent
(2)\ Line bundles $\tilde{E}$ on $\tilde{\SS}$ are classified by $H^1
(\tilde {\SS},{\Cal O}^*)$. Using the exponential exact sequence,
$$
0 \rightarrow {\bold Z} \rightarrow {\Cal O} \rightarrow
{\Cal O}^*\rightarrow 0  ,   \tag{2.10}
$$
we can express infinitesimal variations of $\tilde{E}$ as
elements of $H^1(\tilde{\SS},{\Cal O})$. If we assume  now a  fixed
trivialisation of $\tilde{E}$ at $\lambda =\infty$, then in considering an
infinitesimal variation of an equivalence class represented by a cocycle
$\beta$ over $U_0 \cap U_1$, we can only modify   $\beta$ by
functions on $U_0$, and functions on $U_1$ which vanish
at $\lambda =\infty$. This means  in effect that such variations are
represented by  the cohomology group $H^1 (\SS, {\Cal O} (-1))$.
\hfill$\square$
\enddemo
We can write down explicit formulae as follows.  The
image $\SS$ of $\tilde{\SS}$ lies in the linear system $\mid
{\Cal O} (r m)\mid$ in $T{\bold P}_1$.
If $\SS$ were smooth, sections of the normal bundle of $\SS$
would be sections of ${\Cal O}(r m)$ and hence
represented by polynomials $f(\lambda, z)$ of degree $(r
m)$, with the convention $\text{deg}(\lambda )=1$,
$\text{deg}(z)=m$.  The normal vector field corresponding to $f$
would be:
$$
{f(\lambda, z)\over \partial P/\partial z (\lambda,z)}\
{\partial\over \partial z} . \tag{2.11}
$$
The fact that $\SS$ may be singular at $(\alpha_i, 0)$
forces us to require that $f$ vanish to an appropriate
degree at $(\alpha_i,0)$, in order that (2.11) be finite along
each branch of $\SS$ at $(\alpha_i, 0)$, and thus
represent a section of $N_{\tilde{\SS}}$. (See the discussion
in Section 1d.) Furthermore, since we want $\SS$ to remain
fixed at $\lambda = \alpha_i$ and, in case (ii), at
$\lambda =\infty$, we must require that (2.11) vanish at
$\lambda=\alpha_i$ and, in case (ii), at $\lambda=\infty$.
Thus (2.11) is  divisible by $a(\lambda)$.  The explicit
identification ${\Cal O}(-m-2)\simeq K_{\Cal T}$ used
here then tells us that the $1-$form with a pole at
$\lambda=\infty$ corresponding to (2.11) is:
$$
{f(\lambda, z)\over a(\lambda)\partial P/\partial z
(\lambda, z)}\ d\lambda .   \tag{2.12}
$$

At a point $(\SS,E,\t)$ of ${\Cal U}$, let $\beta (\lambda, z)$
be a cocycle on $U_0 \cap U_1$ representing a class
in $H^1(\SS, {\Cal O} (-1))$ and let $T(\lambda,z)$ be  a transition function
for $E$ over $U_0\cap U_1$ with respect to trivialisations over $U_0,
U_1$ compatible with the trivialisation $\t$. We can write down a one-parameter
family $(E_s, \t_s)$ deforming $(E_0, \t_0):= (E,\t)$, with derivative at
$s=0$ equal to the class $[\beta]$, by choosing for $E_s$ the transition
function (with respect to trivialisations compatible with $\t_s$)
$$
T(\lambda, z,s) = T(\lambda, z) e^{s\beta(\lambda, z)}  . \tag{2.13}
$$

A geometric parametrization of the reduced
orbits can be constructed in a similar fashion.  In reducing, the following
extra constraints must be imposed on the spectrum of $\hat{\LL}(\lambda)$,
and so on the spectral curve $\SS$:
\smallskip
\noindent (3$'$)\  In case (i), $\SS$ intersects infinity ($\tilde{\l}=0$)
at the points $\tilde{z}=
P_i$ , where $P_i$ are distinct constants. In case
(ii), the branches of $\SS$ at infinity have expansions
 $\tilde{z} = Y_i + \tilde{\lambda} P_i + O (\tilde{\lambda}^2)$, $P_i$
constants.
\smallskip
\noindent We then quotient by the action of the diagonal group.
This action can be thought of in terms of changing the trivialisation $\t$,
so we define  ${\Cal U}_{\text{\rm red}}$ to be the variety of equivalence
classes of pairs $(\SS,E)$, where:
\item{(1)} The curve $\SS$  in ${\Cal T}$ satisfies conditions
(1) and (2) of Lemma 2.1 and condition ($3'$), is generic in the sense of
conditions (2) and is such that the
 curve
$\tilde{\SS}$ obtained by desingularising at the $(\a_i, 0)$ is smooth.
\item{(2)}  The torsion free sheaf $E$ satisfies conditions (1) and (2) of
Lemma 2.2 and is generic in the sense of condition (1)
\proclaim{ Theorem  2.5}
There is a biholomorphic equivalence of a non-empty Zariski open subset of
${\QQ}_{\text{\rm red}}$ with the set ${\Cal U}_{\text{\rm red}}$.
\endproclaim
Again, we have a submersion $\sigma_{\text{\rm red}}:
{\Cal U}_{\text{\rm red}}\rightarrow {\Cal W}_{\text{\rm red}}$ onto
a space  ${\Cal W}_{\text{\rm red}}$ of curves,
with ${\Cal U}_{\text{\rm red}}, {\Cal W}_{\text{\rm red}}$  of dimensions
$2\tilde{g},
\tilde{g}$ respectively.  The fibre of this map at $\SS$ is a Lagrangian
submanifold,
and is in
essence the complement of the theta-divisor in the
Jacobian of $\SS$.

For the infinitesimal picture, we can repeat the reasoning of Theorem 2.4.
Since we are imposing one extra order of constraint on the curve at
$\lambda =\infty$, the first order variations of the spectral
curve now correspond to sections of $N\otimes {\Cal O}(-m-2) = K_{\SS}$.
On the other hand, we are just considering infinitesimal variations of
the bundles $E$ since there is no trivialization $\t$ fixed at infinity,
and so  we obtain
\proclaim { Theorem  2.6}
 The tangent space at $\SS$ to ${\Cal W}_{\text{\rm red}}$ is $H^0
(\tilde{\SS}, K_{\tilde{\SS}})$.
Fixing $\SS$, the tangent space of
$\sigma_{\text{\rm red}}^{-1} (\SS)$ is given by $H^1 (\tilde{\SS}, {\Cal
O})$.
\endproclaim
\medskip
\noindent  {\it 2b. \ Symplectic and Algebraic Geometry}

The maps ${\Cal U} \rightarrow {\Cal W},\ {\Cal U}_{\text{\rm red}} \rightarrow
{\Cal W}_{\text{\rm red}}$, the theorems in the preceeding sections, and
Serre duality give us exact sequences, at a point  $p$ of
${\Cal U}$ or ${\Cal U}_{\text{\rm red}}$:
$$
\align
0\rightarrow &H^0 (\tilde{\SS}, K_{\tilde{\SS}}
(1))^* \rightarrow \ T_p {\Cal U}\quad  \rightarrow H^0
(\tilde{\SS}, K_{\tilde{\SS}} (1) )\rightarrow 0     \tag{2.14a}\\
0\rightarrow & \ H^0(\tilde{\SS}, K_{\tilde{\SS}})^*\quad
\rightarrow T_p {\Cal U}_{\text{\rm red}} \rightarrow \ H^0
(\tilde{\SS}, K_{\tilde{\SS}})\ \rightarrow 0  .\tag{2.14b}
\endalign
$$
Given  splittings of these sequences the tangent spaces
can be written as sums of vector spaces of the form $A^*
\oplus A$, where $A = H^0(\tilde{\SS}, K_{\tilde{\SS}}(1))$ in the unreduced
case and $A= H^0(\tilde{\SS}_1, K_{\tilde{\SS}})$ in the reduced case.  A
natural skew form $\omega$  can then be defined
on such a sum:
$$
\omega ((a,v) (b,u))= a(u) - b(v)  .    \tag{2.15}
$$
 A splitting of (2.14a,b) amounts to finding some way,
infinitesimally, of fixing the line bundle (and
trivialisation) while varying the curve.  There is a
natural geometric way of doing this.  At a point
$(\SS,E,\t)$ of ${\Cal U}$ (resp. $(\SS,E)$ of ${\Cal U}_{\text{\rm red}}$), we
extend the bundle $E$ to a neighbourhood of $\SS$ in
$\Cal T$. Similarily, we extend the trivialisation to a
neighbourhood of $\SS \cap F_{\infty}$ in $F_{\infty}$.  This gives
us a ``background'' $(E,\t)$ (resp. $E$) to restrict
to variations of the curve. These extensions, and the splittings of
(2.14a,b) that they define, are not unique.  They do, however,
define the same 2-forms via (2.15). (This will appear as a direct
consequence of the proofs of Theorems 2.7 and 2.8 below).  For the time being,
 let us suppose that some arbitrary choice has been made.  Since these forms
involve Serre duality, we denote them by $\omega_S$ and
$\omega_{S, red}$.  Let $\omega_{orb},\ \omega_{red}$ denote the
Kostant-Kirillov forms on the orbits as in Section 1.
\proclaim{ Theorem 2.7}
Under the identification between ${\Cal U}_{\text{\rm red}}$ and a Zariski
open set in the reduced orbit $\Cal Q_{\text{red}}$ given in Theorem 2.5,
 we have
$$
\omega_{\text{red}} = \omega_{S,red}  . \tag{2.16}
$$
\endproclaim
\demo{Proof}
 Let $(\SS(x,t),E(x,t))$ be a two parameter family in our space,
 with $\SS(x,t)$  defined by  the equation
${{\Cal P}}(\lambda, z, x, t)=0$, and the family of sections
of $E(x,t)$ given over $\{U_i\}_{i=0,1}$ by the functions
$\{s_i(\lambda, z,x,t)\}_{i=0,1}$. We shall now evaluate the symplectic 2-form
on the tangent vectors ${\partial\over \partial t},
{\partial \over \partial x}$ to $U$ at $(x,t) = (0,0)$.
To do this, we use the description given by Theorem 1.4.
If $\sum_\nu (\lambda_{\nu}, z_{\nu})$ is the divisor of $s$ away
from $\lambda=\infty$, so that the points $(\lambda_{\nu},
z_{\nu})$ are given by the simultaneous vanishing of $f$ and
$s_0$, then the reduced Kostant-Kirillov form evaluated on
${\partial \over
\partial t},\ {\partial \over \partial x}$ is
$$
\omega_{\text{red}}\left({\partial\over \partial t},\ {\partial
\over \partial x}\right) =
\sum_{\nu}\ \ {(\lambda_{\nu})_t (z_{\nu})_x - (\lambda_{\nu})_x
(z_{\nu})_t\over a(\lambda_{\nu})}  . \tag{2.17}
$$
where the subscripts ${x,t}$  denote differentiation.  Now define
new variables $\hat{\lambda}_{\nu}, \hat{z}_{\nu}$ by
$$
\align
\hat{\lambda}_{\nu} (0,0)&=\lambda_{\nu} (0,0)\cr
{{\Cal P}}(\hat{\lambda}_{\nu} (x,t), z
(\hat{\lambda}_{\nu}(x,t)),0,0) &=0\cr
s_0 (\hat{\lambda}_{\nu}
(x,t), z(\hat{\lambda}_{\nu}( x,t)), x,t) & =0\cr
\hat{z}_{\nu} (0,0) &=z_{\nu} (0,0)\cr
{\Cal P}(\lambda_{\nu} (0,0), \hat{z}_{\nu} (x,t), x, t) & =0  . \tag{2.18}
\endalign
$$
Here $\hat{\lambda}_{\nu} (x,t)$ is the $\lambda$-coordinate of
the point cut out on $\SS(0,0)$ by the equation  $s_0
(\lambda, z,x,t) = 0$ and $\hat{z}_{\nu}$ represents the variation
of the $z$-coordinate of $\SS(x,t)$ over $\lambda=\lambda_{\nu}
(0,0)$. Implicit differentiation of the equations (2.18)
 and of the corresponding equations for $\lambda_{\nu}, z_{\nu}$,
allows one  to show that at
$(x,t)=(0,0)$,
$$
(\lambda_{\nu})_t (z_{\nu})_x - (\lambda_{\nu})_x (z_{\nu})_t =
(\hat{\lambda}_{\nu})_t (\hat{z}_{\nu})_x - (\hat{\lambda}_{\nu})_x
(\hat{z}_{\nu})_t  . \tag{2.19}
$$
Let $d$ denote exterior differentiation along the curve
$\SS (x, t)$, with $(x,t)$ fixed. We have, for $(x,t)$ small,
$$
\hat{\lambda}_{\nu} = {1\over 2\pi i} \oint_{C_{\nu}} \lambda\
d\ \ln\ s_0    \tag{2.20}
$$
for some suitable contour $C_{\nu}$, and so
$$
(\hat{\lambda}_{\nu})_{t\ \text{or}\ x} = {1\over 2\pi i}
\oint_{C_{\nu}} \lambda (d\ \ln\ s_0)_{t\ \text{or}\ x} \ = {-1\over 2\pi i}
\oint_{C_{\nu}} (\ln\ s_0)_{t\ \text{or}\ x} d\lambda  . \tag{2.21}
$$
At $ (x,t) = (0,0)$
$$
(\hat{z}_{\nu})_{t\ \text{or}\ x} = -{{\Cal P}_{t\ \text{or}\ x}
(\lambda_{\nu}, z_{\nu}, 0, 0)\over {\Cal P}_z (\lambda_{\nu}, z_{\nu}, 0,0)} .
  \tag{2.22}
$$
Let
$$
F=\left[\left( {{\Cal P}_x\over a(\lambda)
{\Cal P}_z} \ d\lambda\right) (\ln\ s_0)_t\right] - \left[ x \leftrightarrow
t\right]    .   \tag{2.23}
$$
Then at  $(x,t)= (0,0)$, we have
$$
{(\lambda_{\nu})_t (z_{\nu})_x - (\lambda_{\nu})_x (z_{\nu})_t\over a
(\lambda_{\nu})}= {1\over 2\pi i} \oint_{C_{\nu}} F  .  \tag{2.24}
$$
Choose a base point $\lambda_0$ and cut open the Riemann
surface $\tilde{\SS}(0,0)$ into a $4\tilde{g}$-gon ($\tilde{g}$ =
 genus$(\tilde{\SS})$) in the
standard fashion. Then,
$$
\align
\omega_{red}\left( {\partial\over \partial t},
{\partial\over \partial x} \right)
&= {1\over 2\pi i} \sum_{\nu} \oint_{C_{\nu}} F\\
&={1\over 2\pi i} \left( \oint _{
\text{edge of } 4\tilde{g}- \text{gon}} F\  -\sum_j \oint_{D_j}
F\right) , \tag{2.25}
\endalign
$$
where the $D_j$ are contours around the points $\infty_j$ over
$\lambda = \infty$. Since the expression $F$ in (2.23) is defined on the curve
itself along the cut locus, the contributions of the two sides of the cut
to (2.25) cancel, so the integral along  the edge is zero.  Also, the 1-form
${{\Cal P}_x\over a(\lambda) {\Cal P}_z}\ d\lambda$ is holomorphic at
$\lambda=\infty$.  As above, we then write the transition function
for the line bundle $E(x,t)$ to first order in $(x,t)$
as $T (\lambda, z)\ e^{t\beta_t +x\beta_x}$, so
that
$$
s_0 (\lambda, z, x, t) = s_1 (\lambda, z, x, t)
\ T (\lambda,z)\  e^{t\beta_t +x\beta_x}\ \
(1 + r (\lambda, z, x, t) {\Cal P}(\lambda, z, x,
t))  \tag{2.26}
$$
for some function $r$ on $U_0 \cap U_1$, where $s_0,s_1$ represent sections
along the curve.
Substituting into (2.25) gives
$$
\omega _{red}\left({\partial\over \partial t},\ {\partial
\over \partial x}\right) = {1\over 2\pi i}\ \sum_j \Biggl(
\oint_{D_j} \left( {{\Cal P}_x
d\lambda\over a(\lambda) {\Cal P}_z}\right) \biggl({\beta_t
+(\ln s_1)_t+\ln (1+r{\Cal P})_t}\biggr)
\Biggr)-(x\leftrightarrow t)   . \tag{2.27}
$$
Of the three terms in the integrand, the second gives zero since
$(\ln s_1)_t$ is holomorphic at $\lambda =\infty$, and the
third vanishes after
antisymmetrization $(x\leftrightarrow t)$. Thus:
$$
\align \omega_{red}\left({\partial\over \partial t} ,
{\partial\over \partial x}\right)
&= {1\over 2\pi i} \sum_j \oint_{D_j}  {{\Cal P}_x\over
a(\lambda) {\Cal P}_z} d\lambda \cdot \beta_t  \ -  \
(x\leftrightarrow t) .  \tag{2.28}
\endalign
$$

To complete the proof, we note that:
\item{1)}  ${{\Cal P}_x d\lambda\over a(\lambda ){\Cal P}_z},
{{\Cal P}_t d\lambda\over a(\lambda){\Cal P}_z}$ are simply the elements of
$H^0 (\tilde{\SS}, K_{\tilde{\SS}})$
corresponding to the variations of the curves in the $x,t$
directions, and  $\beta_t, \beta_x$ are representative cocycles
for the elements of $H^1 (\tilde{\SS}, {\Cal O})$ representing the
variations of the line bundles $E(x,t)$, as in  Theorem 2.4.
\item{2)} The sum of the  contour integrals around the
$D_j$ is the explicit representation of the Serre duality
pairing, when the cocycles are chosen with respect to the
$U_0, U_1$ covering.
\hfill$\square$
\enddemo
For the unreduced case, the corresponding result is:
\bigskip
\proclaim{ Theorem  2.8}
Under the identification between ${\Cal U}$ and a Zariski
open set in the orbit $\Cal Q_{\Cal N_0}$ for case (ii), and the symplectic
submanifold $\Cal Q_{\Cal{N}^0_0}$ for case (i) given in Theorem 2.3,
 we have
$$
\align
\text{Case (i)}\qquad \ \omega_{orb}|_{\QQ^0_{\NN_0}}
 & = \omega_S + {1\over 2}
\sum_{i\neq j} {dP_i \wedge d P_j\over P_i - P_j} \quad  \text{over}
\quad \QQ^0_{\NN_0}  \tag{2.29a}   \\
\text{Case (ii)}\qquad  \qquad  \omega_{orb} &= \omega_S.  \tag{2.29b}
\endalign
$$
\endproclaim
\demo{Proof} Repeating verbatim the
proof of Theorem 2.7, we have, instead of (2.27)
$$
\sum_{\mu} {d\lambda_\mu \wedge d z_\mu\over a
(\lambda_\mu)}
\left({\partial\over \partial t},
{\partial\over \partial x}\right)
 = {1\over 2\pi i}\sum_j \oint_{D_j} {{\Cal P}_x
d\lambda\over a(\lambda) {\Cal P}_z} \Bigl( \beta_t + (\ln
s_1)_t + \left(\ln (1 + r{\Cal P})\right)_t
\Bigr)-(x\leftrightarrow t)  \tag{2.30}
$$
Again, the first term corresponds to $\omega_S$, and the third term
disappears after antisymmetrization.  The second term, however, does not,
since now the $1$-form ${{\Cal P}_x d\lambda\over a(\lambda) {\Cal P}_z}$
can have a pole at $\lambda = \infty$. The residue of this
pole at the point $\infty_i$ is precisely $-(P_i)_x$.

Now let us recall the explicit rebuilding of $\hat{\LL}(\lambda)$ from the
sections of $E$ (see {\bf [AHH1]}).  If $f_j$ is the basis of sections of
$E$, normalised as in (2.5), then for each $\lambda$, we can evaluate $f_j$
in the $U_0$ trivialisation  at the $r$ points $(\lambda, z_i (\lambda))$
of $\SS$ above $\lambda$, and set:
$$
(\psi(\lambda))_{ij} = f_j (\lambda, z_i,
(\lambda))  . \tag{2.31}
$$
Then
$$
\hat{\LL}(\lambda) = \psi^{-1}  \cdot\  {\text{\rm diag}}  (z_i
(\lambda))\cdot\ \psi   . \tag{2.32}
$$
Near $\lambda = \infty$,  $\psi$ can be expanded:
$$
(\psi)_{ij} = \delta_{ij} + \tilde{\lambda} \gamma_{ij} + O(\tilde{\lambda}^2)
{}.
\tag{2.33}
$$
Dividing (2.32) by $\lambda^m$, we have, in case (i)
$$
\align\hat{\LL}(\lambda) \cdot \lambda^{-n+1} &= L_0 +
\tilde{\lambda} L_1 +\cdots\cr
&= \text{\rm diag}(P_i) +\tilde{\lambda}
\bigl(\left[\text{\rm diag}(P_i),\gamma \right] +
\text{\rm diag}\bigl( (L_1)_{ii}\bigr)\bigr)
+O(\tilde{\lambda}^2), \tag{2.34a}
\endalign
$$
and, in case (ii)
$$
\align\hat{\LL}(\lambda)\lambda^{-n}&= Y + \tilde{\lambda}
(L_0 - \sum_{j=1}^n\a_j Y)
+\cdots\cr
&= \text{\rm diag}(Y_i)+ \tilde{\lambda} \bigl(\left[ \text{\rm
diag}(Y_i), \gamma\right] + \text{\rm diag}
\left(P_i - \sum_{j=1}^n\a_j Y_i\right)\bigr) +
O (\tilde{\lambda}^2) . \tag{2.34b}
\endalign
$$

The section $s=(s_0, s_1)$ used in our calculations is
just $f_1$, so that
$$
s_1 (\tilde{\lambda}, \tilde{z}_i (\tilde{\lambda}))=\delta_{i1}
+ \tilde{\lambda} \gamma_{i1} + O(\tilde{\lambda}^2)  . \tag{2.35}
$$
Thus, at $\tilde{\lambda}=0\ (\lambda =\infty)$,
$$
\Bigl(\ln\ s_1 (\tilde{\lambda}, \tilde{z}_1
(\tilde{\lambda}))\Bigr)_t = 0  \tag{2.36}
$$
and for $i\neq 1$, at $\tilde{\lambda} = 0$, in case (i), using (1.63),
$$
\align \bigl(\ln s_1 (\tilde{\lambda}, \tilde{z}_i
(\tilde{\lambda}))\bigr)_t &=
\bigl(\ln(L_1)_{i1}\bigr)_t-
\bigl(\ln ( P_i - P_1 ) \bigr)_t\cr
&=\bigl(q_i -  \frac{1}{2}\sum_{j\neq i,\ j > 1}^r \ln (P_i-P_j) \bigr)_t
 - \bigl(\ln (P_i - P_1)\bigr)_t ,   \tag{2.37a}
\endalign
$$
 while in case (ii), using (1.64),
$$
\align \bigl(\ln s_1(\tilde{\lambda}, \tilde{z}_i
(\tilde{\lambda})\bigr)_t&=\ln \bigl((L_0)_{i1}\bigr)_t\cr
&=(q_i)_t , \tag{2.37b}
\endalign
$$
and similarly for $(\ln s_1)_x$.
Referring to Theorem 1.5, in case (i), the expression (2.30) becomes:
$$
\align \omega_{orb} \left({\partial\over \partial t},
{\partial\over \partial x}\right) &- \sum_{i=2}^m
\biggl((q_i)_t (P_i)_x - (q_i)_x (P_i)_t\biggr) \cr
&= \omega_S \left({\partial\over \partial t},
{\partial\over \partial x}\right) - \sum_{i=2}^m
\bigl((q_i)_t (P_i)_x - (q_i)_x (P_i)_t\bigr)\cr
&\qquad +  {1\over
2} \sum_{i\neq j} {(P_i)_t(P_j)_x -
(P_i)_x (P_j)_t\over P_i - P_j} . \tag{2.38a}
\endalign
$$
and in case (ii):
$$
\align \omega_{orb} \left({\partial\over \partial t},
{\partial\over \partial x}\right) &- \sum_{i=2}^m
\biggl((q_i)_t (P_i)_x - (q_i)_x (P_i)_t\biggr)
\cr
&= \omega_S \left({\partial\over \partial t},
{\partial\over \partial x}\right) - \sum_{i=2}^m
\bigl((q_i)_t (P_i)_x - (q_i)_x (P_i)_t\bigr)  ,   \tag{2.38b}
\endalign
$$
thus proving the theorem. \hfill$\square$
\enddemo
\medskip
\noindent  {\it 2c. Reductions to subalgebras}

Reductions to subalgebras of $\widetilde{\frak{gl}}(r)^+$ can be obtained by
considering the fixed point sets of one or several involutions
$\tilde{\sigma}$ on $\widetilde{\frak{gl}} (r)^+$ (cf. {\bf [AHP, AHH1, HHM]}.
  This procedure can be used
to obtain all the ``classical'' loop algebras. Coadjoint orbits in the
reduced algebras correspond to unions of components of
the fixed point sets on the orbits of the unreduced algebra.
\medskip
\noindent {\it 2c.1. Involutions on Loop Algebras Induced
by Involutions on $\frak{gl}(r, \Bbb C)$}.
\medskip
Let $\sigma:\frak{gl}(r,\Bbb C) \rightarrow \frak{gl} (r, \Bbb C)$ be an
involutive automorphism.  We can define  corresponding linear (resp.
antilinear)
involutions on  $\widetilde{\frak{gl}}(r)^+$ by
$$
\tilde{\sigma}(\hat{\LL})
(\lambda)=\sigma(\hat{\LL}(\lambda))  \quad (\text{resp.\ }
\sigma(\hat{\LL}(\overline{\lambda})) , \tag{2.39}
$$
depending on whether $\sigma$
is linear  or antilinear.  The involutions $\sigma$
that occur in reductions to  the classical algebras fall into three types.
\medskip
 \noindent {\it Type (i)} (antilinear involution)
$$\sigma(x) = \gamma \overline{x} \gamma^{-1},\qquad
\gamma \overline{\gamma} =\pm 1    \tag{2.40}
$$
(This gives, e.g., the reduction to $\widetilde{\frak{gl}}(r, \Bbb R)^+$.)
The  corresponding $\tilde{\sigma}$ induces an antiholomorphic involution $i$
on the ``spectral surface'' $\Cal T$:
$$
i (z, \lambda) = (\overline{z}, \overline{\lambda}) .  \tag{2.41}
$$
A fixed point of $\tilde{\sigma}$ has  its spectrum fixed by $i$. From the
exact sequence (2.12) defining $E$, we have:
$$
0\longrightarrow {\Cal O} (-m)^{\oplus r}
{\buildrel(z{\bold 1} - \overline{\hat{\LL} (\overline{\l})})\over {\hbox to
2cm{\rightarrowfill}}}
{\Cal O}^{\oplus r}
\longrightarrow \overline{i^* E}\longrightarrow 0   , \tag{2.42a}
$$
whereas, the sheaf $\tilde{\sigma}(E)$ corresponding to
$\tilde{\sigma}(\hat{\Cal {L}})$ is given by:
$$
0\longrightarrow {\Cal O} (-m)^{\oplus r}
{\buildrel \gamma(z{\bold 1} -
\overline{\hat{\LL}(\overline{\lambda})})\gamma^{-1}\over
{\hbox to 2cm{\rightarrowfill}}} {\Cal O}^{\oplus r}
\longrightarrow \tilde{\sigma} (E) \longrightarrow 0  . \tag{2.42b}
$$
The maps $\gamma: {\Cal O}^{\oplus r}\rightarrow {\Cal O}^{\oplus r}$,
$\gamma: {\Cal O}(-m)^{\oplus r}\rightarrow {\Cal O}(-m)^{\oplus r}$
then give us, by (2.42a,b), an isomorphism between  $\tilde{\sigma} (E)$
 and $\overline{i^* E}$. If we assume that $\gamma$ preserves the diagonal
form at infinity, we obtain (see also [{\bf HHM}]):
\bigskip
\proclaim {Theorem  2.9} For involutions (2.40) of type (i),
\item{(1)} The fixed point set of $\tilde{\sigma}$  on the orbit
${\QQ_{\NN_0}}$, or on ${\QQ}_{\text{\rm red}}$, is a real symplectic
manifold, with (real) symplectic form given by the restriction of the
symplectic form on the ambient space.
\item{(2)} Under the identification of $\Cal Q_{\Cal N_0}$ with the set
$\Cal U$ of triplets $(\Cal S, E, \tau)$ given in Theorem 2.3,
the action of $\tilde{\sigma}$ is
$$
\tilde{\sigma} (\SS,E,\t) = (i(\SS),\ \overline{i^* E},\
\overline{i^* \t}) . \tag{2.43}
$$
\endproclaim

For this case, real Darboux coordinates may also be obtained. If we choose
an eigenvector $V_0$ in (1.29) which is invariant under the corresponding
involution on $\Bbb C^r$, we obtain an $i$-invariant divisor
$\sum (\l_{\mu},z_{\mu})$.  The points in the divisor can be ordered so that
$$
\align
(\overline{\l}_{2\mu},\overline{z}_{2\mu})
   &= (\l_{2\mu+1},z_{2\mu+1}), \qquad \mu=1,\dots , s \\
(\overline{\l}_{\nu},\overline{z}_{\nu})
   &=(\l_{\nu},z_{\nu}), \qquad \nu=2s+2,\dots,\tilde{g} \tag{2.44}
\endalign
$$
for some $s$. It follows that the real and imaginary parts
$$
(\frac{1}{\sqrt{2}} Re (\l_{2\mu}),\frac{1}{\sqrt{2}} Re(z_{2\mu}),
(\frac{1}{\sqrt{2}} Im (\l_{2\mu}),
\frac{1}{\sqrt{2}} Im(z_{2\mu}))_{ \mu = 1,\dots s},\quad
(\l_{\nu}, z_{\nu})_{\nu= 2s+2,\dots,\tilde{g}} \tag{2.45}
$$
are Darboux coordinates on the real submanifold of fixed points in
 $\Cal Q_{\text{red}}$. For the unreduced case $\Cal Q_{\Cal N_0}$, the
remaining Darboux coordinates are similarly obtained from
$(q_i, P_i)_{i=2, \dots r}$

\noindent  {\it Type (ii) } (linear involution)
$$
\sigma (x) = - \gamma x^T \gamma^T, \qquad
\gamma = \pm \gamma^{-1} = \pm \gamma^T  \tag{2.46}
$$
 (This gives, e.g., reductions to $\frak{o}(r,{\Bbb C})$ and
$\frak{sp} \left({r\over 2}, {\Bbb C}\right)$).
Here, $\tilde{\sigma}$ induces on ${\Cal T}$ the involution:
$$
i (\lambda, z)=(\lambda,- z) , \tag{2.47}
$$
and so  determines a corresponding map on spectral curves. For the
line bundles the map is slightly more complicated.  From
the defining sequence (2.1) for $E$ we have, for
$i^*E,\ \sigma(E)$:
$$
\align
0\longrightarrow {\Cal O} (-m)^{\oplus r}
&{\buildrel(-z{\bold 1} - \hat{\LL} (\lambda))\over {\hbox to
2cm{\rightarrowfill}}}
{\Cal O}^{\oplus r}
\longrightarrow i^* E\longrightarrow 0   \tag{2.48a}
\\
0\longrightarrow {\Cal O} (-m)^{\oplus r}
&{\buildrel \gamma(z{\bold 1} + \hat{\LL}^T(\lambda))\gamma^{-1}\over
{\hbox to 2cm{\rightarrowfill}}} {\Cal O}^{\oplus r}
\longrightarrow \sigma (E) \longrightarrow 0 . \tag{2.48b}
\endalign
$$
Locally, we can use (2.48a,b) to represent sections of
$i^* E,\ \sigma(E)$ by $a, b\ \in {\Cal O}^{\oplus r}$,
respectively.  If $<\ ,\ >$ denotes the standard bilinear pairing ${\Cal
O}^{\oplus r}
\times  {\Cal O}^{\oplus r} \rightarrow {\Cal O}$, consider:
$$
 < \gamma \widetilde{(-z {\bold 1}  - \hat{\LL}(\lambda))} a, b >  ,
\tag{2.49}
$$
where $\widetilde{\quad}$  denotes, as above,
the classical adjoint.  It is easy to check, from (2.48a,b), that this
projects to give a pairing of $i^*E$ with $\sigma (E)$ over $i(\SS)$.
(Remember that $i(\SS)$ has equation
$\text{det} (-z{\bold 1} - \hat{\LL} (\lambda)) =0$.)
Since the entries of $\widetilde{(-z {\bold 1} - \hat{\LL} (\lambda))}$  lie in
$H^0 (\SS, {\Cal O}
((r-1)m)$, this gives a globally defined map:
$$
i^*E\otimes \sigma (E) {\hbox to 1cm{\rightarrowfill}}
{\Cal O} ((r-1)m) . \tag{2.50}
$$
When $i(\SS)$ is smooth, the adjunction formula tells us
that ${\Cal O}((r-1)m)$ is $K_{i(\SS)}(2)$.  Since in this case, the adjoint
matrix $\widetilde{(-z {\bold 1} - \hat{\LL} (\lambda))}$ is everywhere of rank
one,
we can
show directly from the sequences (2.48) that (2.49) is surjective, and so
$$
\sigma(E) \simeq K_{i(\SS)} (2) \otimes (i^* E)^*  . \tag{2.51}
$$
When $i(\SS)$ is not smooth then, in the generic case considered, we
 can identify the canonical bundle
$K_{i (\tilde{\SS})}$ of the desingularisation
$i(\tilde{\SS})$ of $i(\SS)$ with ${\Cal O} ((r - 1) m) [-D]$,
where $D$ is a positive divisor supported by the
singularities of $i(\SS)$ (see [{\bf AHH1}]).  On the other hand,
$\widetilde{(-z{\bold 1} - \hat{\LL} (\lambda))}$ also vanishes at the singular
points of $i(\SS)$ in such a way  that the image of (2.50) in ${\Cal O}
((r-1)m)$ is also ${\Cal O} ((r -1)m)[-D]$.  Therefore again, with a slight
abuse of notation:
$$
\sigma(E) \simeq K_{i(\tilde{\SS})} (2) \otimes (i^*
(E))^*  .  \tag{2.52}
$$
Now  recall that over $\lambda=\infty$, $ \hat{\LL}$ is diagonal.
We assume that $\gamma$ preserves this form.  The
trivialisation of $\sigma (E)$ is then the same as that
of $i^*E$.  Summing up:
\proclaim{ Theorem  2.10} For involutions (2.46) of type (ii),
\item{(1)} The fixed point set of $\tilde{\sigma}$ on the
orbit ${\QQ}$ (or ${\QQ}_{\text{\rm red}}$) is a complex symplectic
manifold.
\item{(2)} Under the identifications of Theorem 2.3, the action of
$\tilde{\sigma}$ on the triplet $(\SS,E,\tau)$ is
$$
\tilde{\sigma}(\SS,E,t) = (i(\SS),\ (i^*(E))^* \otimes
K_{i(\tilde{\SS})} (2), i^* \tau )  .  \tag{2.53}
$$
\endproclaim
\noindent {\it Remarks:}
\item{{\it i)}} If $i(\SS) =\SS$, the
fixed point set of the action of $\tilde{\sigma}$ on the Jacobian
of $\SS$ is a translate of the Prym variety associated to $i$.
\item{{\it ii)}} It is not clear geometrically what the
Darboux coordinates should be in this case.  When $\SS$ is
hyperelliptic (so that there is an extra involution $j$)
such coordinates can be found [{\bf AvM}].

\noindent {\it Type (iii)} (antilinear involution)
 $$
\sigma (x) = -\gamma \overline{x}^T\gamma^{-1},
\qquad \gamma = \pm \overline{\gamma}^{-1} = \pm \overline{\gamma}^T
  \tag{2.54}
$$
 (This gives, e.g., the reduction to $\frak{u}(p,q)$).
 For this case, $\tilde{\sigma}$ induces on $\Cal T$ the involution
$$
i(\lambda,z)=(\overline{\lambda}, -\overline{z})   , \tag{2.55}
$$
and so determines a map on spectral curves.  Proceeding as above,
we have:
\proclaim{ Theorem  2.11} For involution (2.54) of type (iii),
\item{1)} The fixed point set of $\tilde{\sigma}$ in the
orbits ${\QQ}$ (or ${\QQ}_{\text{\rm red}})$ is a real sympletic manifold,
with (real) symplectic form given by restriction of the  symplectic form
on the ambient space.
\item{2)} The action of $\tilde{\sigma}$ on the triplet
$(\SS,E,\tau)$ is
$$
\tilde{\sigma}(\SS,E,\tau)=\left(i(\SS),
K_{i(\tilde{\SS})} (2) \otimes (\overline{i^*E})^*,
 \overline{i^* \tau}\right) . \tag{2.56}
$$
\endproclaim
\medskip
\noindent {\it 2c.2  \ Twisted Involutions}

Given an involution $\sigma:\frak{gl}(r)\rightarrow \frak{gl}(r)$ we
can define a ``twisted'' involution  $\hat{\sigma}$ on
$\widetilde{\frak{gl}}(r)$ by
$$
\hat{\sigma} (\hat{\LL}) (\lambda) = \sigma (\hat{\LL} (-\lambda))  .\tag{2.57}
$$
The case by case study of types (i), (ii) and (iii)
above can be repeated. For each of these, the map $i$ induced by
$\hat{\sigma}$ on ${\Cal T}$ is that induced by
$\tilde{\sigma}$, composed with $(\lambda, z) \mapsto
(-\lambda, z)$.  With this modification of $i$, Theorems
2.9 , 2.10 and 2.11  again hold verbatim.
\bigskip

\subheading{3. Examples}
\medskip
In the following, we examine four applications of the above analysis:
computation of Darboux coordinates on generic coadjoint orbits of
$\frak{sl}(2)$
and $\frak{sl}(3)$; finite dimensional integrable systems involving isospectral
flows in $\wt{\frak{sl}}(2)^{+*}$; the cubically nonlinear Schr\"odinger
equation (NLS) and the coupled 2-component nonlinear Schr\"odinger system
(CNLS). Details on how these systems arise through moment map
embeddings from a space of rank 2 or 3 perturbations of
$N \times N$ matrices may be found in {\bf [AHP]}.

\noindent {\it  3a.  \quad Darboux Coordinates for
$\frak{sl}(2)^*$ and $\frak{sl}(3)^*$ \hfill}

As a first application of the results of Section 1, we compute Darboux
coordinates on generic coadjoint orbits of the algebras
$\frak{sl}(r)$, $r=2,3$ by viewing these
as Poisson subspaces of the corresponding loop algebra
$\wt{\frak{sl}}(r)^{+*}$. Thus, we choose $n=1$ in
eqs.~(1.2), (1.3) and, without loss of generality, $\a_1=0$.
The residue matrix $N_1=L_0$ is identified as an element of $\frak{sl}(r)^*$
and we consider the spectral curve
$$
\Cal P (\l,z)=\det (\hat{\Cal L}(\l) -z I) = 0  \tag{3.1}
$$
for matrices of the form
$$
\hat{\Cal L}(\l) = \l Y + L_0 \in \wt{\frak{sl}}(r)^*, \quad r=2,3 \tag{3.2}
$$
where $Y\in \frak{sl}(r)^*$ is a fixed matrix with simple spectrum.
(Only case (ii), with $Y\neq0$  will be considered, since we want coordinates
on the full coadjoint orbits.)

The Darboux coordinates $\{\l_\mu, \z_{\mu}\}_{\mu=1, \dots g,\dots }$ are
determined
by eq.~(1.29) with $V_0=(1,0,\dots)^T\in \Bbb C^r$, and the ``missing''
coordinates $\{q_i, P_i\}_{i=2,\dots r}$ corresponding to the spectral points
over $\l=\infty$
are given by eqs.~(1.61), (1.64). According to the remark following
Theorem 1.4, for each eigenvalue $\tilde z$ of $Y$ for which
$V_0$ is in the image of $Y-\tilde zI$, one of the $g+r-1$ points in the
spectral divisor will appear over $\l=\infty$, requiring the addition of a
 pair $(q_i,P_i)$ of ``missing'' coordinates to complete the Darboux system. In
particular, if $V_0$ is chosen as an eigenvector of $Y$, as in Theorem 1.5,
there will be $r-1$ such pairs associated to points at $\infty$ and $g$ pairs
of ``finite'' spectral Darboux coordinates.

For $r=2$, the ring of Casimir invariants is generated by $\text{tr}(L_0^2)$
and the generic orbits are $2$-dimensional. Let
$$
Y= \pmatrix
1 & 0 \\ 0 & -1
\endpmatrix,
\qquad
L_0= \pmatrix
-a & r\\
\ u & a
\endpmatrix .  \tag{3.3}
$$
The spectral curve in this case has genus $g=0$ and $V_0$ is an eigenvector
of $Y$; hence, there are no finite
spectral divisor coordinates, only the pair of ``missing'' Darboux coordinates
$$
q_2 = \ln u, \qquad P_2 = a ,  \tag{3.4}
$$
(valid for $u \neq 0$) corresponding to the eigenvalue $\tilde z=-1$.
 Alternatively, choosing $Y$ to be
$$
Y= \pmatrix
0 & 1 \\ 1 & 0
\endpmatrix  ,\tag{3.5}
$$
the curve still has $g=0$, but $V_0$ is not an eigenvector of $Y$, so the
spectral divisor consists of one point at finite $\l$. Eq.~(1.29) reduces
to the linear system
$$
\align
\l + u & = 0\\
z - a & = 0 , \tag{3.6}
\endalign
$$
providing the Darboux coordinates
$$
\l_1=-u,\qquad \z_1= {z_1 \over \l_1} =-{a\over u}  . \tag{3.7}
$$

For $r=3$, the ring of Casimir invariants is generated by
$\text{tr} L_0^2$, $\text{tr}L_0^3$ and the generic orbits are $6$-dimensional.
 Let
$$
  Y=\pmatrix
     0 & 0 & 0 \\
     0 & 1 & 0 \\
     0 & 0 & -1
    \endpmatrix , \qquad
L_0=\pmatrix
     -a-b & r & s \\
     u & a & e \\
     v & f & b
    \endpmatrix .  \tag{3.8}
$$
The spectral curve in this case is generically elliptic ($g=1$) and $V_0$ is
an eigenvector of $Y$. The first pair $(\l_1, \z_1= {z_1 \over \l_1})$ of
Darboux coordinates is obtained by solving the system:
$$
\align
(\l -z + a) (\l + z -b) + ef & =0  \tag{3.9a}\\
u(\l+z  -b) +e v & =0 \tag{3.9b}\\
v (\l - z + a)  - u f & = 0 .  \tag{3.9c}
\endalign
$$
The second and third of these equations imply the first,  and determine
the coordinates:
$$
\align
\l_1 &= {1\over 2} \left(b-a- {e v \over u} + {u f \over v}\right)
\tag{3.10a}\\
\z_1 &= {z_1 \over \l_1}=
 {u v  a + u v  b - e v^2 - f u^2 \over -u v  a + u v  b - e v ^2 + f u^2}  ,
\tag{3.10b}
\endalign
$$
when $u, v$ and $\l_1$ are nonzero. The remaining two points of the spectral
divisor, corresponding to the eigenvalues $\tilde z= 1,-1$ of $Y$,
lie over $\l=\infty$. To complete the system, we must therefore add the two
pairs of ``missing'' coordinates:
$$
\align
q_2 &= \ln u   \qquad P_2 = a \tag{3.11a}
\\
q_3 &= \ln v  \qquad P_3 = b . \tag{3.11b}
\endalign
$$

Alternatively, we may pick $Y$ so that $V_0$ does not lie in the image
of $Y-\tilde z I$ for any eigenvalue $\tilde z$ of $Y$; e.g.
$$
Y= \pmatrix
0 & 1& 0 \\
1 & 0 & 1\\
0 & 1 & 0
\endpmatrix .  \tag{3.12}
$$
The genus is still $g=1$, but the number of finite pairs of spectral divisor
coordinates is now $3$. These may be obtained by solving the pair of equations
$$
\align
  z^2 + z(v - a - b) + \l( u-e) + a b - e f  + f u  - a v &= 0
      \tag{3.13a}\\
 \l z + u z +(v - b) \l  + e v - b u  &= 0 ,
\tag{3.13b}
\endalign
$$
which reduces to a cubic for $z$, with
generically distinct roots $(z_1,z_2,z_3)$. Setting
$$
\l_i=  {z_i^2 +(v- a  - b )z_i  +ab- av)  + fu - fe \over {u-e}},
\qquad \z_i={z_i\over \l_i}, \qquad i=1,2,3 \tag{3.14}
$$
gives the Darboux system.

Continuing similarly for higher $r$, the ring of Casimirs  for $\frak{sl}(r)$
is generated by $\{\text{tr} L_0^l\}_{l=2,\dots,r}$ and the generic orbits are
$r(r-1)$-dimensional. The genus of the generic spectral curve is
 $g={1\over 2}(r-2)(r-1)$ and there
are, in principle, $r$ classes of spectral Darboux coordinates possible,
in which the number of finite coordinate pairs is between $g$ and $g+r-1$.
In each case their determination involves the solution of a polynomial
equation of corresponding degree.

\noindent {\it  3b.  \quad Finite Dimensional Systems and Isospectral Flows
in $\wt{\frak{sl}}(2)^{+ *}$\hfill}

The moment map embedding of finite dimensional integrable systems as
isospectral flows in loop algebras developed in {\bf [AHP]} leads, in the
case $\wt{\frak{sl}}(2)^{+*}$, to the following parametrization. In
eqs.~(1.1-1.3),
let
$$
Y= \pmatrix
a & \quad b \\
c & -a
\endpmatrix \in \frak{sl}(2, \Bbb C) \tag{3.15}
$$
and $ \text{rank} (N_i)= k_i  = 1$. Then
$\Cal{N}_0(\l) \in \wt{\frak{gl}}(2)^{+*}$ may be taken of the form:
$$
\Cal{N}_0(\l)=\l \sum_{i=1}^{n} {G_i^TF_i \over \a_i - \l}
=\l G^T( A - \l I)^{-1}F , \tag{3.16}
$$
where $(F_i, G_i)_{i=1, \dots n}$ are the rows of a pair
$F,G \in M^{n\times 2}$ of $n \times 2$ complex matrices and
$A = \text{diag}(\a_i) \in M^{n \times n}$ is a diagonal matrix with
distinct eigenvalues $(\a_i)_{i=1, \dots n}$. Imposing the
trace-free conditions tr$(G_i^TF_i) = 0$ and using the freedom of replacing
$$
F_i \lmt d_i F_i, \qquad G_i \lmt  d_i^{-1}G_i , \qquad d_i \in \Bbb C -0  ,
\tag{3.17}
$$
we may take $(F,G)$, without loss of generality, to be of the form:
$$
F= {1 \over \sqrt{2}}(\bold x, \bold y), \qquad
G= {1 \over \sqrt{2}}(\bold y, -\bold x)  , \tag{3.18}
$$
where $\bold x, \bold y \in \Bbb C^n$ are viewed as column vectors.
(This amounts to a symplectic reduction with respect to the center of
$\wt{\frak{gl}}(2, \Bbb C)^{+*}$, giving flows in
$\wt{\frak{sl}}(2, \Bbb C)^{+*}$.)  The reduced orbital symplectic form
is then just
$$
\omega = d\bold x^T \wedge d \bold y \tag{3.19}
$$
and $\Cal N(\l)$ has the form
$$
\Cal N (\l) = \l
\pmatrix
a & b \\
c & -a
\endpmatrix +
 {\l \over 2} \pmatrix
-\sum_{i=1}^n{x_iy_i \over \l-\a_i} &
- \sum_{i=1}^n{y_i^2 \over \l-\a_i}\\
 \ \sum_{i=1}^n{x_i^2 \over \l-\a_i} &
\ \sum_{i=1}^n{x_i y_i \over \l-\a_i}
\endpmatrix , \tag{3.20}
$$
where $(x_i, y_i)_{i=1, \dots n}$ are the components of $(\bold x, \bold y)$.
We now also impose the reality conditions
$$
\bold x= \overline{\bold x}, \qquad \bold y= \overline{\bold y},
\qquad Y = \overline{Y}  ,
\tag{3.21}
$$
to obtain flows in $\wt{\frak{sl}}(2, \Bbb R)^{+*}$.
The Hamiltonian systems obtained by pulling back the AKS ring
$\Cal I (\wt{\frak{sl}}(2)^*)$ through the map
$$
\align
\wt J_Y:\Bbb R^n \times \Bbb R^n &\lra \wt{\frak{sl}}(2)^{*}\\
\wt J_Y:\ (\bold x, \bold y)\  &
 \lmt \l Y + \l G^T(A - \l I )^{-1}F = \Cal N(\l) \tag{3.22}
\endalign
$$
are then Poisson commutative and, with the possible addition of certain
quadratic constraints, coincide with those studied by Moser in {\bf [M]}
(cf. also {\bf [AHP, AHH1]}). (The fibres of this map are generated by
the finite group of reflections $(x_i, y_i)\lmt (-x_i,-y_i)$ of the coordinate
axes. Since the points with $(x_i=0, y_i=0)$ are excluded from the inverse
image of the orbit $\Cal Q_{\Cal N}$ by the condition $\text{rank}(N_i) = 1$,
the resulting ambiguity is resolved along the flows by continuity.)

As an illustrative example, consider the C. Neumann system {\bf [N]}.
This has been  amply studied  by a variety of methods in the literature
{\bf [AvM, F,  Kn, M, Sch, Ra1]}. We  include it here to show how our
general  approach reduces to the familiar results for this case, giving a
complete separation of variables in hyperellipsoidal coordinates and
linearization via a hyperelliptic Abel map. To obtain this system, we choose
the matrix $Y$ to be
$$
Y=
\pmatrix
0 & -{1\over 2}\\
0 & \ 0
\endpmatrix, \tag{3.23}
$$
and the Hamiltonian $\phi$ to be:
$$
\align
\phi(\bold x, \bold y) &= - \text{tr}(\Cal N(\l)^2)_0
= {1\over 2}[(\bold x^T \bold x)(\bold y^T \bold y) + \bold x^T A \bold x
- (\bold x^T \bold y)^2]  ,  \tag {3.24}
\endalign
$$
where the subscript $(\ )_0$ signifies the $\l^0$ term in the Laurent
expansion around $\l=0$ for large $\l$. To obtain the appropriate phase space,
we must also add the symplectic constraints:
$$ \bold x^T\bold x = 1, \qquad
\bold y^T \bold x = 0 ,   \tag{3.25}
$$
defining the cotangent bundle $T^*S^{n-1} \sim TS^{n-1} \ss \Bbb R^{2n}$.
The Neumann oscillator Hamiltonian is
$$
H(\bold x, \bold y)={1\over2}[\bold y^T\bold y + \bold x^T A \bold x] ,
\tag{3.26}
$$
which coincides with $\phi(\bold x, \bold y)$ on the constrained manifold.
The constraints (3.25) may be viewed as a Marsden-Weinstein reduction
under the stabilizer $Stab(Y)\ss \frak{sl}(2,\Bbb R)$ (cf. Section 1b),
 in which
$\bold x^T \bold x =1$  defines a  level set of the moment
 map generating the  flow
$$
(\bold x, \bold y ) \lmt (\bold x, \bold y + t \bold x) \tag{3.27}
$$
induced by the one-parameter subgroup $Stab(Y)$, while   $\bold y^T \bold x=0$
 defines  a section over the quotient of the level set  by this flow
(i.e., of the null foliation it generates). It follows  that the $H$-flow of
the constrained system is obtained from the $\phi$-flow of the free system
simply by orthogonal projection of the momentum $\bold y$ relative to $\bold
x$:
$$
(\bold x(t), \bold y(t))_{free}\lmt
(\wh{\bold x}(t), \wh{\bold y}(t))_{constr.}
:= \left( (\bold x(t), \bold y(t)-
\left({\bold x^T(t) \bold y(t)\over \bold x^T(t)\bold
x(t)}\right)\bold x(t) \right)\tag {3.28}
$$
from the invariant manifold defined by $\bold x^T \bold x=1$. The equations
of motion for the unconstrained system are
$$
\align
{d\bold x\over dt} & = (\bold x^T \bold x) \bold y -
(\bold x^T \bold y) \bold x   \tag{3.29a} \\
{d\bold y\over dt} & = - (\bold y^T \bold y) \bold x - A \bold x +
(\bold x^T \bold y) \bold y  .
\tag{3.29b}
\endalign
$$
These are equivalent (within a quotient by the finite group of reflections in
the coordinate axes) to the Lax equation
$$
{d\Cal N \over dt}= [\Cal B, \Cal N]  ,\tag{3.30a}
$$
where
$$\Cal B=d\phi(\Cal N)_+=
\pmatrix
\ \bold x^T \bold y& \l + \bold y^T \bold y\\
- \bold x^T \bold x& -\bold x^T \bold y
\endpmatrix .\tag{3.30b}
$$

The invariant spectral curve is thus given by the characteristic equation
$$
\text{det}\left({\Cal N(\l)\over \l} - \zeta I)\right)=0  ,\tag{3.31}
$$
which, defining as in Section 1
$$
z := a(\l) \zeta, \qquad a(\l) := \prod_{i=1}^n(\l - \a_i)  ,\tag{3.32}
$$
determines a  genus $g=n-1$ hyperelliptic curve defined by
(cf.~ eqs.~(1.15), (1.22):
$$
z^2 - a(\l)\Cal P(\l)=0  ,\tag{3.33}
$$
where
$$
\align
\Cal P(\l) & := -\Cal P_2(\l)
= -{a(\l)\over 4} \sum_{i=1}^{n} {I_i \over \l - \a_i} \\
 &= P_{n-1}\l^{n-1}+ P_{n-2} \l^{n-2} + \dots + P_0\tag{3.34a}\\
P_{n-1} & = -{1\over 4} \sum_{i=1}^n I_i
= -{1\over 4}\bold x^T \bold x \tag{3.34b}\\
P_{n-2}&={1\over 4}\sum_{i=1}^n \a_i I_i = {1\over 2} \phi\tag{3.34c}
\endalign
$$
and
$$I_i := \sum_{j=1, j\neq i}^n {(x_iy_j - x_j y_i)^2\over
\a_i - \a_j} +  x_i^2\tag{3.35}
$$
are the Devaney-Uhlenbeck invariants (cf. {\bf [M]}).

Applying the prescription of Section 1b, with $V_0=(1,0)^T$, the
reduction with respect to the stabilizer of $Y$ gives rise to the
constraints (3.11) as discussed above, and the solutions to
eq.~ (1.34) give us the  Darboux coordinates
$(\l_\mu, \z_\mu)_{\mu = 1, \dots n-1}$ defined by
$$
\align
\sum_{i=1}^n {x_i^2\over \l -\a_i} &=
{\prod_{\mu=1}^{n-1}(\l - \l_\mu)\over a(\l)} \tag{3.36a}\\
\zeta_\mu = {1\over 2} \sum_{i=1}^n{x_iy_i\over\l_\mu - \a_i}
&=\sqrt{{\Cal P(\l_\mu) \over a(\l_\mu)}}  .\tag{3.36b}
\endalign
$$
Thus, the spectral divisor coordinates here are just the usual hyperellipsoidal
coordinates $(\l_\mu)$, together with their conjugate momenta
$(\zeta_\mu)$. The Liouville generating function on the isospectral
foliation thus becomes
$$
S=\sum_{\mu=1}^{n-1} \zeta_\mu d\l_\mu \vert_{P_i=cst.} =
\sum_{\mu=1}^{n-1}\int_0^{\l_\mu}\sqrt{{\Cal P(\l)\over a(\l)}}d\l  ,
\tag{3.37}
$$
and the canonically conjugate coordinates undergoing linear flow are
$$
Q_j:= {\di S\over \di P_j}
= {1\over 2} \sum_{\mu=1}^{n-1}\int_0^{\l_\mu}{\l^j d\l
\over \sqrt{a(\l)\Cal P(\l)}} = b_jt  , \qquad j=0, \dots n-2  ,\tag{3.38}
$$
where, for our Hamiltonian $\phi = 2 P_{n-2}$,
$$
b_{n-2} = 2, \qquad b_j=0, \quad j < n-2  .\tag{3.39}
$$
This reproduces the familiar linearization via the hyperelliptic Abel map
 obtained through the classical methods of Jacobi (cf.~{\bf [M]}).

The other classical systems treated in {\bf [M]} as isospectral
flows of rank $2$ perturbations of a fixed matrix $A$, such as
geodesic flow on hyperellipsoids or the Rosochatius system, follow
identically (cf.~ also {\bf [GHHW, AHP, Ra2]}).
In all these cases, the spectral divisor Darboux coordinates
$(\l_\mu, \zeta_\mu)$ will coincide with the usual hyperellipsoidal
coordinates, or some complexification thereof, the curves will be
hyperelliptic and the spectral invariants will be an analogue of the
Devaney-Uhlenbeck invariants (3.21) encountered in this case.

\noindent {\it  3c.  \quad The NLS System\hfill}

The NLS equation has two distinct forms
\nopagebreak
$$
\align
u_{xx} + {\sqrt{-1}} u_t &= 2|u|^2u \tag{3.40a} \\
u_{xx} + {\sqrt{-1}} u_t &= -2|u|^2u . \tag{3.40b}
\endalign
$$
Here we shall discuss (3.40a), but (3.40b) can be dealt with in a
similar fashion.  The finite genus quasi-periodic solutions of (3.40a) are
determined by a pair of commuting flows in  Poisson submanifolds
of $\wt{\frak{su}}(1,1)^{+*}$ consisting of $\frak{su}(1,1)$-valued
rational or polynomial functions of $\l$. Let  $\hat{\Cal L}(\lambda)$  be
a matricial polynomial of the form
$$
\hat{\Cal L}(\lambda) =\frac{a(\l)}{\l} \NN(\l)
= L_0\lambda^{n-1} + L_1 \lambda^{n-2} + \dots + L_{n-1},
  \tag{3.41}
$$
where $\Cal N(\l) \in \wt{\frak{sl}}(2, \Bbb C)^{+*}$ is an element of a
rational coadjoint orbit of the form (1.2), with  $Y=0$, simple
poles at $(\l=\a_i)_{i=1, \dots n}$ and residues $N_i$ of rank $k_i=1$. The
reality condition implying that $ \Cal N \in \wt{\frak{su}}(1,1)^{+*}$
is equivalent to the conditions $N_i, L_i  \in \frak{su}(1,1)$. The spectral
curve for this case (cf. Lemma 1.1 and Proposition 1.2) is given by
$$
\text{det}(\hat{\Cal L}(\l)- z I) = z^2 + a(\lambda)\PP_2(\lambda) =0   ,
\tag{3.42}
$$
where  $\PP_2(\lambda)$  is a polynomial of degree  $n-2$:
$$
\PP_2(\lambda) = P_{20} + P_{21}\lambda + \dots + P_{2,n-2}\lambda^{n-2}  .
\tag{3.43}
$$
Choosing Hamiltonians of the AKS type (with $Y=0$):
$$
\align
H_x& = \frac{1}{2}\left[ \frac{a(\lambda)}{\lambda^n} \lambda
\ \text{tr}(\Cal N(\lambda)^2)\right]_0
=-P_{2,n-3}   \tag{3.44a}\\
H_t &= \frac{1}{2} \left[\frac{a(\lambda)}{\lambda^n}\lambda^2 \
 \text{tr}
(\Cal N(\lambda)^2)\right]_0
 = -P_{2,n-4} \tag{3.44b}
\endalign
$$
gives the Lax equations
$$
\align
\frac{d}{dx} \hat{\Cal L (\l)} &= [\lambda L_0 + L_1, \hat{\Cal L(\l )}]
\tag{3.45a}\\
\frac{d}{dt} \hat{\Cal L(\l )} &= [\lambda^2 L_0 + \lambda L_1 +
L_2,\hat{\Cal L(\l )}] .
\tag{3.45b}
\endalign
$$
The leading term $L_0$ in $\hat{\Cal L}(\lambda)$ is
the  $\frak{su}(1,1)$  moment map and hence is invariant under all AKS
flows. We choose it to be
$$
L_0 = \frac{i}{2}
\pmatrix
1&0\\
0&-1
\endpmatrix  .
\tag{3.46a}
$$
Parametrizing $L_1, L_2$ as
$$
\align
L_1 &= \left(\matrix s&{\overline{u}}\\u&-s\endmatrix\right)
\tag{3.46b}\\
L_2 &= i\left(\matrix
S&{-\overline{U}}\\ U&-S
\endmatrix\right)  , \tag{3.46c}
\endalign
$$
it is easily verified that $s$ is a constant of motion, which may be
set equal to zero. On this level set we have
$$
U=u_x  \tag{3.47}
$$
and
$S - |u|^2$ is constant, so we may choose
$$
\align
s &= 0  \tag{3.48a} \\
S &= |u|^2. \tag{3.48b}
\endalign
$$
With these values for the matrices $L_0, L_1$ and $L_2$, the commutativity
conditions for the flows determined by eqs.~ (3.45a,b) are equivalent to the
condition that $u(x,t)$ satisfy eq.~ (3.40a), and eq.~ (3.45a) determines
$L_3, \dots , L_{n-1}$ in terms of $u$ and its $x-$derivatives, up to choices
of integration constants.

In {\bf [AHP]} these flows were related to reduced canonical Hamiltonian
flows via rank 2 isospectral perturbations of matrices as follows (cf.~ also
{\bf [AHH1] }).  Consider the subspace
$W \subset M^{n\times 2} \times M^{n \times 2}$ of the space of pairs of
complex $n \times 2$ matrices consisting of  pairs  $(F,G)$ with columns of
the form
$$
F = \frac{1}{{\sqrt{2}}} (\bold z,i \overline{\bold z}),\qquad G =
\frac{1}{{\sqrt{2}}}(-i \overline{\bold z},\bold z) \tag{3.49}
$$
where  $\bold z \in \Bbb C^n$ is a column vector with components
$(z_i)_{i=1, \dots n}$. The space  $W$  is a real symplectic subspace of
$M^{n\times 2}
\times M^{n\times 2}$  that may be identified with $\Bbb C^n$,
with symplectic form
$$
\omega =-i\ d\bold z^T\wedge d\overline{\bold z}
= -i \sum^n_{j=1}  dz_j \wedge d\overline{z}_j .
\tag{3.50}
$$
Fixing $n$ real constants  $\{\alpha_i\}_{i=1, \dots n}$, and defining
as above, $A:=\text{diag}(\a_i) \in M^{n \times n}$, one constructs a
moment map
$$
\align
\wt{J}:  \Bbb C^n &\lra \wt{\frak{su}}(1,1)^{+*} \tag{3.51a}\\
\wt{J}:\bold z & \lra \l G^T(\l I - A)^{-1}F \\
&\qquad  =  \frac{\lambda}{2}
\pmatrix
 i\sum_{j=1}^n
\frac{|z_j|^2}{\lambda - \alpha_j}&-\sum_{j=1}^n
\frac{\overline{z}^2_j}{\lambda - \alpha_j} \\
-\sum_{j=1}^n\frac{z^2_j}{\lambda - \alpha_j}&-i
\sum_{j=1}^n\frac{|z_j|^2}{\lambda-\alpha_j}
\endpmatrix
=: \Cal N(\lambda)   .\tag{3.51b}
\endalign
$$
This is analogous to the map defined by eqs.~ (3.20), (3.22) above, with
different reality conditions. The fibres are again generated by the finite
 group of  reflections in the coordinate hyperplanes $\{z_i=0\}$.
Thus, via $\wt{J}$, $\Bbb C^n$ (minus the coordinate hyperplanes) provides a
canonical model for an orbit
${\QQ}_{\NN}$ in $\widetilde{\frak{su}}(1,1)^{+*}$. This consists of elements
with  simple poles
at $(\a_i)_{i=1,\dots n}$ with rank $k_i=1$ matrix residues, whose kernels and
images are conjugate $2$-vectors, null with respect to the hermitian form
$H(w,z)= |w|^2-|z|^2$. The symplectic form $\omega$ in eq.~ (3.50) is just the
Kostant-Kirillov form $\omega_{orb}$ on this orbit.

The pull back under $\wt{J}$
of the AKS Hamiltonians (3.44a,b) generate commuting flows on the symplectic
space $W$. The condition (3.46a) gives  invariant constraints
$$
\align
\overline{\bold z}^T \bold z &= 1  \tag{3.52a}\\
\bold z^T \bold z &= 0   , \tag{3.52b}
\endalign
$$
while eqs.~(3.48a,b) are equivalent to
$$
\align
\overline{\bold z}^TA\bold z &= \sum_{j=1}^n\alpha_j  \tag{3.52c}\\
|\bold z^TA \bold z|^2 - \overline{\bold z}^TA^2 \bold z
&=  \sum_{j=1}^n \alpha_j^2 -2(\sum_{j=1}^n\alpha_j)^2 .
  \tag{3.52d}
\endalign
$$
The function $u$ entering in eq.~(3.40a) is, by eq.~ (3.46b),
$$
u = -\frac{1}{2} \bold z^TA\bold z  . \tag{3.53}
$$

The coefficients  $P_{20},\dots,P_{2,n-2}$  in the spectral polynomial
(3.43) give  $n-1$  real functions in involution on  $\Bbb C^n \sim W$.
Adding the further invariant
$P_{2} = (L_0)_{22} = -\frac{i}{2}\sum_{j=1}^n|z_j|^2$
gives a completely integrable system on  $W$.

The genus of the spectral curve (3.42) is $g = n - 2$. Since the translation
term $Y$ is zero, the flows on the Jacobi variety of this curve linearize the
above completely integrable systems on  $\Bbb C^n$  reduced by the
$\frak{su}(1,1)$  action.  The reduction is given in this case by fixing the
value (3.46a)  of the  $\frak{su}(1,1)$  moment map $L_0$, and then dividing by
 the isotropy group of  $L_0$.  The isotropy group is  $S^1$, whose action
$$
e^{i\psi}:\bold z \rightarrow e^{i\psi}\bold z  \tag{3.54}
$$
is generated by the Hamiltonian  $P_2$.  With  $u$  given by (3.53), we see
that
this action maps  $u$  to  $e^{2i\psi}u$ and hence integration of the reduced
system only determines  $u$  up to a phase factor.

By considering instead the constrained submanifold,
$\QQ^0_{\NN} \subset \QQ_{\NN}$, described for case (i) of Section 1d,
we can obtain the linearization of the flows implicitly in terms of complex
hyperelliptic coordinates and determine $u$ explicitly without any
arbitrariness of phase. Since $\QQ^0_{\NN}$ is given by setting  the
off-diagonal terms in $L_0$ equal to zero, the two real constraints given by
eq.~ (3.52b) give a symplectic submanifold  $X \ss \Bbb C^n$(minus the
coordinate hyperplanes) which models $\QQ^0_{\NN}$. The functions
$P_{20},\dots,P_{2,n-2}$  give a complete set of integrals on the  $2n - 2$
 dimensional space  $X$. (On this space, we have $P_{2,n-2} = -P^2_{2}$.)
The constraint (3.52a)is equivalent to choosing
$$
P_{2, n-2} = {1\over 4}  , \tag{3.55a}
$$
while (3.52c,d) are then equivalent to choosing
$$
\align
P_{0,n-3} &= \frac{1}{4}(\sum_{j=1}^n \alpha_j)   \tag{3.55b}\\
P_{0,n-4} &= {1\over 8} \left[\sum_{j=1}^n \a_j^2 +
(\sum_{j=1}^n\alpha_j)^2\right]  .
\tag{3.55c}
\endalign
$$

Following the prescription of Section 1b, again choosing $V_0=(1,0)^T$ and
requiring it to be in the kernel of  the matrix of cofactors of
$(\hat{\Cal L}(\l) - z I)$, the divisor coordinates
$\{\l_\mu,\z_\mu\}_{\mu=1, \dots n-2}$ are
given by
$$
 \sum_{i=1}^n \frac{z^2_i}{\lambda - \alpha_i} = -{2 u
\prod^{n-2}_{\mu=1} (\lambda - \lambda_\mu)\over a(\lambda)} \tag{3.56}
$$
and
$$
\z_\mu =
-{i\over 2} \sum_{i=1}^{n} {|z_i|^2\over \l_\mu - \a_i}
=\sqrt{ -{\Cal P_2(\l_\mu)\over a(\l_\mu)}}  , \tag{3.57}
$$
which define complex hyperelliptic coordinates $(\l_{\mu})_{\mu=1, \dots n-2}$
and their canonically conjugate momenta $(\z_{\mu})_{\mu=1, \dots n-2}$.

The function $u$ in equation (3.57) is defined in eq.~(3.53) and gives
the off-diagonal term in $L_1$. By the results of Theorem 1.5,
 $q_2 = \ln(u)$ must be included, along with its canonical conjugate
$P_{2}$, to complete the Darboux coordinate system on $X$. Thus, the
restriction
 of the orbital symplectic form (3.50) to $X$ is just
$$
\omega = \sum_{\mu=1}^{n-2} d\l_{\mu} \wedge d \z_{\mu} + dq_2 \wedge dP_2  .
\tag{3.58}
$$
Notice that $(\l_\nu, \z_\nu)_{\nu=1,\dots n-2}$, $q_2, P_{2}$ appear to
give $2n - 2$ complex functions on $X$, which has real dimension $2n - 2$.
However, since we have reduced the loop algebra to  $\wt{\frak{su}}(1,1)$ there
 are reality conditions satisfied by these functions (see Section 2c,
case (iii)) which will be preserved by the flows. Similarly, the
conditions (3.52a-d)  (or equivalently, (3.52b), (3.55a-c)) may be imposed on
the initial data, and will be preserved under the flows.

{}From (3.57), we see that the Liouville generating function (1.76)
on the isospectral foliation  is
$$
S=\sum_{\mu=1}^{n-1} \zeta_\mu d\l_\mu \vert_{P_i=cst.} + q_2 P_2=
\sum_{\mu=1}^{n-1}\int_0^{\l_\mu}\sqrt{-{\Cal P_2(\l)\over a(\l)}}d\l
+ P_2 \ln u .
\tag{3.59}
$$
and hence (cf.~ eqs.~ (1.79a,b)), the coordinates canonically
conjugate to the conserved quantities $P_{2,0},\dots P_{2,n-2}$ are given by
$$
\align
Q_{2,i}  &= \frac{1}{2}
\sum^{n-2}_{\mu=1} \int^{\lambda_\mu}_0 \frac{\lambda^i
d\lambda}{{\sqrt{-a(\lambda)\PP_2(\lambda)}}} ,\qquad i = 0,\dots,n-3
\tag{3.60a}\\
Q_{2,n-2}  &= \frac{1}{2}
\sum^{n-2}_{\mu=1} \int^{\lambda_\mu}_0
\frac{\lambda^{n-2}d\lambda}{{\sqrt{-a(\lambda)\PP_2(\lambda)}}} -
\frac{\ln  u}{2 P_{2}} ,
\tag{3.60b}
\endalign
$$
where the final term is derived using the relation  $P^2_{2} =
-P_{2,n-2}$  on  $X$. Thus, up to normalization of the $n-2$ holomorphic
differentials
appearing in eq.~ (3.60a), this represents the hyperelliptic Abel map
(cf.~Theorem 1.6)
 It follows from Hamilton's equations for $h=H_x$ and $H_t$ given in
(3.44a,b) that the  $x$  and $t$  dependence of the  $Q_i$'s is given by
$$
\align
Q_{2,i} &= c_i, \qquad i < n - 4 \tag{3.61a} \\
Q_{2,n-4} &= c_{n-4} - t \tag{3.61b} \\
Q_{2,n-3} &= c_{n-3} - x  \tag{3.61c} \\
Q_{2,n-2} &= c_{n-2}  ,  \tag{3.61d}
\endalign
$$
where  $c_0,\dots c_{n-2}$  are constants.  Expressing the singular (3rd kind)
abelian differential appearing in (3.60b) in terms of the Abel map and the
appropriate hyperelliptic $\theta$-function, we obtain, through the procedure
described in Corollary 1.7, the following explicit formula for $u(x,t)$
$$
u(x,t) =\text{exp}(q_2)= \tilde{K}\ \text{exp}(bx + ct)
\frac{\theta(\bold A(\infty_2,p)  + t\bold U + x\bold V -
\bold K)}{\theta(\bold A(\infty_1,p)  + t\bold U + x\bold V - \bold K)}  ,
\tag{3.62}
$$
where $\bold U, \bold V \in \Bbb C^{\wt g}$ are determined as in Theorem 1.6
from the Hamiltonians $h=H_x,\ H_t$, respectively, $b,\ c$ are determined as in
Corollary 1.7, and the integration constants
$\tilde{K}\in \Bbb C,\ p\in \Cal S$ are determined by initial data.
This agrees with the standard $\theta$-function solution of NLS (cf. {\bf
[P]}).

\noindent {\it  3d. \quad The CNLS System and Flows in}
 $\wt{\frak{su}}(1,2)^{+*}$ \hfill

   The CNLS system is a $2$-component generalization of eqs.~ (3.40a,b).
The real form we consider involves  two complex functions $u(x,t), \ v(x,t)$
satisfying the coupled system of equations:
$$
\align
iu_t + u_{xx} &=2u(|u|^2 + |v|^2) \tag{3.63a}
\\
iv_t + v_{xx} &=2v(|u|^2 + |v|^2)  . \tag{3.63b}
\endalign
$$
These can be obtained as the compatibility conditions for a pair of Lax
equations of the type (1.1) with $\Cal N(\lambda) \in
\widetilde{\frak{su}}(1,2)^{+*}$
of the form given in eq.~ (1.2) and $Y=0$, corresponding, as above,
to the Hamiltonians
$$
\align
 H_x(\Cal N) & =\frac12{\Big[\frac{a(\lambda)}{\lambda^{n-1}}
\rom{tr}(\Cal N(\lambda)^2) \Big]}_0 \tag{3.64a}\\
H_t(\Cal N) & =\frac12{\Big[\frac{a(\lambda)}{\lambda^{n-2}}
\rom{tr}(\Cal N(\lambda)^2)\Big]}_0  . \tag{3.64b}
\endalign
$$
If, as above, we set:
$$
\hat{\Cal L}(\lambda) =
 \frac{a(\lambda)}{\lambda}\Cal N(\lambda)
 = L_0\l^{n-1} + L_1\lambda^{n-2} + L_2\lambda^{n-3} + \dots + L_{n-1}  ,
\tag{3.65}
$$
then Hamilton's equations again take the Lax form
$$
\align
{d \over dx}\hat{\Cal L}(\l) = & [\l L_0 + L_1, \hat{\Cal L}(\l)]
\tag{3.66a} \\
{d \over dx} \hat{\Cal L}(\l)
= & [\l^2 L_0 + \l L_1 + L_2, \hat{\Cal L}(\l)] , \tag{3.66b}
\endalign
$$
and the flows commute.
If the following invariant constraints are imposed:
 $$
\align
 L_0 &= \frac{i}{3}\pmatrix
2&0&0\\0&-1&0\\0&0&-1
\endpmatrix \tag{3.67a}\\
\\
L_1 &=\pmatrix
0&\overline u&\overline v\\u&0&0\\v&0&0
\endpmatrix \tag{3.67b}\\
\\
L_2 &= i\pmatrix
\mid u \mid^2 + \mid v \mid^2&-\overline{u_x}&-\overline{v_x}\\
u_x &-\mid u\mid^2&-\overline {v}u\\
v_x &- {\overline u}v&-\mid v \mid^2
\endpmatrix  , \tag{3.67c}
\endalign
$$
the CNLS equations are obtained as the compatibility conditions for the
equations (3.66a,b).

It is possible, similarly to the NLS equation discussed above, to obtain
an intrinsic characterization of the orbit corresponding to residue matrices
$N_i$ of rank $k_i=1$ as an open, dense subset of $\Bbb C^{2n}$, viewed as a
real symplectic space, (cf.~ {\bf [AHP, AHH1]}). However, the approach
developed in
Section 1 allows us to treat all orbits of the type (1.3) on the same footing,
regardless of the rank $k_i=1,2,3$. Only the explicit formulae (1.27) for the
genus $\wt g$ of the spectral curve and (1.25a) for the degrees  of the
invariant polynomials will change.

By  Lemma 1.1 and Proposition 1.2,  the invariant spectral curve
is given by a polynomial equation  of the general form
$$
\text{det}(\hat{\Cal L}(\l) - z I)=
\Cal P(\lambda,z)= \Cal P_R (\lambda,z) + p(\lambda, z)=0 , \tag{3.68a}
$$
where
$$
\Cal P_R (\lambda,z) :=-z^3 +  z\Cal A_2(\l)\Cal P_{R2} (\l) +
\Cal A_3(\l)\Cal P_{R3} (\l) \tag{3.68b}
$$
defines a reference curve $\Cal S_R$, determined by the initial data for a
 particular solution of the CNLS system, and
$$
p(\lambda, z):= a(\lambda)
\left(z a_2 (\lambda) \sum_{a=0}^{\rho}\
P_{2a} \lambda^{a} + a_3 (\lambda)
\sum_{a=0}^{\sigma} P_{3a}
\lambda^{a}\right) \tag{3.68c}
$$
is of the form given by Proposition 1.2 for neighbouring curves, with
 $\rho = \delta_2,\ \sigma = \delta_3$, given by formulae (1.25a  ) and
$(P_{2a},\ P_{3b})_{a=0,\dots \rho, b=0,\dots \sigma}$ are the Poisson
commuting
spectral invariants. It will be convenient to reparametrize the space of
polynomials $p(\l,z)$ as follows. Set
$$
\align
a_2 (\lambda) \sum_{a=0}^{\rho}P_{2a} \lambda^{a}&= \hat{P}_{2,\rho}\l^{n-2}
+ \hat{P}_{2,\rho-1}\l^{n-3} + \cdots + \hat{P}_{2,0}\l^{n-2-\rho} +
(\text{lower\
order}) \tag{3.68d}\\
a_3 (\lambda) \sum_{a=0}^{\sigma}
P_{3a}\lambda^{a}&=\hat{P}_{3,\sigma}\l^{2n-3}
+ \hat{P}_{3,\sigma-1}\l^{2n-4} + \cdots + \hat{P}_{3,0}\l^{2n-3-\sigma} +
(\text{lower\
order}).\tag{3.68e}
\endalign
$$
Note that the lower order terms are completely determined by the terms
 $\hat P_{2a}, \hat P_{2a}$ explicitly appearing.

There can be singularities in general over the points with $\l=\a_i$, which are
assumed to be resolved as indicated in Section 1.
Due to the normalizations (3.67a-c) the spectral curve of
$\hat{\Cal L}(\lambda)$ also has singularities at $\lambda =\infty$.  In
terms of the coordinates $\tilde{\lambda} = 1/\lambda$,
the eigenvalues of $\lambda^{-m} \hat{\Cal L}(\lambda)$ have the
expansion around $\tilde{\lambda}=0$
$$\align
\tilde z_1(\tilde{\lambda}) &= {2i\over 3} - (m_2 +m_3)
\tilde{\lambda}^3 +\cdots\cr
\tilde z_2 (\tilde{\lambda})&= -{i\over 3} + m_2\tilde{\lambda}^3
+\cdots\cr
\tilde z_3 (\tilde{\lambda}) &= -{i\over 3} + m_3\tilde{\lambda}^3
+\cdots \tag{3.69}
\endalign
$$
where for generic $\hat{\Cal L}$, $ m_2\neq m_3$.  This gives the curve
$\SS$ a triple singularity (tacnode of order 3) at $\lambda
=\infty$, and the desingularisation $\tilde{\SS}$ is
3-sheeted over $\lambda =\infty$.  Furthermore,
$\hat{\Cal L}(\lambda)$ is diagonalisable in a neighbourhood of
$\lambda=\infty$.  In terms of the algebro-geometric
constructions of Section 2, this means that the sheaf $E$
of (2.1) is a direct image of a line bundle $\tilde{E}$
on $\tilde{\SS}$.

The genus  of the curve $\tilde{\SS}$ is $\wt{g}'=\rho +\sigma -3$, three
less than the genus $\wt{g}=\rho + \sigma$ given in formulae (1.26), (1.27)
for the generic spectral curve in the coadjoint orbit.  Similarily, the
degree of $\tilde{E}$ is $\rho + \sigma -1$, three less
than the generic case.  Accordingly, we have six fewer
divisor coordinates than in the generic case, and hence an insufficient
number to provide a Darboux system on the orbit $\Cal Q_{\Cal N}$.

The solution to this problem is to impose six extra
constraints on elements of the orbit, so that the
dimensions of the constrained submanifold coincides with
the number of coordinates.  This must be done in an
invariant way.  Note that, under the Lax equations
(3.66a,b), the matrix $\hat{\Cal L}(\lambda)$, $\lambda$ fixed,
evolves by conjugation.  In $Gl(n, {\bold C})$, the
spectrum is not the only invariant under conjugation;
when there are multiple eigenvalues, we also have the
different Jordan canonical forms.  For example, among
matrices with one double eigenvalue, the generic
coadjoint orbits have nondiagonal canonical form, and there are
orbits that are two dimensions smaller, consisting of diagonalisable
matrices.  With this model in mind, we impose the following further
invariant constraints, defining a $2(\rho + \sigma -1)$-dimensional symplectic
submanifold  $\Cal Q_s \ss \Cal{Q}_{\Cal{N}_0}^0$:

\noindent {\it (i)} The spectral curve has genus three
less than the generic curve in the orbit, and so has
three extra singularities, counted with multiplicity.\hfill (3.70a)

\noindent {\it (ii)}  At these extra singular points, $\hat{\Cal L}(\lambda)$
is
diagonalisable.\hfill (3.70b)

If ${\Cal U}$ denotes the space of triples (curves, sheaves, trivialisations
over infinity) obtained from the orbit $\Cal{Q}_{\Cal{N}_0}^0$, and ${\Cal W}$
is the corresponding  space of curves, with the projection map
$\delta: {\Cal U} \rightarrow {\Cal W}$, condition (3.70a)
restricts us to the inverse image under $\delta$ of a
codimension three subvariety of curves in ${\Cal W}$.  For
curves $\SS$ in this variety, $\delta^{-1} (\SS)$ is a
stratified space, with a generic stratum consisting of
line bundles (+ trivialisations at $\lambda =\infty)$ on
the singular curve $\SS$, and other strata
corresponding to direct images of line bundles on various
desingularisations of the curve at subsets of the three
singular points.  Constraint (3.70b) then restricts us
to the codimension three stratum in $\delta^{-1} (\SS)$ of
direct images of line bundles on the curve $\tilde{\SS}$
obtained from $\SS$ by desingularising all three points.

As remarked in section 2, it is irrelevant, for the purpose of integrating
the AKS flows, whether one uses the Kostant-Kirillov form
$\omega_{orb}$ or the Serre duality form $\omega_S$.
Let $(\lambda_{\mu}, \zeta_{\mu})_{\mu =1,...,\wt{g}'}$, be the divisor
coordinates and let $(q_i, P_i)_{i=2,3}$ be the ``extra'' Darboux coordinates
as in Section 1c.
\bigskip
\proclaim{Theorem 3.1}
The restriction to ${\Cal Q}_s$ of the form $\omega_S$ is
given, over a suitable dense set, by
$$\omega_S = \sum_{\mu=1}^{\tilde{g}'} d\lambda_\mu
\wedge d \zeta_\mu +\sum_{i=2}^3 dq_i \land dP_i
+ {1\over 2} \sum_{i\neq j}^{3} {dP_i \wedge d P_j\over P_i - P_j}
\tag{3.71}
$$
\endproclaim
\demo{Proof} One begins by noting that ${\Cal Q}_s$ can be described,
as in section 2, as a variety of generically smooth curves $\tilde{\SS}$,
along with line bundles $\tilde{E}$ (with trivialisations
at $\lambda =\infty$) defined over $\tilde{\SS}$.
Proceeding as in Section 2, we can define a symplectic
form $\omega_{S,s}$ on this space, using Serre duality.
As in the proof of Theorem 2.8, it follows that the
explicit form of $\omega_{S,s}$ is given by (3.71).
There remains only to show that $\omega_{S,s}$ is the
restriction to ${\Cal Q}_s$ of $\omega_S$.

 This is fairly easy to see.  Any line bundle $\tilde{E}$
over $\tilde{\SS}$ is the pull-back of a line bundle $E$ on
$\SS$.  Extending this bundle to a neighbourhood of $\SS$
gives a splitting of the map (curves, bundles, trivializations over
$\l=\infty$) $\lra$ (curves), both on the constrained submanifold $\Cal Q_s$
and on the ambient space ${\Cal Q}_{N_0}^0$.  Using this
splitting, we now write a pair of vector fields $\{V_i\}_{i=1,2}$ on
${\Cal Q}_{N_0}^0$ as $(v_i, e_i),\ v_i \in H^0 (\SS, K(1))$,
  $e_i \in H^1 (\SS, {\Cal O}(-1))$, where $\{V_i\}$ are tangent to
${\Cal Q}_s$ along ${\Cal Q}_s$.  The
$v_i$ are represented in Dolbeault cohomology at $\SS$
by holomorphic forms, with a simple pole at infinity.
Along ${\Cal Q}_s$, these stay finite at the
singularities; this is  the condition of
tangency.  The $e_i$, in turn are represented by $(0,1)$
forms, which vanish at infinity. On $V_1, V_2,\  \omega_S,\ \omega_{S,s}$ are
both given by
$$
\int_{\SS} v_1 \wedge e_2 - v_2 \wedge e_1  ,
$$
which is obviously well behaved in a neighbourhood of
$\Cal Q_s$.
\hfill$\square$
\enddemo

   The constraints (3.70a,b) and the relations between $P_2,P_3$ and the
coefficients of the polynomial $p(\l,z)$ imply that the coefficients
$P_{2,\rho}, P_{3,\sigma}$, $\ P_{3,\sigma -1}$,
$P_{3,\sigma -2}$, $\ P_{3 ,\sigma -3}$
in eq.~ (3.68c) can be expressed in terms of the lower coefficients
$(P_{2a}, P_{3b})$ $ _{a=0, \dots \rho -1,}$ $ _{b= 0, \dots \sigma -4}$ and
$P_2, P_3$. To apply the Liouville method, we only need to know the
constraints to order $1$ at the CNLS curve $\SS_R$ corresponding to the
spectral polynomial $\Cal P_R(\l, z)$.  The
constraint (3.70a), requiring  the neighboring curve to have the same degree of
singularity as $\SS_R$, is equivalent to first order
to requiring  the induced section of the normal bundle along $\SS_R$:
$$
{p(\lambda,z)\over \partial \Cal P_R /\partial z} \cdot
{\partial\over \partial z}  . \tag{3.72}
$$
to remain finite at the singular points of $\SS_R$.

At the CNLS curve this means that, passing to the
coordinates $\tilde{z}, \tilde{\lambda}$, the three first
terms in the Taylor expansion at $\tilde{\lambda}=0$ of the expression
$$
\tilde{\lambda}^{2n-3} \left(\tilde{z}
\tilde{\lambda}^{-n+1} a_2 (\tilde{\lambda}^{-1}) p_2
(\tilde{\lambda}^{-1}) + a_3 (\tilde{\lambda}^{-1}) p_3
(\tilde{\lambda}^{-1})\right)
=: \tilde{z} f_2 (\tilde{\lambda}) + f_3
(\tilde{\lambda})\tag{3.73}
$$
must vanish when one substitutes $\tilde{z} = -{i\over
3}$.  This yields the linearized constraints:
$$
\align
\hat P_{3,\sigma} &= {i\over 3} \hat P_{2,\rho} \tag{3.74a}\\
\hat P_{3,\sigma -1} &= {i\over 3} \hat P_{2,\rho-1}  \tag{3.74b}\\
\hat P_{3,\sigma -2} & = {i\over 3} \hat P_{2,\rho-2}. \tag{3.74c}
\endalign
$$

 Setting
$$
P_j =: -{i\over 3} + \hat{P}_j, \qquad
j=2,3\tag{3.75}
$$
 the linear variation of the $(\hat{P}_j)$ at the CNLS
curve can  be computed by evaluating the limits of the
normal vector field (3.72) along the two branches of
the curve as one approaches the
singular point. To first order:
$$
\hat{P}_j = \lim_{\tilde{\lambda} \rightarrow 0}
{\tilde{z}_j (\tilde{\lambda}) f_2 (\tilde{\lambda}) +
f_3 (\tilde{\lambda})\over (-\tilde{z}_j
(\tilde{\lambda}) + \tilde{z}_k (\tilde{\lambda}))
(-2\tilde{z}_j (\tilde{\lambda}) - \tilde{z}_k
(\tilde{\lambda}))} \tag{3.76}
$$
where $(j,k) = (2,3)$ or (3,\,2), and $\tilde{z}_j(\tilde{\lambda})$ is as in
(3.69). This gives
 $$
\hat{P}_2 + \hat{P}_3=if_2(0) = iP_{2,\rho}\tag{3.77}
$$
and, if $m_2, m_3$ are as in (3.69),
$$
\align
m_3 \hat{P}_2 + m_2 \hat{P}_3 &=
\lim_{\tilde{\lambda}\rightarrow 0} -{i\over
\tilde{\lambda}^3} \left(-{i\over 3} f_2 + f_3\right)
\\
{}&=-i
[-{i\over 3} (\hat P_{2,\rho-3}) +  \hat P_{3,\sigma-3}]
 , \tag{3.78}
\endalign
$$
where $c_i, d_i$ are constants.  Isolating $\hat{P}_{3,\sigma-3}$
above, and using (3.77-78) to rewrite (3.74a-c) we obtain
 $$
\align
\hat P_{2,\rho}&=-i \left(\hat{P}_2
+\hat{P}_3\right) \tag{3.79a}\\
\hat P_{3,\sigma}&= {1\over 3} (\hat{P}_2 + \hat{P}_3) \tag{3.79b}\\
\hat P_{3,\sigma -1} &= {i\over 3} \hat P_{2,\rho-1}   \tag{3.79c}\\
\hat P_{3,\sigma -2}&= {i\over3} \hat P_{2,\rho-2} \tag{3.79d} \\
\hat P_{3,\sigma -3} &= {i\over 3} \hat P_{2\rho-3}  +i (m_3 \hat{P}_2 +
 m_2 \hat{P}_3 ) . \ \tag{3.79e}
\endalign
$$
This expresses the terms
on the left in terms of the independent complete set of
integrals of motion $\hat P_{2,0},..,\hat P_{2,\rho-1},
\hat P_{3,0},.. , \hat P_{3,\sigma -4}$, $\hat P_2, \hat P_3$ (at least to
first order around the reference curve, which is all we need to
apply the Liouville method).  Since
$(\lambda_{\nu},z_{\nu})_{\nu =1,..,\wt{g}'}$, $(q_i, P_i)_{i=2,3}$
form a Darboux coordinate system ``up to constants of motion'', if we define
our generating function $S(\lambda_\nu, q_2, q_3, P_{2a}, P_{3b}, P_2, P_3)$
as in (1.76), then the canonically conjugate coordinates
$$
\bigg\{\hat Q_{2a} = {\partial S\over \partial
\hat P_{2a}} \bigg\}_{ a  = 0,..,\rho-1},\
\bigg\{\hat Q_{3b} = {\partial S\over \partial \hat P_{3b}}
\bigg\}_{ b = 0,..,\sigma -4},\ \bigg\{\hat Q_i = {\partial
S\over \partial \hat P_i}\bigg\}_{ i=2,3}.\tag{3.80}
$$
undergo linear flow:
$$
\align
\hat Q_{2a} &= c_{2a} -
\delta_{a,\rho-1} x - \delta_{a,\rho-2} t \\
\hat Q_{3b} &= c_{3b}\\
\hat Q_i &= c_i  , \tag{3.81}
\endalign
$$
where $c_{2a}, c_{3b}, c_i$ are constants. (Up to additive constants,
$H_x= -\hat P_{2,\rho-1},\ H_t = -\hat P_{2,\rho-2}$.)
Evaluating the derivatives (3.82), taking (3.79a-e) into
account, we obtain:
$$
\align
\hat Q_{2a}& =\sum_{\nu=1}^{\tilde{g}}
\int_0^{\lambda_\nu} {z \lambda^{n-2-\rho+a} + {i\over 3}
(\delta_{a,\rho-3} \lambda^{2n-6} +
\delta_{a,\rho-2} \lambda^{2n-5} +
 \delta_{a,\rho-1} \lambda^{2n-4})
 \over -3 z^2 + \Cal A_2(\lambda)
\Cal P_{R2} (\lambda)}
\ d\lambda  \tag{3.82a}\\
\hat Q_{3b} &=\sum_{\nu=1}^{\tilde{g}}
\int_0^{\lambda \nu} { \lambda^{2n-3-\sigma+b} \over
-3 z^2 + \Cal A_2 (\lambda ) \Cal P_{R2} (\lambda)} \  d\lambda
\tag{3.82b}\\
\hat Q_2 &= \ln u + I + im_3J  \tag{3.82c}\\
\hat Q_3 &= \ln v + I + im_2J ,  \tag{3.82d}
\endalign
$$
where $I, J$ are defined by
$$
\align
I&=\sum_{\nu=1}^{\tilde{g}} \int_0^{\lambda_\nu}
{-i z  \lambda^{n-2}  + {1\over 3} \lambda^{2n-3} \over
-3 z^2 + \Cal A_2 (\lambda) \Cal P_{R2} (\lambda)}\  d\lambda
\tag{3.83a}\\
J&= \sum_{\nu=1}^{\tilde{g}} \int_0^{\lambda_\nu}
{ \lambda^{2n-6}\over -3 z^2 +\Cal A_2 (\lambda)
\Cal P_{R2} (\lambda) }\ d\lambda . \tag{3.83b}
\endalign
$$
 We can check explicitly that the integrands of (3.82a,b)  form a
basis for the holomorphic differentials on the curve $\tilde S$ which
has been desingularized over $\l=\a_i$ and over $\l=\infty$. If
$\infty_1,\infty_2,\infty_3$ are the three points of $\tilde S$ over
$\l=\infty$
corresponding to $\tilde z= \frac{2i}3 ,\frac{-i}3, \frac{-i}3$ respectively,
the integrands of $I+im_3J$, $I+im_2J$ have only simple poles over $\l =\infty$
with residues $(-1,1,0)$ and $(-1,0,1)$ respectively at $(\infty_1,
\infty_2,\infty_3)$. Proceeding as in Corollary 1.7, we obtain the
$\theta$-function formulae
$$
\align
u(x,t) &= \text{exp}(q_2) = \tilde{K_2}\ \text{exp}(e_2x + d_2t)
\frac{\theta(\bold A(\infty_2,p)  + t\bold U + x\bold V -
\bold K)}{\theta(\bold A(\infty_1,p)  + t\bold U + x\bold V - \bold K)}
 \tag{3.84a}\\
v(x,t) &= \text{exp}(q_2) = \tilde{K_3}\ \text{exp}(e_3x + d_3t)
\frac{\theta(\bold A(\infty_3,p)  + t\bold U + x\bold V -\bold K)}
{\theta(\bold A(\infty_1,p)  + t\bold U + x\bold V - \bold K)}  ,
\tag{3.84b}
\endalign
$$
with $\bold U,\ \bold V \in \Bbb C^{\wt{g}'}$ determined from the
Hamiltonians $h= H_x,\ H_t$ as in Theorem 1.6, $(e_i,d_i)_{i=2,3}$ as
in Corollary 1.7, and the remaining integration constants
 determined to satisfy the appropriate initial conditions.

\newpage
\centerline{\smc References}
\bigskip
{\smaller{
\item{\bf [A]} Adler, M., ``On a Trace Functional for Formal
Pseudo-Differential Operators and the Symplectic Structure of the Korteweg-
de Vries Equation'', {\it Invent\. Math\.} {\bf 50}, 219-248, (1979).
\item{\bf [AHH1]} Adams, M.R., Harnad, J. and Hurtubise, J., ``Isospectral
Hamiltonian Flows in Finite and Infinite Dimensions II.  Integration of
Flows'',
 {\it Commun\. Math\. Phys\.} {\bf 134}, 555-585 (1990).
\item{\bf [AHH2]} Adams, M.R., Harnad, J. and Hurtubise, J.,
``Dual Moment Maps to Loop Algebras'', {\it Lett\. Math\. Phys\.} {\bf 20},
 294-308 (1990).
\item{\bf [AHH3]} Adams, M.R., Harnad, J. and Hurtubise, J.,
 ``Integrable Hamiltonian Systems on Rational Coadjoint
Orbits of Loop Algebras'', in: {\it Hamiltonian Systems, Transformation
Groups and Spectral Transform Methods}, ed. J. Harnad and J. Marsden,
Publ. C.R.M., Montr\'{e}al  (1990).
\item{\bf [AHH4]} Adams, M.R., Harnad, J. and Hurtubise, J.,
``Liouville Generating Function for Isospectral Hamiltonian
Flow in Loop Algebras'', in:  {\it Integrable and
Superintergrable Systems}, ed. B. Kuperschmidt, World Scientific,
Singapore (1990).
\item{\bf [AHH5]} Adams, M.R., Harnad, J. and Hurtubise, J.,
``Coadjoint Orbits, Spectral Curves and Darboux Coordinates'',
in: {\it The Geometry of Hamiltonian Systems}, ed\. T. Ratiu,
Publ. MSRI Springer-Verlag, New York (1991).
\item{\bf [AHP]} Adams, M.R., Harnad, J. and Previato, E., ``Isospectral
Hamiltonian Flows in Finite and Infinite Dimensions I. Generalised Moser
Systems
and Moment Maps into Loop Algebras'',  {\it Commun\. Math\. Phys\.} {\bf 117},
451-500 (1988).
 \item{\bf[AvM]} Adler, M. and van Moerbeke, P., ``Completely Integrable
Systems, Euclidean Lie Algebras, and Curves'', {\it Adv\. Math\.}
{\bf 38}, 267-317 (1980); ``Linearization of Hamiltonian Systems, Jacobi
Varieties and Representation Theory'', {\it ibid\.} {\bf 38}, 318-379 (1980).
\item{\bf [B]} Beauville, A., ``Jacobiennes des courbes spectrales et
syst\`emes
Hamiltoniens  compl\`etement \newline
int\'egrables'', {Acta\. Math\.} {\bf 164},
 211-235 (1990).
\item{\bf [BS1]} Beals, R. and  and Sattinger, D.H.,
``On the Complete Integrability of Completely Integrable Systems'',
{\it Commun\. Math\. Phys\.} {\bf 38}, 409-436 (1991).
\item{\bf [BS2]} Beals, R. and  and Sattinger, D.H.,
``Action--Angle Variables for the Gel'fand--Dikii Flows'', preprint (1991).
\item{\bf [C]} Clebsch, A., {\it Le\c cons cur la G\'eom\'etrie}, Vol.3,
Gauthier-Villars, Paris (1883).
\item{\bf [D]} Dickey, L.A., ``Integrable Nonlinear Equations and Liouville's
Theorem I, II'', {\it Commun\. Math\. Phys\.}
{\bf 82}, 345-360 (1981); {\it ibid.} {\bf 82}, 360-375 (1981).
\item{\bf [Du]} Dubrovin, B.A., ``Theta Functions and Nonlinear Equations'',
 {\it Russ\. Math\. Surv\.} {\bf 36}, 11-92 (1981).
\item{\bf [F]} Flaschka, H., ``Toward an Algebro-Geometrical Interpretation
of the Neumann System'',
{\it Tohoku Math\. J\.} {\bf 37}, 407-426 (1984).
\item{\bf [FRS]} Frenkel, I.B., Reiman, A.G. and Semenov-Tian-Shansky,
M.A., ``Graded Lie Algebras and Completely Integrable Hamiltonian Systems'',
{\it Sov\. Math\. Doklady} {\bf 20}, 811-814 (1979).
\item{\bf [FT]} Faddeev, L.D\. and Takhtajan, L.A.,
  {\it Hamiltonian Methods in the Theory of Solitons},
 Springer-Verlag, Heidelberg (1987).
\item {\bf [GHHW]}  Gagnon, L., Harnad, J., Hurtubise, J. and  Winternitz, P.,
 ``Abelian Integrals and the Reduction Method for an
Integrable Hamiltonian System''
{\it J. Math. Phys\.} {\bf 26}, 1605-1612 (1985).
 \item {\bf [GD]} Gel'fand, I.M. and
Dik'ii, L.A., ``Integrable Nonlinear Equations and the Liouville Theorem'',
{\it Funct\. Anal\. Appl\.} {\bf 13}, 8-20 (1979).
\item{\bf [GH]} Griffiths, P. and Harris, J., {\it Principles of Algebraic
Geometry},  Wiley, New York (1978).
\item{\bf [H]} Harnad, J., ``Uses of Infinite Dimensional Moment Maps'',
in: {\it Proceedings of XVIIth International Colloquium on Group Theoretical
Methods in Physics}, pp\. 68-89, ed. Y\. Saint-Aubin and L\. Vinet,
World Scientific, Singapore (1989).
\item{\bf [HHM]} Harnad, J.,  Hurtubise, J. and Marsden, J.E.,
``Reduction of Hamiltonian Systems with Discrete Symmetries'',
preprint CRM (1991).
\item{\bf [Hi]} Hitchin, N.,  ``On the Construction of Monopoles''
{\it Commun\. Math\. Phys\.} {\bf 89}, 145-190 (1983).
\item{\bf [Kn]} Kn\"orrer, ``Geodesics on Quadrics and a  Mechanical Problem of
C. Neumann'', {\it J\. Reine Angew\. Math\.} {\bf 334}, 69-78 (1982).
 \item{\bf [K]} Krichever, I.M.,
``Methods of Algebraic Geometry in the Theory of Nonlinear Equations'', {\it
Russ\. Math\. Surveys} {\bf 32}, 185-213 (1977).
\item{\bf [KN]} Krichever, I.M.and  Novikov, S.P.,  ``Holomorphic Bundles over
Algebraic Curves and Nonlinear Equations'', {\it Russ\. Math.~ Surveys}
{\bf 32}, 53-79 (1980).
\item{\bf [Ko]} Kostant, B., ``The Solution to a Generalized Toda Lattice and
Representation Theory'', {\it Adv\. Math\.} {\bf 34}, 195-338 (1979).
\item{\bf [vMM]} van Moerbeke, P. and Mumford, D.,
 ``The Spectrum of Difference Operators and Algebraic Curves'',
{\it Act Math\.} {\bf 143}, 93-154 (1979).
\item{\bf [M]}  Moser, J.,  ``Geometry of Quadrics and Spectral Theory", {\it
The Chern Symposium, Berkeley, June 1979}, 147-188, Springer, New York, (1980).
\item{\bf [N]} Neumann, C., ``De problemate quodam mechanico, quod ad primam
integralium ultraellipticorum classem revocatur'',
 {\it J. Reine Angew\. Math\.} {\bf 56}, 46-63 (1859).
\item{\bf [P]} Previato, E., ``Hyperelliptic quasi-periodic and soliton
solutions of the nonlinear Schr\"odinger equation.'' {\it Duke Math\. J\.}
{\bf 52}, 329-377 (1985).
 \item{\bf [Ra1]} Ratiu, T.
``The C. Neumann Problem as a Completely Integrable System on an Adjoint
Orbit'',
{\it Trans\. A.M.S\. }{\bf 439}, 321-329 (1981).
\item{\bf [Ra2]} Ratiu, T. ``The Lie Algebraic Interpretation of the
Rosochatius System'' in: {\it  Mathematical Methods in Hydrodynamics and
Integrability in Dynamical Systems, A.I.P\. Conf. Proc\.} {\bf 88},
 La Jolla (1981).
 \item{\bf [RS]} Reiman, A.G., and Semenov-Tian-Shansky,
M.A., ``Reduction of Hamiltonian systems, Affine Lie algebras and Lax Equations
I,
II'', {\it Invent\. Math\.} {\bf 54}, 81-100 (1979);  {\it ibid\.}  {\bf 63},
423-432
(1981).
\item {\bf [S]} Symes, W., ``Systems of Toda Type, Inverse Spectral Problems
and Representation Theory'', {\it Invent\. Math\.} {\bf 59}, 13-51(1980).
\item {\bf [Sch]} Schilling, R., ``Generalizations of the Neumann
System - A Curve Theoretic Approach'',
{\it Commun\. Pure Appl\. Math\.} {\bf 40}, 455-522 (1987).
\medskip}}
\vfil\eject
\enddocument